\documentclass[pra,twocolumn]{revtex4}
\usepackage{amsfonts,amsmath,amssymb,float}
\usepackage{bm,dsfont}
\usepackage{color}
\usepackage{bbm}
\usepackage[dvips]{epsfig}
\usepackage{amsmath,amssymb,lscape,float}
\usepackage{hyperref}
\usepackage{listings}
\usepackage{lipsum}
\usepackage{diagbox}

\usepackage{graphicx}
\usepackage{epstopdf}
\usepackage{dcolumn}
\usepackage{bm}
\usepackage{amsfonts,amsmath,amssymb
}
\usepackage{todonotes}
\usepackage{braket}
\usepackage{verbatim}
\usepackage{siunitx}
\usepackage{xcolor}

\renewcommand{\vec}[1]{\mathbf{#1}}
\renewcommand{\vec}[1]{\ensuremath{\boldsymbol #1 }}

\newcommand{\mrm}{\mathrm}

\DeclareSIUnit\betaz{\beta_\mrm{z}}
\DeclareSIUnit\Er{\mrm{E_{\textrm{R}}}}
\DeclareSIUnit\SystemEnergy{\hbar\mrm{\omega_z}}
\DeclareSIUnit\SystemTime{\mrm{\omega_z}^{-1}}

\begin{document}
\title{Coherent Atom Transport via Enhanced Shortcuts to Adiabaticity: 
Double-Well Optical Lattice}

\author{Sascha H. Hauck and Vladimir M. Stojanovi\'c} 
\affiliation{Institut f\"{u}r Angewandte Physik, Technical
University of Darmstadt, D-64289 Darmstadt, Germany}

\date{\today}
\begin{abstract}
Theoretical studies of coherent atom transport have as yet mainly been restricted to
one-dimensional model systems with harmonic trapping potentials. Here we investigate this
important phenomenon -- a prerequisite for a variety of quantum-technology applications based
on cold neutral atoms -- under much more complex physical circumstances. More specificially
yet, we study fast atomic transport in a moving {\em double-well optical lattice}, whose
three-dimensional (anharmonic) potential is nonseparable in the $x-y$ plane. We first propose 
specific configurations of acousto-optic modulators that give rise to the moving-lattice effect 
in an arbitrary direction in this plane. We then determine moving-lattice trajectories that enable
single-atom transport using two classes of quantum-control methods: shortcuts to adiabaticity (STA),
here utilized in the form of inverse engineering based on a quadratic-in-momentum dynamical
invariant of Lewis-Riesenfeld type, and their recently proposed modification termed enhanced
STA (eSTA). Subsequently, we quantify the resulting single-atom dynamics by numerically solving
the relevant time-dependent Schr\"{o}dinger equations and compare the efficiency of STA- and
eSTA-based transport by evaluating the respective fidelities. We show that -- except for the
regime of shallow lattices -- the eSTA method consistently outperforms its STA counterpart.
This study has direct implications for neutral-atom quantum computing based on collisional 
entangling two-qubit gates and quantum sensing of constant homogeneous forces via guided-atom 
interferometry.
\end{abstract}

\maketitle
\section{Introduction} \label{Intro}
Fast, nearly lossless, transport of cold neutral atoms~\cite{Couvert+:08,Murphy+:09,Masuda+Nakamura:10,Chen+:10,
Torrontegui+:11,Chen+:11,Stefanatos+Li:14,Ness+:18,Lu+:20,Hickman+Saffman:20,Ding+:20,Lam+:21} is of paramount 
interest for emerging quantum technologies~\cite{Adams+:20}, such as quantum sensing based on  
atom interferometry~\cite{Navez+:16,Dupont-Nivet+:16,Rodriguez-Prieto+:20} and neutral-atom quantum 
computing (QC)~\cite{Jaksch+:99,Briegel+:00,Weitenberg+:11,MohrJensen+:19,Nemirovsky+Sagi:21,Henriet+:20,Morgado+Whitlock:21,
ShiREVIEW:22}. It is also a prerequisite for analog simulation~\cite{Morgado+Whitlock:21} and quantum-state 
engineering in neutral-atom ensembles~\cite{XFShi:20,StojanovicPRA:21,HaaseEtAl}. In particular, coherent atom 
transport~\cite{Qi+:21} enabled by moving the confining optical trap -- either an optical lattice or a 
tweezer~\cite{Kuhr+:03,Sauer+:04,Lengwenus+:10,Barredo+:18} -- has attracted considerable attention quite recently, 
both theoretically and experimentally~\cite{Hickman+Saffman:20,Lam+:21,Hauck+:21,Bluvstein+:22}. While the renewed interest 
on the experimental side is largely interwoven with the recent progress pertaining to the scalability of optically-trapped 
neutral-atom systems~\cite{deMello+:19,Browaeys+Lahaye:20}, increased theoretical activity is motivated in part by 
the availability of a powerful family of control protocols based on STA~\cite{STA_RMP:19} and their recently proposed
modification known as eSTA~\cite{Whitty+:20,Whitty+:22}.

In single-atom transport one aims for a final atomic state that closely matches the initial one
in the rest frame of the moving trap (up to an irrelevant global phase), which is equivalent to demanding
minimization (or, in the ideal case, complete absence) of vibrational excitations at the end of transport. This
last requirement, however, does not necessarily rule out the existence of transient excitations at intermediate
times~\cite{Torrontegui+:11}, a circumstance that motivates one to design moving-trap trajectories using STA-based 
control protocols~\cite{STA_RMP:19}. Namely, STA protocols typically lead to the same final states as their adiabatic 
counterparts, but require significantly shorter times to accomplish that. This, in turn, alleviates the debilitating 
effects of noise and decoherence. Given that adiabatic changes of control parameters typically leave some dynamical 
properties of the system invariant, inverse-engineering techniques based on Lewis-Riesenfeld invariants 
(LRIs)~\cite{Lewis+Riesenfeld:69} proved to be the most pervasive ones among STA methods.

Theoretical investigations of single-atom transport have as yet mostly been restricted to one-dimensional 
(1D) systems~\cite{Torrontegui+:11,ZhangEtAl:15} with purely harmonic confining potentials~\cite{Murphy+:09,
Chen+:11}. Yet, such simplifications often do not apply in realistic systems; either there is a significant coupling
between the longitudinal and transverse degrees of freedom or the relevant confining potential is anharmonic.
To bridge the gap between realistic experimental features and state-of-the-art theoretical studies, we have quite
recently addressed atom transport~\cite{Hauck+:21} in moving optical lattices (conveyor belts)~\cite{Schrader+:01,Kuhr+:03,
Sauer+:04,Dotsenko+:05,Fortier+:07} using both STA- and eSTA methods. Taking into account the full anharmonic,
3D optical potential of conveyor belts, we have demonstrated the superiority of the eSTA-based atom transport 
protocols compared to STA-based ones for all but the lowest optical-lattice depths~\cite{Hauck+:21}.

Attempting to demonstrate the possibility of a time-efficient atom transport under even more complex physical
circumstances, in this paper we investigate an optical lattice characterized by a {\em nonseparable} potential~\cite{Blakie+Clark:04}.
More precisely, we study both STA- and eSTA-based single-atom transport in a moving {\em double-well optical
lattice} (DWOL), a 3D system whose corresponding potential in the $x-y$ plane cannot be written
as a sum of $x$- and $y$-dependent contributions~\cite{Strabley:06,Anderlini:06}. Initially introduced with
the aim of realizing a controlled entanglement of atoms in isolated pairs~\cite{StrableyDW:07,LeeDW:07,Anderlini:07},
DWOLs have subsequently been proposed as a testbed for a number of interesting quantum-coherent phenomena with
cold atoms. Examples include metastable incommensurate superfludity~\cite{Stojanovic+:08,Paul+Tiesinga:13}, 
superfluid-to-Mott-insulator quantum phase transitions~\cite{Danshita:07,Zhou++:11}, superfluid-drag phenomena~\cite{Hofer+:12}, 
nonlinear looped band structure of Bose-Einstein condensates~\cite{Koller+:16}, to name but a few. The nontrivial geometrical
structure of DWOLs gives rise to unconventional features in the single-particle energy spectrum. For instance, a
parameter regime exists where the first excited Bloch band of a DWOL has degenerate energy minima at nonzero 
quasimomenta~\cite{Stojanovic+:08}. Such band minima at nonzero quasimomenta are quite a rare occurrence in 
other physical systems, where they typically do not have a purely kinetic origin but are instead brought about 
by the presence of interactions~\cite{Stojanovic+:14,Stojanovic+Salom:19,Stojanovic:20,StojanovicPRL:20}.

We first propose specific configurations of acousto-optic modulators (AOMs) that give rise to the moving-lattice effect
in an arbitrary direction in the $x-y$ plane of a DWOL. We subsequently model coherent atom transport using two classes
of control protocols: STA-type ones, here obtained using inverse-engineering techniques based on LRIs, as well as those
originating from eSTA. Having determined the trajectories of the moving lattice using these two state-of-the-art methods,
we compute the resulting single-atom dynamics by numerically solving the corresponding time-dependent Schr\"{o}dinger
equations using the Fourier split-operator method. Finally, we quantify and compare the efficiency of STA- and eSTA-based
atomic transport by evaluating the corresponding fidelities for a broad range of values of the relevant parameters (lattice
depths, transport distances, etc.).

We demonstrate that both STA and eSTA allow time-efficient atom transport in DWOLs. In particular, while STA provides somewhat
faster transport in shallow lattices (i.e. for relatively low lattice depths), eSTA becomes superior to it for larger lattice
depths. Our numerical calculations show that this superiority of eSTA approach compared to its STA counterpart is more pronounced
for shorter transport distances. In other words, for larger transport distances one also needs larger lattice depths for eSTA-based
atom transport to be faster than its STA-based counterpart. This is an important conclusion from the methodological standpoint
as -- due to its heuristic character -- the eSTA method cannot {\em a priori} be expected to consistently outperform STA.

The outline of the remainder of this paper is as follows. In Sec.~\ref{system} we first provide the basic background 
of the atom-transport problem, specifying the full Hamiltonian governing such transport in a DWOL and the relevant 
distance, time, and energy scales. We then introduce DWOLs, specifying first their full 3D optical potential and then 
proposing configurations of AOMs that enable single-atom transport in this system. In Sec.~\ref{MovingTrap_STA} we
discuss the methodology for obtaining STA-based trajectories of a moving DWOL and present typical resulting trajectories. 
The following Sec.~\ref{MovingTrap_eSTA} addresses the same problem using the eSTA method. In Sec.~\ref{AtomDynamics}, 
after briefly introducing our chosen computational method (the Fourier split operator method) for evaluating single-atom
dynamics, we describe the use of the same method in conjunction with the imaginary-time evolution approach for obtaining 
the ground state of the system. We then discuss the methodology for solving the time-dependent Schr\"{o}dinger 
in the comoving reference frame, followed by the relevant details of our numerical implementation thereof. The results 
obtained for the atom-transport fidelity within both STA and eSTA methods are presented and discussed in Sec.~\ref{ResultsDiscussion}. 
The quantum-technology implications of fast atom transport in DWOLs are discussed in detail in Sec.~\ref{QuantTechImpl}. 
Before closing, we provide a short summary of the paper -- along with the statement of the main conclusions drawn -- in 
Sec.~\ref{SummaryConclusions}. Some cumbersome mathematical details are relegated to Appendices~\ref{SecDWOPGn}, 
\ref{SecDWOPKn} and \ref{derivationTrp}.

\section{DWOL and moving-lattice effect} \label{system}
In the following, we will be concerned with the problem of transporting an atom of mass $m$ (e.g.
of $^{87}$Rb) located at the time $t=0$ in one of the lattice-potential minima of a DWOL, to another  
minimum at a distant location in the $x-y$ plane. In Sec.~\ref{SingleAtomTransportDWOL} below we 
briefly describe the background of this problem and introduce the relevant spatial, temporal, 
and energy scales. We then introduce the main features and characteristic parameters of DWOLs 
in Sec.~\ref{DWOLpotential}, while in Sec.~\ref{AOMconfig} we discuss specific configurations
of AOMs in this system that give rise to the moving-lattice effect in different spatial directions.
\subsection{Single-atom transport in DWOLs} \label{SingleAtomTransportDWOL}
In line with the conventional atom-transport phenomenology, the transport distance $d$ will be assumed
in what follows to be at least an order of magnitude larger than the size $l_0$ of the atomic ground-state 
wave packet, i.e. $d\gtrsim 10\:l_0$. The relevant Hamiltonian describing single-atom transport in the 
$x-y$ plane of a DWOL is given by
\begin{equation}\label{FullSingleAtomHamiltonian}
H_{\textrm{D}}(t)=-\frac{\hbar^2\nabla^2}{2m} + 
U_{\mathrm{D}}\left[\mathbf{r}-\mathbf{q}_0(t)\right]\:,
\end{equation}
where $U_{\mathrm{D}}(x,y,z)$ is the full optical potential of a DWOL (cf. Sec.~\ref{DWOLpotential} below)
and $\mathbf{q}_0(t)\equiv\left[q_\mathrm{0,x}(t),q_\mathrm{0,y}(t),0\right]^\intercal$ the time-dependent 
vector describing the moving-DWOL trajectory. In Secs.~\ref{MovingTrap_STA} and \ref{MovingTrap_eSTA} below 
we discuss the design of such trajectories using STA- and eSTA methods, respectively, that enable fast atomic 
transport in this system. While our approach -- as emphasized in Sec.~\ref{AOMconfig} -- enables transport 
along an arbitrary direction in the $x-y$ plane, we will mostly be interested in the $x$, $y$, and diagonal 
directions. In what follows, the transport distances in these three directions will be denoted by $d_\mathrm{x}$, 
$d_\mathrm{y}$, and $d_\mathrm{r}$, respectively.

To facilitate further discussion, we specify at this point the characteristic time-, length-, and energy scales
in the problem at hand. The oscillation periods $T_\mathrm{x}\equiv 2\pi/\omega_\mathrm{x}$ and $T_\mathrm{y}\equiv 
2\pi/\omega_\mathrm{y}$ that correspond to the harmonic potential approximating the full DWOL optical potential [cf. Sec.~\ref{DWOLpotential}] 
in the $x$- and $y$ directions will serve as the relevant time scales characterizing atom transport in the respective 
directions. At the same time, the corresponding harmonic-oscillator zero-point lengths $l_{\textrm{x}}\equiv\sqrt{\hbar
/(2m\omega_\mathrm{x})}$ and $l_{\textrm{y}}\equiv\sqrt{\hbar/(2m\omega_\mathrm{y})}$ will serve as the characteristic 
lengthscales. Finally, all energies in the problem under consideration will be expressed in units of the recoil energy 
$E_{\textrm{R}}\equiv\hbar^2 k_{\textrm{L}}^2/(2m)$, where $k_{\textrm{L}}$ is the magnitude of the relevant laser wave 
vector $\mathbf{k}_{\textrm{L}}$.

It is worthwhile to mention that atom transport in DWOLs was investigated more than a decade ago using the idea
of adjustable optical potentials in conjunction with optimal-control methods~\cite{DeChiara++:08}. Yet, this previous 
study only considered atom transport between neighboring sites of a DWOL. In contrast to this previous study, our 
present approach is designed so as to facilitate coherent atom transport over an -- in principle -- arbitrary distance
in the $x-y$ plane of a DWOL. In addition, this approach is also based on a completely different physical mechanism 
(the moving-lattice effect enabled by AOMs) and entails different theoretical methodologies for designing specific 
transport trajectories (the STA and eSTA methods instead of the optimal-control techniques).

\subsection{DWOL: geometric structure and underlying optical potential} \label{DWOLpotential}
The DWOL is constructed from two 2D lattices with different periods, thus its unit cell -- whose orientation 
can be changed -- contains two sites. Importantly, the height of the barrier between the two sites and their 
relative depth (``tilt'') are controllable. DWOLs entail a nonseparable optical double-well potential in the 
$x - y$ plane and an independent, conventional optical potential in the $z$-direction that provides 3D confinement. 
In particular, the potential in the $x - y$ plane consists of two components, which originate, respectively, 
from ``in-plane''-polarized light (with intensity $I_{\textrm{xy}}$) and light polarized in the $z$-direction 
(with intensity $I_{\textrm{z}}$).

Using the general expression for the light intensity~\cite{ShoreBOOK:90}
\begin{equation}\label{eqIntensity}
I(x,y,z) = c\epsilon_0 \:|\mathbf{E}(x,y,z)|^2 \:,
\end{equation}
and adopting the paraxial approximation $w_\mathrm{0,x/y}, \gg k_\mathrm{z}^{-1}$
as well as the assumption that $|z|\gg Z_\mathrm{R}^\mathrm{x/y}$, one finds the 
total intensity in the form 
\begin{eqnarray}\label{eqIntensityDWOL}
I_\mathrm{r}(x,y,z)/I_\mathrm{xy} &=& \cos^2\left(\frac{\beta}{2}\right)\:
\tilde{I}_\parallel + \sin^2\left(\frac{\beta}{2}\right)\:\tilde{I}_\perp \nonumber\\
&+& \xi_\mathrm{z} \, \tilde{I}_\mathrm{z}+ \sqrt{\xi_\mathrm{z}}\:
\cos\left(\frac{\beta}{2}\right)\:\tilde{I}_\parallel^z  \:.
\end{eqnarray}
Here $I_\mathrm{xy}\equiv 2\:c\epsilon_0\:|E_\mathrm{xy}|^2$, with $E_\mathrm{xy}$ being 
the electric-field amplitude of the lasers forming the 2D DWOL in the $x - y$ plane.
Different contributions to the total intensity in Eq.~\eqref{eqIntensityDWOL} are given by
\begin{eqnarray}
\tilde{I}_\parallel &=& \cos\big[k_\mathrm{L}\left(x+\frac{\pi}
{2k_{\mathrm{L}}}\right)-2 \theta_\mathrm{r}-2\phi_\mathrm{r}\big] \nonumber \\
&+& \cos\big(2k_\mathrm{L} y +2\phi_\mathrm{r}\big)+2  \:, \label{eqIntens1}\\
\tilde{I}_\perp &=&  2\:\Big\{\cos\big[k_\mathrm{L}\left(x+ \frac{\pi}{2 k_\mathrm{L}}
\right)-\theta_\mathrm{z}-\phi_\mathrm{z}\big] \nonumber \\
&+& \cos\big(k_\mathrm{L} y + \phi_\mathrm{z}\big)\Big\}^2  \:, \label{eqIntens2}\\
\tilde{I}_\mathrm{z} &=& \frac{w_\mathrm{0,x} w_\mathrm{0,y} }{ w_\mathrm{y}(z)w_\mathrm{x}(z) } \,
\cos^2(k_\mathrm{z} z) \nonumber \\
&\times& \mathrm{exp}\left(-2\left[\frac{x^2}{w_\mathrm{x}^2(z)}+\frac{y^2}{w_\mathrm{y}^2(z)}
\right] \right) \:, \label{eqIntens3} \\
\tilde{I}_\parallel^z  &=& 2 \, \sqrt{\frac{w_\mathrm{0,x} w_\mathrm{0,y} }{ w_\mathrm{y}(z)
w_\mathrm{x}(z) }} \, \,
\exp \left( - \left[ \frac{ x^2 }{ w_\mathrm{x}^2(z) } + \frac{ y^2 }{ w_\mathrm{y}^2(z) }
\right] \right) \nonumber \\
\hspace{0.5cm} &\times& \cos\left(k_\mathrm{z} z\right) \left[\sin\left(\frac{\phi}{2}\right) \:
\tilde{E}^\mathrm{x}_\parallel+\cos \left(\frac{\phi}{2}\right)\:\tilde{E}^\mathrm{y}_\parallel 
\right] \label{eqIntens4} \:.
\end{eqnarray}
The $z$-dependent transverse beam waists $w_{u}(z)$ ($u=x,y$) in Eqs.~\eqref{eqIntens3} and 
\eqref{eqIntens4} are given by
\begin{equation}\label{zDependentWaists}
w_{u}(z)=w_{u,0}\:\sqrt{\displaystyle
1+\left(\frac{z}{Z_{R,u}}\right)^2} \:,
\end{equation}
with $Z_{R,x}$ and $Z_{R,y}$ being the respective Rayleigh lengths. At the same time, $\tilde{E}^\mathrm{x}_\parallel 
\equiv E^\mathrm{x}_\parallel/E_\mathrm{xy}$ and $\tilde{E}^\mathrm{y}_\parallel \equiv E^\mathrm{y}_\parallel/
E_\mathrm{xy}$ in Eq.~\eqref{eqIntens4} stand for the dimensionless electric-field components in the $x$- and $y$ 
directions, respectively, corresponding to the the $x-y$-plane polarization:
\begin{eqnarray}\label{eqElecticFields}
\tilde{E}^\mathrm{x}_\parallel &=&
\exp \left(2 i \theta_\mathrm{r}\right) \, \exp \left[- i k_\mathrm{L} \left(x +
\frac{\pi}{2 k_\mathrm{L}}\right)\right]  \, \exp \left(2 i \phi_\mathrm{r}\right) \nonumber \\
&+&
\exp \left[ i k_\mathrm{L}  \left(x+ \frac{\pi}{2 k_\mathrm{L}}\right)\right] \:,
\\
\tilde{E}^\mathrm{y}_\parallel &=&
\exp \left( i \theta_\mathrm{r}\right) \left[ \exp \left(-i k_\mathrm{L} y\right)
+
\exp \left( i k_\mathrm{L} y\right) \, \exp \left(2 i \phi_\mathrm{r}\right)\right] \nonumber\:.
\end{eqnarray}

\begin{figure}[t!]
\includegraphics[width=0.95\linewidth]{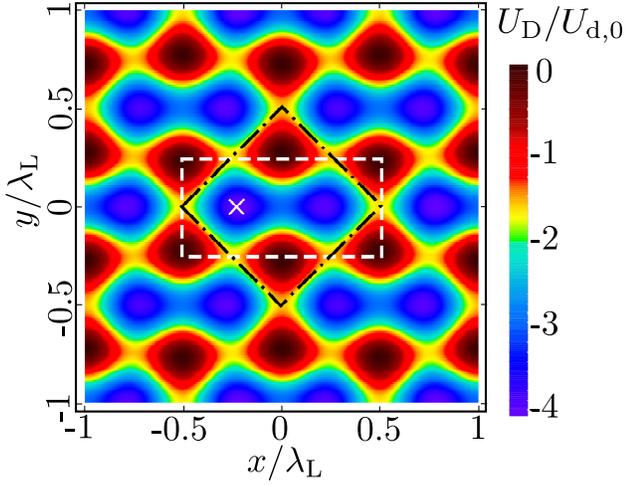}
\caption{\label{DWOLtransport_Fig1}(Color online) Density plot of the optical potential
of a DWOL in the $x-y$ plane. Two different choices for the unit cell of this lattice
are indicated with the dashed black- and white lines. The point around which the harmonic
approximation of the full DWOL potential is developed was denoted by ``X.''}
\end{figure}

The angles $\phi_\mathrm{r/z}$ and $\theta_\mathrm{r/z}$ in Eqs.~\eqref{eqIntens1}, \eqref{eqIntens2} and 
\eqref{eqElecticFields} correspond to phase-shifts and represent a remnant of generally polarization-dependent 
paths, as indicated by the subscripts $r$ and $z$ for path differences pertaining to the in- and out-of-plane 
light, respectively. These phases can most easily be understood from Fig.~\ref{fig:Setup1} below. While the angle 
$\theta_\mathrm{r/z}$ is due to the additional path that either the beam described by $\boldsymbol{k}_3$ or 
the one denoted by $\boldsymbol{k}_4$ needs to travel before passing through the origin again, the 
combination $2 \phi_\mathrm{r/z} + \phi_\mathrm{r/z}$ corresponds to the phase-shift introduced due to the path 
difference between the the two incoming laser beams $\boldsymbol{k}_1$ and $\boldsymbol{k}_2$. This definition 
becomes clearer when comparing Fig.~\ref{fig:Setup1} with the general setup used for a static DWOL. In the latter 
case the wave-vector $\boldsymbol{k}_2$ is not introduced separately to $\boldsymbol{k}_1$ but rather as a result
of an additional reflection of $\boldsymbol{k}_4$ after passing through the origin.

\begin{figure}[t!]
\includegraphics[width=0.8\linewidth]{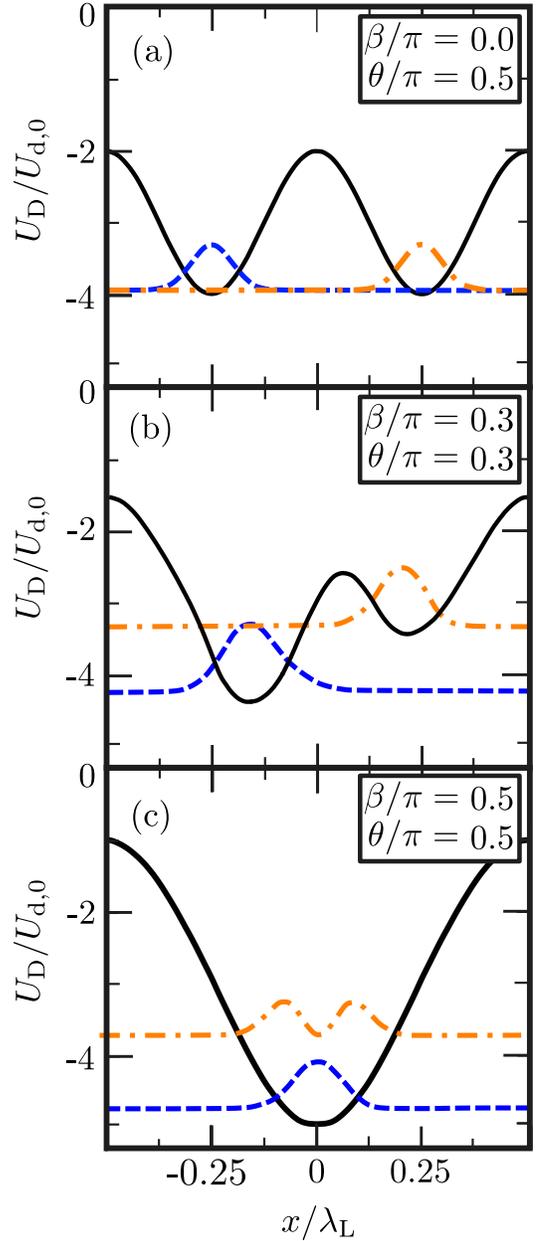}
\caption{\label{DWOLtransport_Fig2}(Color online) Illustration of the inter-well barrier
in a DWOL for different values of angles $\beta$ and $\theta$: (a) Double well without a tilt 
between the two individual wells, (b) double well with a tilt between the two wells,
and (c) single-well configuration. The dashed- and dash-dotted lines represent, respectively, 
the ground-state wave functions of individual wells in (a) and (b). In the single-well configuration 
[(c)] they represent the wave functions corresponding to the two lowest eigenstates (the ground 
state and the first excited state, respectively).}
\end{figure}

Without loss of generality, we hereafter restrict our discussion to a DWOL oriented
along the $x$ axis, which corresponds to the values $\phi_\mathrm{z}=\phi_\mathrm{r}=\theta_\mathrm{r}=0$
and $\theta\equiv\theta_\mathrm{z}$ of the relevant parameters. The total DWOL potential
in that case (for an illustration, see Fig.~\ref{DWOLtransport_Fig1}) is given by
\begin{eqnarray}\label{eqDWOLPot}
U_\mathrm{D}(x,y,z) &=& - U_\mathrm{d,0}\big[\tilde{U}_\parallel(x,y)+\tilde{U}_\perp(x,y) \nonumber \\
&+& \tilde{U}_\mathrm{z}(x,y,z)+\tilde{U}_\mathrm{cr}(x,y,z)\big] \:,
\end{eqnarray}
where $U_\mathrm{d,0}\equiv\alpha_\mathrm{S} I_{\mathrm{xy}}/(2c)$, with $\alpha_\mathrm{S}$ being the 
polarizability corresponding to the scalar light shift~\cite{Deutsch+Jessen:10}, and different (dimensionless) 
contributions to this total potential are 
\begin{eqnarray}\label{eqDWOLPterms}
\tilde{U}_\parallel &=& \cos^2 \left(\frac{\beta}{2}\right)\Big[\cos
\left(2 k_\mathrm{L} y\right)-\cos\left(2k_\mathrm{L}x\right)+2\Big] \:,\nonumber \\
\tilde{U}_\perp &=& 2\sin^2 \left(\frac{\beta}{2}\right)\Big[\cos
\left(k_\mathrm{L}y\right)-\sin\left(k_\mathrm{L} x-\theta\right)\Big]^2 \:, \nonumber\\
\tilde{U}_\mathrm{z} &=&
\xi_\mathrm{z} \, \frac{ w_\mathrm{0,x} w_\mathrm{0,y} }{ w_\mathrm{y}(z)
w_\mathrm{x}(z) } \, \cos^2(k_\mathrm{z} z) \nonumber \\
&\times& \exp\left(-2\left[ \frac{ x^2 }{w_\mathrm{x}^2(z)}+\frac{y^2}
{w_\mathrm{y}^2(z)}\right] \right) \:, \\
\tilde{U}_\mathrm{cr} &=& 2\:\sqrt{\xi_\mathrm{z}}\:\cos\left(\frac{\beta}{2}\right) \,
\sqrt{\frac{w_\mathrm{0,x} w_\mathrm{0,y} }{ w_\mathrm{y}(z) w_\mathrm{x}(z)
}}  \nonumber \\
&\times& \exp\left( - \left[ \frac{ x^2 }{ w_\mathrm{x}^2(z) } +
\frac{ y^2 }{ w_\mathrm{y}^2(z) }\right] \right)\cos \left(k_\mathrm{z}z
\right) \nonumber \\
&\times&\left[\cos \left(\frac{\phi}{2}\right)
\cos \left(k_\mathrm{L} y\right) - \sin \left(\frac{\phi}{2}\right)
\sin \left( k_\mathrm{L} x\right)\right]\:. \nonumber
\end{eqnarray}
In the above equations, $\xi_\mathrm{z} \equiv I_\mathrm{z}/I_{xy}$ is the relative
intensity between the lasers in the $z$ direction and the ones in the $x-y$ plane,
the angle $\beta\in\{0,\,\pi/2\}$ controls the barrier height of adjacent double-well
minima, while $\theta\in\{-\pi,\,\pi\}$, controls the energy offset (tilt) between
these minima. In particular, as illustrated in Fig.~\ref{DWOLtransport_Fig2}, tuning the
angle $\beta$ from $0$ to $\pi/2$ results in a vanishing intermediate barrier between
adjacent minima, thus resulting in a single central minimum.

Having presented the full optical potential of a DWOL, it is pertinent to find its approximate
form $V(x,y,z)$ (with terms up to quadratic order) that is of interest for our further 
discussion. Given that the DWOL configuration can be tuned by varying $\beta$ and $\theta$ (for 
an illustration, see Fig.~\ref{DWOLtransport_Fig2}) we will approximate the full DWOL potential around 
the point $\left[-\lambda_{\mathrm{L}}/4,0,0\right]^\intercal$ in the $x-y$ plane. We do so assuming 
that $\xi_\mathrm{z}\ll w_{\mathrm{0,x}}/l_\mathrm{x}\:,w_\mathrm{0,y}/l_\mathrm{y}$, which ensures 
that the cross term $\tilde{U}_\mathrm{cr}$ in Eq.~\eqref{eqDWOLPterms} gives rise to only a small 
disturbance of the DWOL in the $x-y$ plane. In order to determine the approximated potential $V(x,y,z)$
we introduce the new variable $\tilde{x}\equiv x + \pi/(2 k_\mathrm{L})$. We also assume that
$|\tilde{x}|/Z_\mathrm{R,x}\ll 1$, $|y|/Z_\mathrm{R,y}\ll 1$, $|\tilde{x}|/w_\mathrm{0,x}\ll 1$,
$|y|/w_\mathrm{0,y}\ll 1$, $k_\mathrm{z}\:|z|\ll 1$, $k_\mathrm{L}\:|\tilde{x}|\ll 1$,
and $k_\mathrm{L}\:|y|\ll 1$. Based on these assumptions, we straightforwardly obtain
\begin{equation}\label{eqApproxPottDWOL}
V_\mathrm{D}= -V_{\mathrm{d},0} + m a_{\mathrm{x}} x+\frac{m}{2} \left(
\omega_{\mathrm{x}}^2 x^2+\omega_{\mathrm{y}}^2 y^2+\omega_{\mathrm{z}}^2
z^2 \right) \:,
\end{equation}
where
\begin{equation}
\frac{V_{\mathrm{d},0}}{U_\mathrm{d,0}} = 4\left[\cos^2
\left(\frac{\beta}{2}\right)+2\:\cos^4\left(\frac{\theta}{2}
\right)\:\sin^2\left(\frac{\beta}{2}\right)\right] \:.
\end{equation}
The squared frequencies in Eq.~\eqref{eqApproxPottDWOL} are given by
\begin{eqnarray}
\omega_{\mathrm{x}}^2 &=& \frac{4 \, U_\mathrm{d,0}\, k_\mathrm{L}^2}{m}
\Bigg\{\left[\cos\theta+\cos\left(2\theta\right)\right]\sin^2\left(
\frac{\beta}{2}\right) \nonumber \\
&+& \cos^2\left(\frac{\beta}{2}\right)\Bigg\} \:,  \nonumber \\
\omega_{\mathrm{y}}^2 &=& \frac{4 U_\mathrm{d,0}\, k_\mathrm{L}^2}{m}
\left[1 +\cos\theta\:\sin\left(\frac{\beta}{2}\right) \right] \:,\\
\omega_{\mathrm{z}}^2 &=& \frac{2 \,U_\mathrm{d,0}\, k_\mathrm{z}^2}{m}
\Bigg\{\xi_\mathrm{z} + \sqrt{\xi_\mathrm{z}}\left[\cos\left(\frac{\phi}
{2}\right)+\sin\left(\frac{\phi}{2}\right)\right]\Bigg\}  \:, \nonumber
\end{eqnarray}
while the expression for the factor $a_{\mathrm{x}}$, which has dimensions
of acceleration, reads
\begin{equation}\label{eqDWOLacceleration}
a_{\mathrm{x}}= -\frac{4 U_\mathrm{d,0}\:k_\mathrm{L}}{m}\sin^2
\left(\frac{\beta}{2}\right)\left(1 +\cos\theta\right)\sin\theta \:.
\end{equation}

It is pertinent at this point to underscore two important properties of the approximated
potential $V_\mathrm{D}$ of interest for our further discussion. Firstly, by contrast to
the total DWOL potential that is nonseparable in the $x-y$ plane, this approximated potential
is separable, which simplifies the treatment of atom transport governed by it. Secondly, in
addition to quadratic terms, characteristic of linear harmonic oscillators, this potential
also contains a term linear in $x$.
\subsection{AOM configurations for atom transport} \label{AOMconfig}
\begin{figure}[t!]
\includegraphics[clip,width=0.95\linewidth]{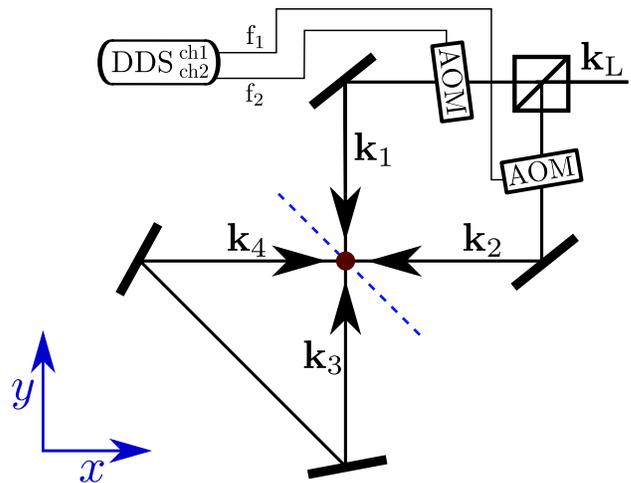}
\caption{\label{fig:Setup1}(Color online) Schematic of a setup that enables
a 2D lattice formed from a folded retroreflected beam. The two AOMs tune
the frequency up and down, respectively, resulting in a moving DWOL.
Two counterpropagating beams in the $z$ direction, not shown here, form an 
optical lattice.}
\end{figure}

\begin{figure}[t!]
\includegraphics[clip,width=0.95\linewidth]{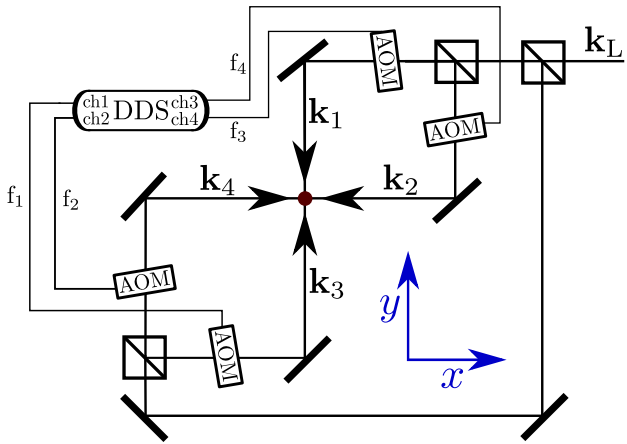}
\caption{\label{fig:Setup2}(Color online) Schematic of a setup that enables
a 2D lattice formed by four different beams. The two AOMs for one pair of
counterpropagating beams tune the frequency up and down, respectively,
resulting in a moving DWOL within the $x-y$ plane. They are controlled
separately through the DDS system. An additional pair of counterpropagating 
laser beams, not shown here, forms an optical lattice in the $z$ direction.}
\end{figure}

To set the stage for further discussion of atom transport in DWOLs, we first specify the
relevant configurations of AOMs that enable transport along $x$, $y$, and diagonal ($r$) directions
(for a schematic illustration, see Figs.~\ref{fig:Setup1} and \ref{fig:Setup2}).

To understand the proposed configurations, it is instructive to first recall how the use of an
AOM engenders the basic moving-lattice effect. Generally speaking, AOMs can shift an incoming
frequency by applying an ultrasonic frequency onto a crystal resulting in an density modulation
along that crystal. Consequently, the incoming laser light encounters the density modulations
that effectively play the role of a grating, thus leading to diffraction. Therefore, the application
of the AOM leads to a frequency shift by the amount equal to the frequency of the ultrasonic
sound wave~\cite{Huang+:14}. The AOMs are connected to different channels of the direct-digital-synthesizer 
(DDS), which control the
frequencies $f_1$ and $f_2$ applied to the crystals in the AOM. While one AOM ramps the frequency
upwards, the other one turns it down by the same amount, thus leading to a detuning $\Delta f=f_1
-f_2=c\left(|\mathbf{k}_+|-|\mathbf{k}_-|\right)/\left(2\pi\right)$ between the counterpropagating
laser beams with wave-numbers $|\mathbf{k}_+|$ and $|\mathbf{k}_-|$, respectively. This results
in an optical lattice moving with velocity $v =\pi \Delta f/ k_\mathrm{L}$, where $k_\mathrm{L}=
|\mathbf{k}_\mathrm{L}|$ is the laser wave number~\cite{Schrader+:01}.

Tuning one of the AOMs up in frequency and the other one down gives rise to a frequency
difference between the beams with wave vectors $\mathbf{k}_1$ and $\mathbf{k}_3$. The same
can be done with the pair of beams with wave vectors $\mathbf{k}_2$ and $\mathbf{k}_4$. Therefore,
a moving-lattice effect in the corresponding directions is achieved. However, with the above
realization one needs to tune both AOMs to achieve a displacement. Hence, the two pairs
of laser beams will always undergo the same displacement in their respective directions.
As a result, atom transport with this configuration of AOMs is only possible in the direction
represented by the dashed line in Fig.~\ref{fig:Setup1} in the trivial case of equal frequencies 
in the $x$- and $y$ directions.

To provide implementations for a broader set of displacement schemes and different values of the
parameters $\beta$ and $\theta$, a more sophisticated experimental setup is required. As it turns out,
to enable transport in an arbitrary direction within the $x-y$ plane one ought to forgo the retroreflected 
beams and the ensuing intrinsic phase stability, introducing instead two additional AOMs to the system. 
This approach, illustrated in Fig.~\ref{fig:Setup2}), allows one to separately address the beams in the 
$x$- and $y$ directions by tuning the corresponding pairs of AOMs. In this manner, it possible to have
different lattice displacements in those two directions.

It should be stressed that, unlike DWOLs formed from a folded retroreflected beam, the chosen setup 
requires active phase stabilization to counteract the effects of mirror-induced phase noise. Namely, 
it is known that a $D$-dimensional optical lattice, created with no more than $D+1$ independent light 
beams is topologically stable to arbitrary changes of the relative phases of the $D+1$ beams~\cite{Grynberg+:93}. 
However, the DWOL in the $x-y$ plane ($D=2$) is created by $4$ beams, which is more than $D+1=3$, 
hence active stabilization of the relative phase between standing waves is needed here. 

\section{STA-based trajectories} \label{MovingTrap_STA}
Inverse engineering based on LRIs is the most widely used approach for modelling atomic transport.
In particular, the basic LRI-based transport theory, developed in Ref.~\cite{Torrontegui+:11},
makes use of a particular family of quadratic-in-momentum invariants~\cite{Lewis+Riesenfeld:69}.
One of the crucial implications of that theory is that the harmonic-trapping case and that of
an arbitrary trap entail different treatments. Namely, the perfect transport in the latter case
necessitates, in principle, the use of compensating forces in the comoving reference frame of
the trap (cf. Sec.~\ref{ComovingFrameFSOM}).

In the following, we apply the STA theory of Ref.~\cite{Torrontegui+:11} to determine the classical
path of the potential minima in a moving DWOL. We do that using a single-atom Hamiltonian with the
harmonically approximated DWOL potential of Eq.~\eqref{eqApproxPottDWOL}. To begin with, we briefly
summarize the main concepts behind the inverse-engineering approach based
on LRIs (for a more detailed introduction, see, e.g., Ref.~\cite{STA_RMP:19}).
\subsection{Inverse-engineering approach to atom transport}
For a time-dependent Hamiltonian $H(t)$, any operator $I(t)$ that satisfies the equation
\begin{equation}\label{eqLewisRiesenfeldInvariant}
\frac{\partial }{\partial t} I(t) + \left[H(t),I(t)\right] = 0
\end{equation}
constitutes a dynamical invariant of this Hamiltonian. One immediate consequence of the
last equation is that the eigenvalues $\lambda_n$ of $I(t)$ are time-independent. If
these eigenvalues are also non-degenerate, the corresponding eigenstates $\ket{\Phi_n(t)}$
and the instantaneous eigenstates $\ket{\Psi_n(t)}$ of the Hamiltonian $H(t)$, referred to
as transport modes of $H(t)$, are connected through the relation $\ket{\Psi_n(t)}=e^{i\theta_{\textrm{LR}}
(t)}\ket{\Phi_n(t)}$; here $\theta_{\textrm{LR}}(t)=\hbar^{-1}\int_0^t\bra{\Phi_n(t')}
\left[i\hbar\partial_{t'}-H(t')\right]\ket{\Phi_n(t'} dt'$ stands for the Lewis-Riesenfeld
phase. As a result, the general solution of the Schr\"{o}dinger equation for $H(t)$ can
succinctly be written in the form
\begin{equation}\label{eqGeneralSolutionSchroedinger}
|\Psi(t)\rangle=\sum_{n}C_n\: e^{i\theta_{\textrm{LR}}(t)}|\Phi_n(t)\rangle \:.
\end{equation}

An important class of Hamiltonians used in modelling atom transport are those
of the Lewis-Leach type~\cite{Lewis+Leach:82}. The latter are given by
\begin{equation}\label{LewisLeachHamiltonians}
H(t)=\frac{p^2}{2m} + V(q,t)  \:,
\end{equation}
where the potential $V(q,t)$ in the most general case reads
\begin{equation}\label{eqPotentialLewisLeach}
V(q,t) = -F(t)q +\frac{m}{2}\:\omega^2(t) q^2 +\frac{1}{\rho^2 (t)}
U\left(\frac{q-\alpha(t)}{\rho(t)}\right)  \:,
\end{equation}
with $\rho(t)$, $\alpha(t)$, $\omega(t)$, and $F(t)$ being arbitrary functions
of time that satisfy the auxiliary equations~\cite{Torrontegui+:11}
\begin{eqnarray}
\ddot{\rho}+\omega^2(t)\rho = \frac{\omega^2_0}{\rho^3} \:, \label{AuxEq1}\\
\ddot{\alpha}+\omega^2(t)\alpha = \frac{F(t)}{m} \:, \label{AuxEq2}
\end{eqnarray}
where $\omega_0$ is a constant. The quadratic-in-momentum invariant corresponding
to the general Hamiltonian of Eq.~\eqref{LewisLeachHamiltonians}, up to a constant
factor, is given by~\cite{Torrontegui+:11}
\begin{eqnarray}
I(t) &=& \frac{1}{2m}\big[\rho(p-m\dot{\alpha})-m\dot{\rho}(q-\alpha)]^2 \\
&+& \frac{m}{2}\:\omega^2_0\left(\frac{q-\alpha}{\rho}\right)^2 +U\left(
\frac{q-\alpha}{\rho}\right) \nonumber \:.
\end{eqnarray}
For a broad class of transport problems it suffices to take $\rho(t)\equiv 1$
and $\omega(t)\equiv\omega_0$, which renders Eq.~\eqref{AuxEq1} irrelevant,
with Eq.~\eqref{AuxEq2} remaining the only auxiliary equation. In particular,
if those transport problems involve a rigid harmonic oscillator driven by the
``transport function'' $q_0(t)$ -- which corresponds to choosing $F(t)=m\omega^2_0
q_0(t)$ in Eq.~\eqref{eqPotentialLewisLeach} -- then the remaining auxiliary
equation [cf. Eq.~\eqref{AuxEq2}] has the form characteristic of a forced harmonic
oscillator and $\alpha(t)$ can be identified with a classical-particle trajectory
$q_c(t)$ that satisfies $\ddot{q}_c+\omega_0^2[q_c-q_0(t)]=0$.

\subsection{LRI treatment for atom transport in DWOLs and STA moving-lattice trajectory} \label{STAtrajectory}
In the following, we apply the theory of LRIs to the approximate DWOL Hamiltonian
\begin{equation}\label{simpleDWOL_Hamiltonian}
H_{\textrm{D},0}(t)=-\frac{\hbar^2\nabla^2}{2m}+V(\mathbf{r}-\mathbf{q}_0(t))\:,
\end{equation}
i.e. a single-atom Hamiltonian where $V(\mathbf{r})\equiv V(x,y,z)$ is given by
the simplified DWOL potential $V_{\mathrm{D}}$ of Eq.~\eqref{eqApproxPottDWOL}.
This approximated potential, used in what follows to determine the STA trap trajectory, 
has the form
\begin{eqnarray}\label{eqDWOLSTAHamiltonian1}
& & V_\mathrm{D}\left[\mathbf{r}-\mathbf{q}_0(t)\right]= -V_\mathrm{d,0} +
m a_\mathrm{x} \big[x - q_\mathrm{0,x}(t)\big] \\
&+& \frac{m}{2} \big\{ \omega_\mathrm{x}^2 \big[x - q_\mathrm{0,x}(t)\big]^2
+\omega_\mathrm{y}^2\big[y-q_\mathrm{0,y}(t)\big]^2 +\omega_\mathrm{z}^2 z^2
\big\} \nonumber \:.
\end{eqnarray}

It is important to notice at this point that -- owing to the separability of this last
potential -- displacements in the $x$- and $y$ directions can be treated independently.
Moreover, after restricting the above potential to one direction, i.e. making the replacement
$V_\mathrm{D} \rightarrow V^\mathrm{x}_\mathrm{D}\left(x - q_\mathrm{0,x}(t)\right)
\:\delta_{x,u}+ V^\mathrm{y}_\mathrm{D} \left(y - q_\mathrm{0,y}(t)\right)\:\delta_{y,u}
+ V_\mathrm{d,0}$ (where $u = x,\,y$), this potential -- up to an additive constant -- has
the form characteristic of the Lewis-Leach type Hamiltonians [cf. Eqs.~\eqref{LewisLeachHamiltonians}
and \eqref{eqPotentialLewisLeach}].

In the problem at hand, we have a Hamiltonian of the Lewis-Leach type with $U=0$
and $\rho(t)\equiv 1$. As regards the remaining time-dependent parameters of the
general potential in Eq.~\eqref{eqPotentialLewisLeach}, we can make the following
identifications. For the case of transport in the $x$ direction, i.e. the potential
$V_\mathrm{D}^\mathrm{x}$, we have $\omega_\mathrm{x}(t) = \omega_\mathrm{x}$ and
\begin{equation}\label{eqLewisLeach1}
F_\mathrm{x}(t)= m\left[\omega^2_\mathrm{x} q_{0,x}(t) - a_\mathrm{x}\right]  \:.
\end{equation}
Similarly, for transport in the $y$ direction, i.e. potential $V_\mathrm{D}^\mathrm{y}$,
we have $\omega_\mathrm{y}(t) = \omega_\mathrm{y}$ and
\begin{equation}\label{eqLewisLeach2}
F_\mathrm{y}(t)= m\omega^2_\mathrm{y} q_{0,y}(t) \:.
\end{equation}
In line with the standard practice~\cite{Torrontegui+:11}, we further add to $V_\mathrm{D}^\mathrm{x}$
and $V_\mathrm{D}^\mathrm{y}$ the irrelevant time-dependent global terms $m\omega^2_\mathrm{x} q^2_{0,x}(t)$
and  $m\omega^2_\mathrm{y} q^2_{0,y}(t)$, respectively, which do not give rise to any force. As a result,
the auxiliary equations of the forced-harmonic-oscillator type are here given by
\begin{align}
&\ddot{q}_\mathrm{c,x}(t) + \omega_\mathrm{x}^2
\left[ q_\mathrm{c,x}(t) - q_\mathrm{0,x}(t) \right]=-a_\mathrm{x} \label{FHOeq1}\:,
\\
&\ddot{q}_\mathrm{c,y}(t) + \omega_\mathrm{y}^2 \left[ q_\mathrm{c,y}(t)
- q_\mathrm{0,y}(t) \right]=0 \:, \label{FHOeq2}
\end{align}
where $\mathbf{q}_\mathrm{c}(t)\equiv\left[q_\mathrm{c,x}(t),q_\mathrm{c,y}(t),0\right]^\intercal$
stands for the corresponding 2D classical-particle trajectory.

The most interesting aspect of the last auxiliary equations is the presence of $a_x$ on the
right-hand-side (RHS) of Eq.~\eqref{FHOeq1}, which results from the existence of terms linear in
$x$ in the approximated potentials $V_\mathrm{D}$ [cf. Eq.~\eqref{eqApproxPottDWOL}]. The
presence of $a_x$ makes the problem at hand analogous to transport problems that involve a
constant force (e.g., gravity)~\cite{Torrontegui+:11}. It is straightfoward to see that
Eq.~\eqref{FHOeq1} can be reduced to the form of Eq.~\eqref{FHOeq2} by a slight redefinition
of $q_\mathrm{c,x}(t)$, i.e. by introducing $\tilde{q}_\mathrm{c,x}(t)\equiv q_\mathrm{c,x}(t)
+a_x/\omega_\mathrm{x}^2$. In this way, we reduce the problem of atom transport in the $x$ direction, 
equally like the one in the $y$-direction, to a known class of atom-transport problems solvable 
via inverse engineering~\cite{Torrontegui+:11}.

\begin{figure}[b!]
\includegraphics[clip,width=0.875\linewidth]{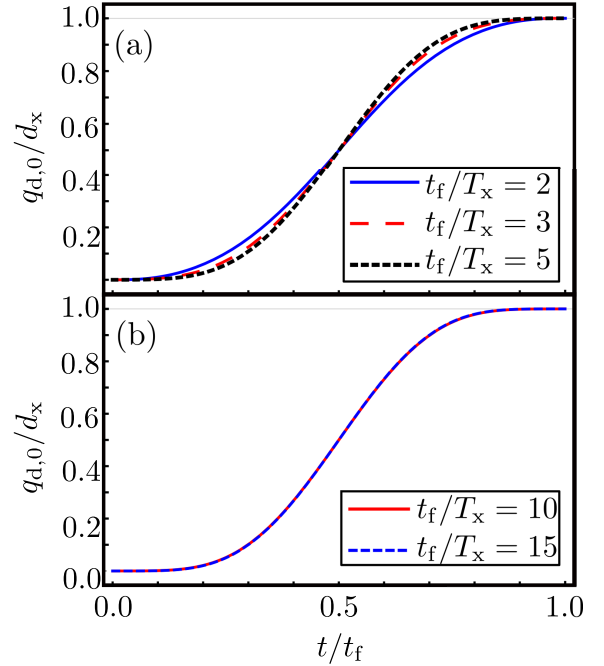}
\caption{\label{fig:STAminimum}(Color online) Path of a DWOL potential minimum as a function
of time, obtained using the STA approach, for transport times $t_f$ comparable to [(a)], and
an order of magnitude longer than [(b)] the internal timescale $T_\mathrm{x}$. The chosen parameter
values are the following: lattice depth $U_{d,0} = 1500\:E_{\textrm{R}}$, transverse waists
$w_{0,x} = w_{0,y} = 4.2\times 10^3\:l_\mathrm{x}$, $\beta = 3\pi/20$, $\theta = pi/2$, 
$\phi_z = \pi/2$, and the transport distance is set to $d_\mathrm{x} = 5\pi/k_{\textrm{L}}\approx\:160\:l_\mathrm{x}$.}
\end{figure}

By imposing appropriate boundary conditions for $q_{0,\mathrm{x}}(t)$ and $q_{0,\mathrm{y}}(t)$ -- along
with the corresponding conditions for $q_\mathrm{c,x}(t)$ and $q_\mathrm{c,y}(t)$ consistent with
Eqs.~\eqref{FHOeq1} and \eqref{FHOeq2} -- the general solution for the path of the potential minimum 
can be sought in the form of the ninth-degree polynomial~\cite{Hauck+:21}
\begin{equation}\label{STAsolution}
q_{0,u}(t) = d_u \sum_{n=3}^9 b_{n,u}\left(\frac{t}{t_\mathrm{f}}
\right)^n \quad (u=x,y)\:,
\end{equation}
where $d_u$ is the transport distance (cf. Sec.~\ref{SingleAtomTransportDWOL}). By taking into account 
the relevant initial/boundary conditions~\cite{Hauck+:21}, the following solution for constants $b_{n,u}$ 
is obtained:
\begin{eqnarray}
b_{3,u} &=& 2520\:(t_\mathrm{f} \omega_u)^{-2} \quad,\quad b_{4,u}=-12600\:(t_\mathrm{f} \omega_u)^{-2}\:, \nonumber\\
b_{5,u} &=& 22680\:(t_\mathrm{f} \omega_u)^{-2} +126 \:, \nonumber \\
b_{6,u} &=& -17640\:(t_\mathrm{f} \omega_u)^{-2} -420 \:, \\
b_{7,u} &=& 5040\:(t_\mathrm{f} \omega_u)^{-2} +540 \:, \nonumber \\
b_{8,u} &=& -315 \quad,\quad b_{9,u}=70 \:. \nonumber
\end{eqnarray}

The obtained solution for the STA-based trap trajectory, i.e. path of the potential
minimum, is illustrated by Fig.~\ref{fig:STAminimum} which shows its $x$ component $q_{0,x}$
as a function of the transport time $t_\mathrm{f}$.

\section{$\textrm{e}$STA-based trajectories} \label{MovingTrap_eSTA}
Despite the fact that STA have proven their worth in a variety of quantum systems~\cite{STA_RMP:19},
their modification -- known as eSTA -- has recently been proposed~\cite{Whitty+:20}. The principal motivation
behind this method, which is inspired by optimal-control techniques~\cite{Werschnik+Gross:07},
is to enable the design of efficient control protocols for systems that are not directly amenable to an
STA-type treatment. The main idea behind eSTA is to approximate the Hamiltonian of such a system by a
simpler one for which an STA-based protocol can straightforwardly be found. Under the assumption that
this STA-based protocol is nearly optimal even when applied to the original system Hamiltonian, its
sought-after eSTA counterpart is obtained using a gradient expansion in the control-parameter space.
This last assumption underscores the heuristic character of eSTA, which -- in principle -- does not
guarantee its superiority over STA~\cite{Whitty+:20}. However, it was already demonstrated that eSTA
outperforms STA in certain classes of quantum-control problems~\cite{Whitty+:20}.

Having found the STA solution for the simplified harmonic DWOL potential in Sec.~\ref{MovingTrap_STA},
in what follows we apply the eSTA scheme to the same problem. To start with, we briefly review the basic
concepts and assumptions behind this method.
\subsection{Basics of the eSTA method} \label{eSTAbasics}
The first step required for the application of the eSTA method to a system described by the Hamiltonian
$H_{\textrm S}$ amounts to finding an STA solution, parameterized by a vector $\vec{\lambda}_0\in\mathbb{R}^n$,
for a ``close'' Hamiltonian $H_0$~\cite{Whitty+:20}. It is assumed that there exists a parameter
$\mu_{\textrm S}$, such that a series expansion of the type
\begin{equation} \label{eqCloseHamiltonian}
H_{\textrm S} = \sum_{k=0}^\infty \mu_{\textrm S}^k \, H^{(k)} \:,
\end{equation}
involves $H_0$ as its zeroth-order term [i.e. $H^{(0)}\equiv H_0$]. In the system under consideration,
the role of $H_{\textrm S}$ is played by $H_{\textrm{D}}(t)$ of Eq.~\eqref{FullSingleAtomHamiltonian}.
At the same time, $H_0$ is represented by the Hamiltonian $H_{\textrm{D},0}(t)$ of Eq.~\eqref{simpleDWOL_Hamiltonian}.

Aiming to find an optimal solution for $H_{\textrm S}$ based on a previously obtained STA solution for $H_0$,
the general control vector $\vec{\lambda}_{\mathrm{S}}$ of the full system can be expressed as a sum of the
STA control vector $\vec{\lambda}_0$ and an auxiliary control vector $\vec{\alpha}$, i.e. $\vec{\lambda}_{\mathrm{S}}
=\vec{\lambda}_0 +\vec{\alpha}$. The special value of $\vec{\alpha}$ that corresponds to the sought-after
optimal solution, i.e. the optimal eSTA correction vector, will be denoted by $\vec{\epsilon}$ in what follows.

The main assumption underlying the eSTA scheme is that the STA-based protocol corresponding to the simplified
Hamiltonian $H_0$ is close to being optimal when applied to the full Hamiltonian $H_{\textrm S}$~\cite{Whitty+:20}.
Another crucial assumption of the eSTA approach is that the deviation of the fidelity $\mathcal{F}$ around its maximal 
value depends quadratically on the difference $\alpha-\epsilon$. In other words, the fidelity satisfies the approximate
relation~\cite{Whitty+:20}
\begin{equation} \label{eqESTAFidelity}
\mathcal{F}\left(\mu_{\textrm{S}},\vec{\lambda}_0 + \alpha \frac{\vec{\nabla}\mathcal{F}(\mu_{\textrm{S}},\vec{\lambda}_0)}
{\rVert\vec{\nabla}\mathcal{F}(\mu_{\textrm{S}},\vec{\lambda}_0)\rVert}\right)\approx 1 - c\left(\alpha-\epsilon
\right)^2 \:,
\end{equation}
where $\epsilon\equiv\rVert\vec{\epsilon}\rVert$, $\alpha\equiv\rVert\vec{\alpha}\rVert$, and $c$
is a positive constant. Beased on the above assumptions and a Taylor expansion of the left-hand-side
of Eq.~\eqref{eqESTAFidelity} around $\epsilon=\alpha$, one straightforwardly obtains~\cite{Whitty+:20}
\begin{equation}\label{eqInitialEpsilon}
\vec{\epsilon}\approx\frac{2\left[1- \mathcal{F}(\mu_{\textrm{S}},\vec{\lambda}_{\textrm{S}})\right]
\vec{\nabla}\mathcal{F}(\mu_{\textrm{S}},\vec{\lambda}_0)}{\rVert\vec{\nabla}\mathcal{F}(\mu_{\textrm{S}},
\vec{\lambda}_0)\rVert^2} \:.
\end{equation}
Up to second order in $\mu_{\textrm S}$ the fidelity is given by~\cite{Whitty+:20}
\begin{equation}\label{eqFidelityGn}
\mathcal{F}(\mu_{\textrm{S}},\vec{\lambda}_{\textrm{S}}) \approx
1-\frac{1}{\hbar^2}\sum_{n=1}^\infty |G_n|^2 \:,
\end{equation}
where $G_n$ is an auxiliary scalar function, given in terms of the transport modes of $H_0$
[cf. Sec.~\ref{MovingTrap_STA}] by
\begin{equation}\label{eqExpressionG}
G_n = \int_0^{t_f} dt \braket{\Psi_n(t)|
\left[H_{\textrm{S}} (\vec{\lambda}_0; t) - H_0
(\vec{\lambda}_0; t)\right]|\Psi_0(t)} \:.
\end{equation}
Similarly, up to second order in $\mu_{\textrm S}$ the gradient of $\mathcal{F}(\mu_{\textrm{S}},\vec{\lambda}_0)$
is given by an analogous approximate expression: ~\cite{Whitty+:20}
\begin{equation}\label{eqGradientFidelityKn}
\vec{\nabla}\mathcal{F}(\mu_{\textrm{S}},\vec{\lambda}_0) \approx
-\frac{2}{\hbar^2} \sum_{n=1}^\infty \textrm{Re}
\left(G_n\:\vec{K}_n^*\right) \:.
\end{equation}
Here $\vec{K}_n$ is an auxiliary vector function:
\begin{align}\label{eqExpressionK}
\vec{K}_n = \int_0^{t_f} dt \braket{\Psi_n(t)|\nabla_\lambda H_{\textrm{S}}
(\vec{\lambda};t)\big|_{\vec{\lambda}=\vec{\lambda}_0}|\Psi_0(t)} \:.
\end{align}

The optimal eSTA correction vector $\vec{\epsilon}$ can be expressed in terms of
$G_n$ and $\vec{K}_n$ as
\begin{equation}\label{eqEpsilonDefinition}
\vec{\epsilon}= -\frac{\left(\sum_{n=1}^{N} |G_n|^2 \right) \sum_{n=1}^{N}
\textrm{Re} \left( G_n^\ast \vec{K}_n \right)}{\left\rVert \sum_{n=1}^{N} \textrm{Re}
\left( G_n^\ast \vec{K}_n \right) \right\rVert^2} \:,
\end{equation}
where $N$ is the cut-off parameter. Based on this expression, $\vec{\epsilon}$ can straightforwardly
be calculated numerically provided that $G_n$ and $\vec{K}_n$ are previously obtained by evaluating the
integrals in Eqs.~\eqref{eqExpressionG} and \eqref{eqExpressionK}, respectively. In the problem under
consideration, where $\ket{\Psi_n(t)}$ are the transport modes of a time-dependent 3D harmonic-oscillator
Hamiltonian, the evaluation of those integrals is highly nontrivial and entails the use of various
properties of Hermite polynomials~\cite{ChowBOOK:00} (for detailed derivations, see Appendices~\ref{SecDWOPGn}
and \ref{SecDWOPKn}).
\subsection{eSTA moving-lattice trajectories} \label{eSTAtrajectory}
In the transport problem at hand, the $x$- and $y$ components of the trap-trajectory
vector, $q_{0,\mathrm{x}}(\vec{\lambda}_{\mathrm{x}};t)$ and $q_{0,\mathrm{y}}
(\vec{\lambda}_{\mathrm{y}};t)$, are parametrized by the twelve-component real-valued control
vector $\vec{\lambda}=\left(\vec{\lambda}_\mathrm{x},\vec{\lambda}_\mathrm{y}\right)^\intercal$.
Here $\vec{\lambda}_{\mathrm{x}}\equiv [\lambda^{(1)}_{\mathrm{x}},\ldots,\lambda^{(6)}_{\mathrm{x}}]$
and $\vec{\lambda}_{\mathrm{y}}\equiv[\lambda^{(1)}_{\mathrm{y}},\ldots,\lambda^{(6)}_{\mathrm{y}}]$
are the reduced six-component control vectors corresponding to the displacements in the $x$- and $y$ 
directions, respectively, which are assumed to satisfy the following conditions:
\begin{eqnarray}
q_{0,\mathrm{x}}(\vec{\lambda}_{\mathrm{x}};\: j t_f/7)
&=& \lambda^{(j)}_{\mathrm{x}} \quad (j=1,\ldots, 6) \:,\nonumber\\
q_{0,\mathrm{y}}(\vec{\lambda}_{\mathrm{y}};\: j t_f/7)
&=& \lambda^{(j)}_{\mathrm{y}} \:.
\end{eqnarray}
The STA-based trajectory corresponds to the control vector $\vec{\lambda}=\left(\vec{\lambda}_{\mathrm{x},0}
,\vec{\lambda}_{\mathrm{y},0}\right)^\intercal$, with components $\lambda^{(j)}_{\mathrm{x},0}$ and
$\lambda^{(j)}_{\mathrm{y},0}$ $(j=1,\ldots, 6)$. Thus, this trajectory will be denoted by $q_{0,\mathrm{x}}
(\vec{\lambda}_{\mathrm{x},0};\:t)$ and $q_{0,\mathrm{y}}(\vec{\lambda}_{\mathrm{y},0};\:t)$.
At the same time, the sought-after optimized (eSTA) trajectory corresponds to the
control vectors $\vec{\lambda}_{\mathrm{x,S}}$ and $\vec{\lambda}_{\mathrm{y,S}}$,
hence it will be denoted by $q_{0,\mathrm{x}}(\vec{\lambda}_{\mathrm{x,S}};\:t)$
and $q_{0,\mathrm{y}}(\vec{\lambda}_{\mathrm{y,S}};\:t)$.

The optimized (eSTA) paths of the potential minimum can be expressed through their STA
counterparts as
\begin{eqnarray}\label{eqControlVectorRedefined}
q_{0,\mathrm{x}}(\vec{\lambda}_{\mathrm{x,S}}; \, t) &=&
q_{0,\mathrm{x}}(\vec{\lambda}_{\mathrm{x},0}; \,t) +
f_{\mathrm{x}}(\vec{\alpha}_{\mathrm{x}}; \, t) \:, \\
q_{0,\mathrm{y}}(\vec{\lambda}_{\mathrm{y,S}}; \, t) &=&
q_{0,\mathrm{y}}(\vec{\lambda}_{\mathrm{y},0}; \,t) +
f_{\mathrm{y}}(\vec{\alpha}_{\mathrm{y}}; \, t) \:,
\end{eqnarray}
where $\vec{\alpha}:=\left(\vec{\alpha}_{\mathrm{x}},\vec{\alpha}_{\mathrm{y}}\right)^\intercal$ 
is an auxiliary twelve-component control vector. These eSTA paths are assumed to satisfy the conditions 
expressed by $q_{0,\mathrm{x}}(\vec{\lambda}_{\mathrm{x,S}}; \, jt_f/7)=\lambda^{(j)}_{\mathrm{x},0}+
\alpha^{(j)}_{\mathrm{x}}$ and $q_{0,\mathrm{y}}(\vec{\lambda}_{\mathrm{x,S}}; \, jt_f/7) =
\lambda^{(j)}_{\mathrm{y},0} + \alpha^{(j)}_{\mathrm{y}}$ ($j=1,\ldots, 6$). The auxiliary functions 
$f_u(\vec{\alpha}_u;\:t)$  ($u=x,y$) have to obey the following boundary conditions:
\begin{equation}
\begin{split}
& f_u(\vec{\alpha}_{u};0) = f_u(\vec{\alpha}_u ;t_f) = 0\:, \\
& f_u(\vec{\alpha}_{u}; jt_f/7 ) = \alpha^{(j)}_u \quad (\:j=1,\ldots,6\:) \:,\\
& \frac{\mrm{d^{(n)}} }{\mrm{d}t^{(n)}}f_u(\vec{\alpha}_u ;t')
|_{t'=\lbrace 0,t_f \rbrace} = 0 \quad (\:n=1,\ldots,4\:) \:.
\end{split}
\end{equation}
The latter conditions are chosen such that $f_{u}(\vec{\alpha}_{u};\:t)$
can be controlled through $\vec{\alpha}_{u}$, obeying at the same time the continuity
conditions. Thus, we choose the following polynomial Ansatz of eleventh degree:
\begin{equation}\label{eqSolutionvectorF}
f_{u}(\vec{\alpha}_{u};\:t) = \sum_{n=0}^{11}\sum_{j=1}^6
\tilde{a}^{(j)}_{u,n}\alpha^{(j)}_{u} \left( \frac{t}{t_f} \right)^n \:.
\end{equation}
The values of the coefficients $\tilde{a}^{(j)}_{\mathrm{x},n}$ and $\tilde{a}^{(j)}_{\mathrm{y},n}$
in the last equations are given in Table I of Ref.~\cite{Hauck+:21}.

The auxiliary control vector $\vec{\alpha}$ that corresponds to the sought-after eSTA
solution is given by the twelve-component optimal correction vector $\vec{\epsilon}\equiv\left(
\vec{\epsilon}_{\mathrm{x}}, \vec{\epsilon}_{\mathrm{y}}\right)^\intercal$, where 
$\vec{\epsilon}_{\mathrm{x}}$ and $\vec{\epsilon}_{\mathrm{y}}$ are evaluated using the 
general expression in Eq.~\eqref{eqEpsilonDefinition}. Given that in our 3D problem the transport 
modes can be enumerated by three 1D quantum numbers $\{n_x,n_y,n_z\}$, we can recast the sum 
in Eq.~\eqref{eqEpsilonDefinition} in terms of the main quantum number $n$ and $\{n_x,n_y,n_z\}$. 
For the cut-off parameter $N$ we take the value $N=2$, despite the fact that our numerical 
evaluations show that it suffices to take $N=1$.

\begin{figure}[t!]
\includegraphics[clip,width=0.875\linewidth]{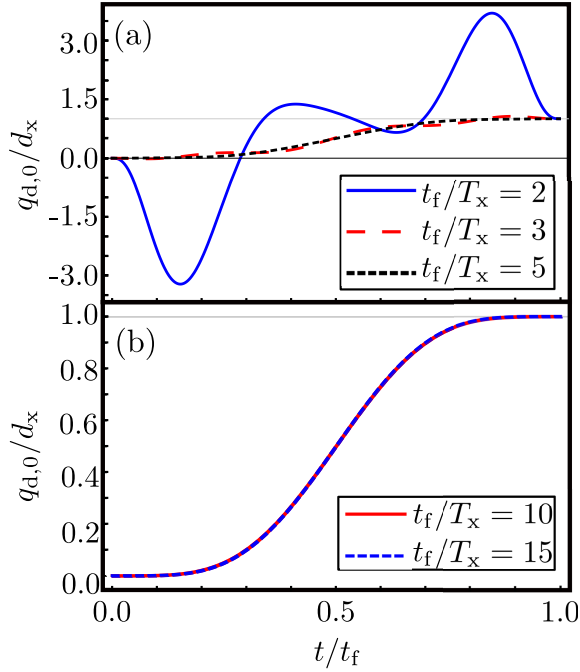}
\caption{\label{fig:eSTAminimum}(Color online)Path of a DWOL potential minimum as a function of time,
obtained using the eSTA approach, for transport times $t_f$ which are (a) comparable to-, and (b) an 
order of magnitude longer than, the internal timescale $T_\mathrm{x}$. The parameters chosen are the following:
lattice depth $U_{d,0} = 1500\:E_{\textrm{R}}$, transverse waists $w_{0,x} = w_{0,y} = 4.2\times 10^{3}
\:l_\mathrm{x}$, $\beta = 3\pi/20$, relevant angles $\theta = \pi/2$, $\phi_z = \pi/2$, and the transport 
distance is set to $d_\mathrm{x} = 5\pi/k_{\textrm{L}}\approx\:160\:l_\mathrm{x}$.}
\end{figure}

The $x$ component of a typical eSTA-based trajectory is depicted in Fig.~\ref{fig:eSTAminimum}. By 
comparing this trajectory to the STA-based one (cf. Fig.~\ref{fig:STAminimum}) it can be inferred 
that their shapes differ significantly only for short transport times.

\section{Single-atom dynamics} \label{AtomDynamics}
Having described the design of STA-based trap trajectories in Sec.~\ref{MovingTrap_STA} and their eSTA-based
counterparts in Sec.~\ref{MovingTrap_eSTA}, in the following we briefly present our chosen approach for evaluating
the resulting single-atom dynamics. We start by reviewing the basic aspects of the use of the Fourier split operator
method (FSOM) for solving time-dependent Schr\"{o}dinger equations (TDSEs) in Sec.~\ref{FSOMbasics} and for computing
ground-state properties in Sec.~\ref{FSOMgroundState}. We then discuss our approach for numerically solving the
relevant comoving-frame TDSE (Sec.~\ref{ComovingFrameFSOM}), followed by the relevant details of the numerical
implementation (Sec.~\ref{DetailsNumImpl}).
\subsection{Basic aspects of the FSOM} \label{FSOMbasics}
The FSOM has established itself as the method of choice for solving Cauchy-type
initial-value problems $\partial_t f(\mathbf{r},t)=\hat{A}(\mathbf{r},t)f(\mathbf{r},t)$,
where $\hat{A}(\mathbf{r},t)$ is an operator, possibly time-dependent, and the initial condition
reads $f(\mathbf{r},t)=f_0(\mathbf{r})$. This method is particularly convenient to use
in cases where the operator $\hat{A}$ can be written as a sum of two operators, one
of which can straightforwardly be diagonalized in real space and the other one in Fourier
space. Typical example of such problems is a TDSE~\cite{Feit+:82,Bandrauk+Shen:93}. The
central idea behind the use of the FSOM for solving TDSEs is to approximate the time-evolution
operator of the system by a product of operators that are diagonal either in real- or in momentum space.

In what follows, we consider the TDSE describing the motion of a particle of mass $m$ in the potential 
$U(\vec{r})$. The relevant single-particle Hamiltonian is given by
\begin{equation}\label{SingleAtomHamiltonian}
H=-\frac{\hbar^2\nabla^2}{2m}+U(\vec{r})\:.
\end{equation}
To conveniently express the time-evolution operator that corresponds to this last Hamiltonian we invoke 
the following identity, obtained using the Baker-Campbell-Hausdorff formula~\cite{GilmoreBOOK:12}, which 
holds for arbitrary Hermitian operators $A$ and $B$:
\begin{equation}\label{SymmetricTrotter}
e^{i(A+B)\delta t}=e^{iA\delta t/2}e^{iB\delta t}e^{iA\delta t/2}
+\mathcal{O}(\delta t^3)\:.
\end{equation}
As a special case of the last identity for $A=U(\vec{r})$ and $B=-\hbar^2\nabla^2/(2m)$,
we obtain an accurate second-order time-stepping scheme for the propagation
of the wave-function $\Psi(\vec{r},t)$~\cite{Pechukas+Light:66}:
\begin{eqnarray}\label{eqFSOMBCH}
\Psi(\vec{r},t&+&\delta t) = \exp\left[-\frac{i}{\hbar}\:U(\vec{r})
\frac{\delta t}{2}\right]\exp\left(i\frac{\hbar\nabla^2}{2m}\:\delta t\right) \nonumber \\
&\times& \exp\left[-\frac{i}{\hbar}\:U(\vec{r})\frac{\delta t}{2}\right]
\Psi(\vec{r},t)+ \mathcal{O}(\delta t^3)  \:.
\end{eqnarray}
In particular, the last equation allows one to treat the different exponential terms
independently. Thus, it is pertinent to Fourier-transform the kinetic term to momentum
space, noticing at the same time that the RHS of Eq.~\eqref{eqFSOMBCH} can be recast 
using the identity
\begin{eqnarray}\label{eqFourierRepresentation}
&&\exp\left(i\frac{\hbar\nabla^2}{2m}\:\delta t\right)\exp
\left[-\frac{i}{\hbar}\:U(\vec{r})\frac{\delta t}{2}\right]\Psi(\vec{r},t)  \\
&=& F^{-1}\left[e^{-i\frac{\hbar k^2}{2m}\delta t}\:F\left\{\exp\left[-\frac{i}{\hbar}
\:U(\vec{r})\frac{\delta t}{2}\right]\Psi(\vec{r},t)\right\}\right] \nonumber \:,
\end{eqnarray}
where $F[\ldots]$ denotes the Fourier transform of the argument and $F^{-1}[\ldots]$ its inverse.

A numerical solution of a TDSE at time $t'=t+N_t\delta t$ is obtained by applying $N_t$ times
in succession the time-propagation scheme based on Eq.~\eqref{eqFSOMBCH} on an initial wave-function
$\Psi(\vec{r},t)$. In numerical implementations of the FSOM, this wave-function is discretized
on a rectangular spatial grid with $N_s$ points and the exact, continuous Fourier transform is
approximated by its discrete counterpart. The computational burden of propagating the function
$\Psi(\vec{r},t)$ is dominated by the needed spatial Fourier transformation and its inverse.
When carried out using the fast Fourier transform (FFT) algorithm~\cite{NRcBook}, an elementary
step in these transformations within the framework of the FSOM requires $\mathcal{O}(N_s\log_2 N_s)$
operations.
\subsection{Ground-state calculation: Imaginary-time evolution approach} \label{FSOMgroundState}
Apart from our use of the FSOM for the treatment of single-atom dynamics, we also utilize this
method within the framework of the imaginary-time evolution (ITE) approach to determine the ground
state of our DWOL optical potential. In particular, the ground-state wave function found in this
manner represents the initial ($t=0$) atomic wave packet -- the initial condition for single-atom
dynamics. In what follows, we briefly review the ITE approach, which represents one of the most robust
computational approaches for obtaining the ground state of a quantum system~\cite{Anderson:75,Feit+:82}.
We do so using the abstract, representation-independent notation.

Assuming that $H$ is the Hamiltonian whose ground state one aims to determine, one expands
the quantum state $|\phi\rangle$ in terms of the eigenstates $\{|\psi_n\rangle,\:n=0,1,\ldots\}$
of $H$, i.e. $|\phi\rangle=\sum_{n}c_n|\psi_n\rangle$. It is assumed that the sought-after ground
state $|\psi_0\rangle$ has a nonvanishing contribution to this expansion, i.e. $c_0 \equiv \langle
\psi_0|\phi\rangle \neq 0$. Assuming that the system is initially ($t=0$) in the state $|\phi\rangle$,
its states at later times $t$ are given by
\begin{equation}\label{eqRealTimeEvolution}
|\phi(t)\rangle=\sum_{n} c_n e^{-iE_n t/\hbar}|\psi_n\rangle \:,
\end{equation}
with $E_n$ being the eigenvalue of $H$ corresponding to the eigenstate $|\psi_n\rangle$.

By switching from real to imaginary time $t\rightarrow \tau=it$, i.e. performing a Wick rotation
into the complex plane, Eq.~\eqref{eqRealTimeEvolution} adopts the form
\begin{equation}\label{eqImagTimeEvolution}
|\phi(\tau)\rangle=\sum_{n}c_n e^{-E_n\tau/\hbar}|\psi_n\rangle\:.
\end{equation}
The last equation expresses $|\phi(\tau)\rangle$ as an exponentially-decaying superposition of
the eigenstates $|\psi_n\rangle$ with the decay rates given by the corresponding eigenvalues
$E_n$. Because the exponential-decay rate of the ground state is the smallest one, in the large-$\tau$
limit one obtains
\begin{equation}\label{LargeTauEvolution}
|\phi(\tau)\rangle \approx c_0 e^{-E_0\tau/\hbar}|\psi_0\rangle\:.
\end{equation}
This implies that, given a trial state $|\phi(\tau=0)\rangle\equiv|\phi\rangle$ with a nonzero
overlap with the actual ground state, the ITE approach engenders a monotonously decreasing functional 
of $\tau$ that converges to the ground-state energy as $\tau\rightarrow\infty$. In other words,
$E_0 = \lim_{\tau\rightarrow\infty}\langle\phi(\tau)|H|\phi(\tau)\rangle$.

In the problem at hand, we make use of the coordinate representation, thus the counterpart of 
Eq.~\eqref{LargeTauEvolution} of interest for the present work then reads
\begin{equation}\label{eqImagTimeEvolution}
\phi(\vec{r},\tau)=\sum_{n} c_n e^{-E_n\tau/\hbar}\psi_n(\vec{r})\:,
\end{equation}
where $\phi(\vec{r},\tau)\equiv \langle\mathbf{r}|\phi(\tau)\rangle$ and $\psi_n(\vec{r})
\equiv\langle\mathbf{r}|\psi_n\rangle$ are the relevant wave functions.

For completeness, it is worthwhile to mention that the ground state of the DWOL potential can also be
obtained to requisite accuracy using the plane-wave-expansion method (see, e.g., Ref.~\cite{Stojanovic+:08}).
This approach is based on expanding the underlying periodic potential and the periodic part of single-particle
Bloch wave functions in terms of reciprocal-lattice vectors.

\begin{figure}[b!]
\includegraphics[width=0.95\linewidth]{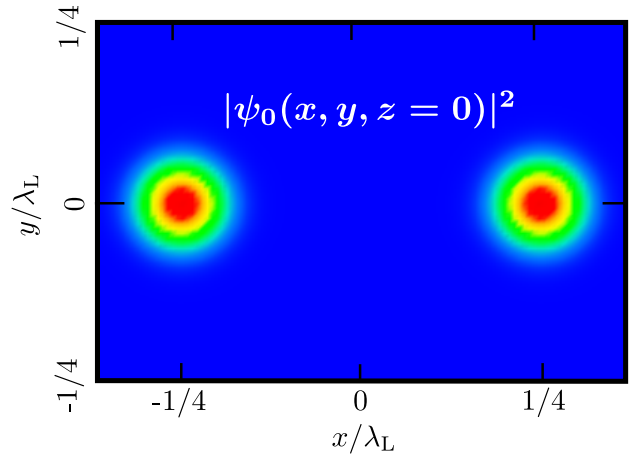}
\caption{\label{GroundStateWF}(Color online) The squared modulus $|\psi_0(x,y,z=0)|^2$ of the 
ground-state wave function (i.e. the ground-state probability density) in the $x-y$ plane for 
$U_{d,0}=300\:E_{\textrm{R}}$. The transverse beam waists are set to $w_\mathrm{x/y,0}=4.2\times 
10^3\:l_\mathrm{x}$. At the same time, the relevant angles are chosen such that $\beta=3\pi/20$ 
and $\theta=\phi=\pi/2$.}
\end{figure}

The squared modulus $|\psi_0(x,y,z=0)|^2$ of the ground-state wave function, i.e. the ground-state probability 
density, in the $x-y$ plane -- computed using the ITE approach -- is shown for one specific choice of system 
parameters in Fig.~\ref{GroundStateWF}. The obtained bimodal shape of this probability density originates from 
the characteristic form of the ground-state wave function in a double well.

\subsection{Relevant TDSE in the comoving frame and its FSOM-based numerical solution} \label{ComovingFrameFSOM}
The character of the problem at hand makes it prudent to switch from the lab frame to the
one that corresponds to the DWOL ({\em comoving} frame). This change is effected through
a generalized Galilean transformation~\cite{GottfriedYanBOOK:03}. The latter is represented
by the unitary operator (see, e.g. Ref.~\cite{SLKannals:08})
\begin{equation}\label{eqUnitaryOperatorTrapFrame}
\hat{\mathcal{U}} = e^{i \hat{\mathbf{p}}\cdot \mathbf{q}_0(t)/\hbar}
\:e^{-i m \hat{\mathbf{r}}\cdot\dot{\mathbf{q}}_0(t)/\hbar} \:,
\end{equation}
which transforms the relevant lab-frame TDSE
\begin{equation}
i\hbar\:\frac{\partial}{\partial t}\Psi(\mathbf{r},t)=
\left[\frac{\hat{\mathbf{p}}^2}{2m}+U_\mathrm{D}(\mathbf{r}-
\mathbf{q}_0(t))\right]\Psi(\mathbf{r},t)
\end{equation}
with the single-atom wave-function $\Psi(\mathbf{r},t)$ into its counterpart in the
comoving frame with its corresponding wave-function $\Phi(\mathbf{r},t)\equiv\hat{\mathcal{U}}
\Psi(\mathbf{r},t)$:
\begin{eqnarray}\label{eqSchroedingerequationTrapFrame}
i\hbar\:\frac{\partial}{\partial t}\Phi(\mathbf{r},t) &=&
\Big[\frac{\hat{\mathbf{p}}^2}{2m}+\frac{m}{2}\:\dot{\mathbf{q}}_{0}^2(t)
+ m \ddot{\mathbf{q}}_0\cdot \mathbf{q}_0 \nonumber \\
&+& m \mathbf{r}\cdot\ddot{\mathbf{q}}_0(t)+U_\mathrm{D}(\mathbf{r})
\Big]\Phi(\mathbf{r},t) \:.
\end{eqnarray}
It is important to stress that the terms $m\dot{\mathbf{q}}_{0}^2(t)/2$ and $m\ddot{\mathbf{q}}_0(t)
\cdot\mathbf{q}_0(t)$ in Eq.~\eqref{eqSchroedingerequationTrapFrame} lead only to time-dependent global
phase factors, which are immaterial for the atom-transport problem at hand. Therefore, these terms
can henceforth be safely neglected and the total potential experienced by a transported atom 
in the comoving frame is given by $W(\mathbf{r},t)\equiv U_\mathrm{D}(\mathbf{r})+
m\mathbf{r}\cdot\ddot{\mathbf{q}}_0(t)$.

In the atom-transport problem at hand, we make use of the FSOM to compute the final atomic state after 
center-of-mass displacement of the atomic wave packet by a certain distance. Because the potential 
$W(\mathbf{r},t)$ carries an explicit time dependence, the exact time-evolution operator of the system 
is given by the most general expression that involves a time-ordered product. Yet, by making use of the 
identity in Eq.~\eqref{SymmetricTrotter} for $\hat{A}=U_\mathrm{D}(\hat{\mathbf{r}})$ and 
$\hat{B}=\hat{\mathbf{p}}^2/(2m)+m\hat{\mathbf{r}}\cdot\ddot{\mathbf{q}}_0(t)$ the counterpart 
of the time-stepping scheme of Eq.~\eqref{eqFSOMBCH} in the atom-transport problem under consideration 
reads
\begin{eqnarray}\label{eqFSOM_DWOL}
\Psi(\vec{r},t&+&\delta t) = \exp\left[-\frac{i}{\hbar}\:U_\mathrm{D}(\vec{r})
\frac{\delta t}{2}\right]T_{\mathbf{r},\mathbf{p}}(\delta t)\nonumber\\
&\times& \exp\left[-\frac{i}{\hbar}\:U_\mathrm{D}(\vec{r})\frac{\delta t}{2}\right]
\Psi(\vec{r},t)+ \mathcal{O}(\delta t^3)  \:,
\end{eqnarray}
where $\hat{T}_{\mathbf{r},\mathbf{p}}(\delta t)$ is defined as 
\begin{equation}\label{TrpDef}
\hat{T}_{\mathbf{r},\mathbf{p}}(\delta t)\equiv\exp\left[-\frac{i}{\hbar} 
\left(\frac{\hat{\mathbf{p}}^2}{2m}\:\delta t+m\hat{\mathbf{r}}\cdot\int_t
^{t+\delta t}\ddot{\mathbf{q}}_0(t')\mathrm{d}t'\right)\right] \:.
\end{equation}
Using the Baker-Campbell-Hausdorff formula~\cite{GilmoreBOOK:12} and its immediate implications,
we can recast $\hat{T}_{\mathbf{r},\mathbf{p}}(\delta t)$ as (for a detailed derivation, see Appendix~\ref{derivationTrp})
\begin{eqnarray} \label{TrpFinal}
\hat{T}_{\mathbf{r},\mathbf{p}}(\delta t) &=& \exp\left[-\frac{i}{\hbar} 
\frac{\hat{\mathbf{p}}^2}{2m}\:\delta t \right] \,
\exp\left[-\frac{i}{2\hbar}\:\hat{\mathbf{p}}\cdot\delta\dot{\mathbf{q}}_0(t)\:\delta t\right] \, \nonumber\\
&\times& \exp\left[-\frac{i}{\hbar}\:m\hat{\mathbf{r}}\cdot\delta\dot{\mathbf{q}}_0(t)\right] \:,
\end{eqnarray}
where $\delta\dot{\mathbf{q}}_0(t)\equiv \dot{\mathbf{q}}_0(t+\delta t)-\dot{\mathbf{q}}_0(t)$.
It is worthwhile mentioning that another two multiplicative terms -- which do not depend on the momentum- 
and coordinate operators and lead to irrelevant time-dependent global phase factors -- have been omitted 
on the RHS of the last equation, by analogy to the omitted terms of Eq.~\eqref{eqSchroedingerequationTrapFrame}
above.

\begin{figure}[t!]
\includegraphics[width=0.95\linewidth]{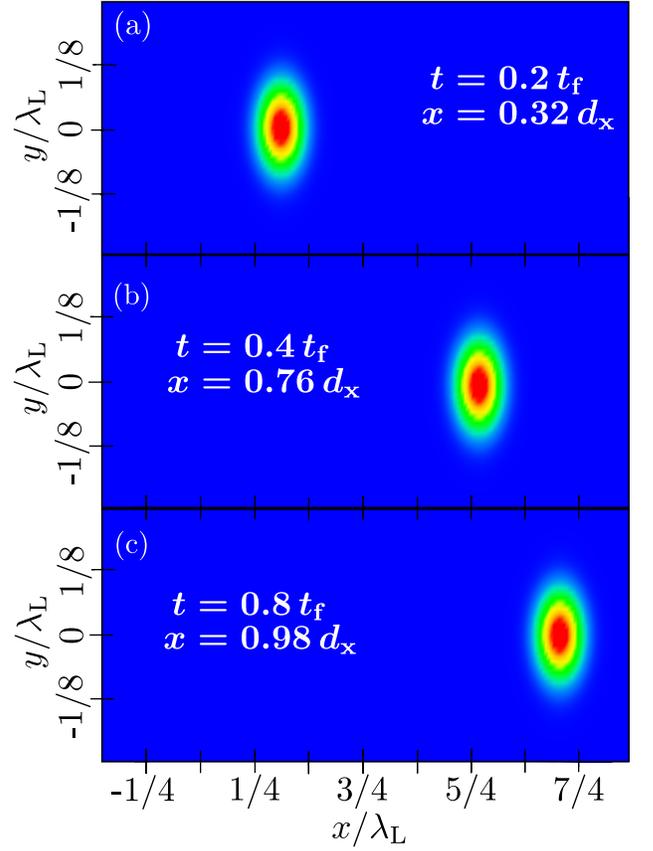}
\caption{\label{WavePacketMotion}(Color online) Illustration of the motion of a single-atom wave 
packet located in the left well of a double well at $t=0$ and transported to the distance $d_\mathrm{x}=30.2\:l_\mathrm{x}$ 
at $t=t_{\mathrm{f}}$. The snapshots of the wave-packet motion, which corresponds to an eSTA-based 
moving-DWOL trajectory, are shown for (a) $t=0.2\:t_{\mathrm{f}}$, (b) $t=0.4\:t_{\mathrm{f}}$, 
and (c) $t=0.8\:t_{\mathrm{f}}$. The values of the system parameters $(U_{d,0},w_{0,x}, w_{0,y},
\beta,\phi,\theta)$ are the same as in Fig.~\ref{GroundStateWF}.}
\end{figure}

By making use of an analog of Eq.~\eqref{eqFourierRepresentation} and performing a Fourier transformation of 
the $\hat{\mathbf{p}}$-dependent terms in \eqref{TrpFinal} [switching also to the coordinate representation 
of the momentum operator $\hat{\mathbf{p}}\rightarrow (\hbar/i)\nabla$], we finally find that the FSOM-based 
second-order time-stepping scheme for the atom-transport problem at hand is given by:
\begin{eqnarray}\label{eqFSOMFinal}
\Phi(\mathbf{r},t&+&\delta t)= \exp\left[-\frac{i}{\hbar}\:U_\mathrm{D}(\mathbf{r})
\frac{\delta t}{2}\right]\mathrm{F}^{-1}\Bigg[e^{-i\frac{\hbar k^2}{2m}
\delta t} \nonumber \\
&\times& \exp\left[-\frac{i}{2}\:\mathbf{k}\cdot\delta\dot{\mathbf{q}}_0\delta t\right]
\mathrm{F}\Bigg\{\exp\left[-\frac{i}{\hbar}\:m\mathbf{r}\cdot\delta\dot{\mathbf{q}}_0\right] \nonumber\\
&\times& \exp\left[-\frac{i}{\hbar}\:U_\mathrm{D} (\mathbf{r}) \frac{\delta t}{2}\right]
\Phi(\mathbf{r},t)\Bigg\}\Bigg] + \mathcal{O}(\delta t^3)  \:.
\end{eqnarray}

The typical dynamics of an atomic wave-packet moving in the $x$-direction in a DWOL, evaluated by numerically solving 
the above TDSE in the comoving frame based on Eq.~\eqref{eqFSOMFinal} with an eSTA-based moving-DWOL trajectory, are 
illustrated in Fig.~\ref{WavePacketMotion}. 
The three snapshots of the motion shown in this plot illustrate the characteristic incipient acceleration and 
the terminal slowing down in atom transport. Namely, as can be inferred from Fig.~\ref{WavePacketMotion}, the 
distance covered in the first fifth of the total transport time $t=t_{\mathrm{f}}$ constitutes nearly one third 
of the total distance $d_\mathrm{x}$, while the one corresponding to the final fifth amounts to only $0.02\:d_\mathrm{x}$.
\subsection{Details of the numerical implementation} \label{DetailsNumImpl}
For the sake of alleviating the computational burden in the problem under consideration, it is pertinent to
make use of the discrete translational symmetry of the system and restrict ourselves to the displacement of
one unit cell of the original DWOL. Needless to say, the appropriate choice of the unit cell has to be consistent
with the transport direction. For instance, in order to evaluate transport in the $x$- and $y$ directions
we make use of a rectangular unit cell (indicated by white dashed lines in Fig.~\ref{DWOLtransport_Fig1}).
At the same time, describing transport in the diagonal direction necessitates the use of a $45$-degree rotated
square-shaped unit cell (black dashed lines in Fig.~\ref{DWOLtransport_Fig1}).

Another problem-specific circumstance that allows one to further reduce the computational burden in the problem
at hand is intimately related to its description in the comoving frame [cf. Sec.~\ref{ComovingFrameFSOM}]. Namely,
from the form of the relevant TDSE [cf. Eq.~\eqref{eqSchroedingerequationTrapFrame}] it can straightforwardly
be inferred that there is no need to calculate the potential term after each time step. Instead, we need only
evaluate the correction term $m\mathbf{r}\cdot\ddot{\mathbf{q}}_0$ [cf. Eq.~\eqref{TrpDef}] that depends on the 
acceleration $\ddot{\mathbf{q}}_0$ of the potential minimum.

To carry out the spatial Fourier transformation, here implemented using the FFT algorithm [recall the discussion
in Sec.~\ref{FSOMbasics}], we make use of a discrete three-dimensional $N_x\times N_y\times N_z$ grid with $N_x=200$, 
$N_y = 300$, and $N_z=100$ points in the computational window whose respective dimensions (in units of $l_\mathrm{x}$, 
$l_\mathrm{y}$, and $l_\mathrm{z}$) are $100$, $100$, and $500$. As for the numerical time-domain propagation of 
the relevant TDSE in the comoving frame [cf. Eq.~\eqref{eqSchroedingerequationTrapFrame}], we utilize the adaptive 
approach~\cite{NRcBook} with a maximal relative error of $10^{-4}$, which could -- in principle -- require additional 
(smaller) substeps in time. However, in all our runs, it was sufficient to take $N_t$ between $20$ and $100$.

\section{Transport fidelity: Results and discussion} \label{ResultsDiscussion}
In the following, we discuss the efficiency of single-atom transport in a DWOL resulting
from the STA- and eSTA-based trap trajectories [cf. Secs.~\ref{STAtrajectory} and \ref{eSTAtrajectory}, respectively].
The central figure of merit that quantifies the efficiency of atom transport is the transport fidelity $\mathcal{F}(t_f)
=|\braket{\Psi_{\textrm{target}}|\Psi(t_f)}|^2$. The dependence of $\mathcal{F}$ on the transport time $t_f$ is determined
by the overlap of the target state $|\Psi_{\textrm{target}}\rangle$ (the ground state of the displaced DWOL potential)
and the final atomic state $|\Psi(t_f)\rangle$, obtained using the FSOM.

\begin{figure}[t!]
\includegraphics[width=0.85\linewidth]{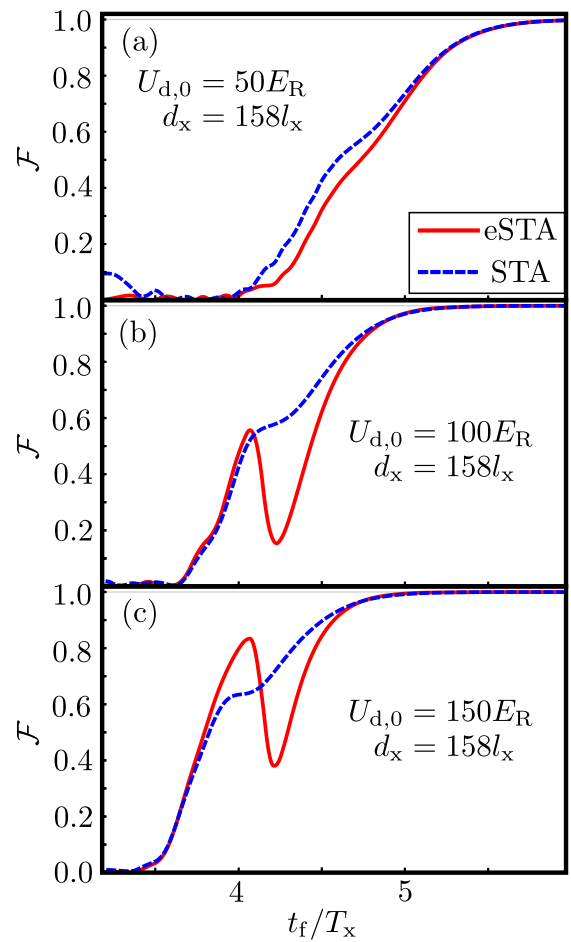}
\caption{\label{FidelityResults1}(Color online) The dependence of the atom-transport fidelity on the transport time $t_\mathrm{f}$, 
for a lattice depth $U_\mathrm{d,0}$ of (a) $50\:E_{\textrm{R}}$, (b) $100\:E_{\textrm{R}}$ and (c) $150\:E_{\textrm{R}}$. 
The transport distance $d_\mathrm{x}$ along the $x$ axis was set to $158\:l_\mathrm{x}$, while the transverse beam waists 
are $w_\mathrm{x/y,0}=4.2\times 10^3\:l_\mathrm{x}$. The relevant angles are chosen such that $\beta = 3\pi/20$ and 
$\theta=\phi=\pi/2$.}
\end{figure}

\begin{figure}[b!]
\includegraphics[width=0.85\linewidth]{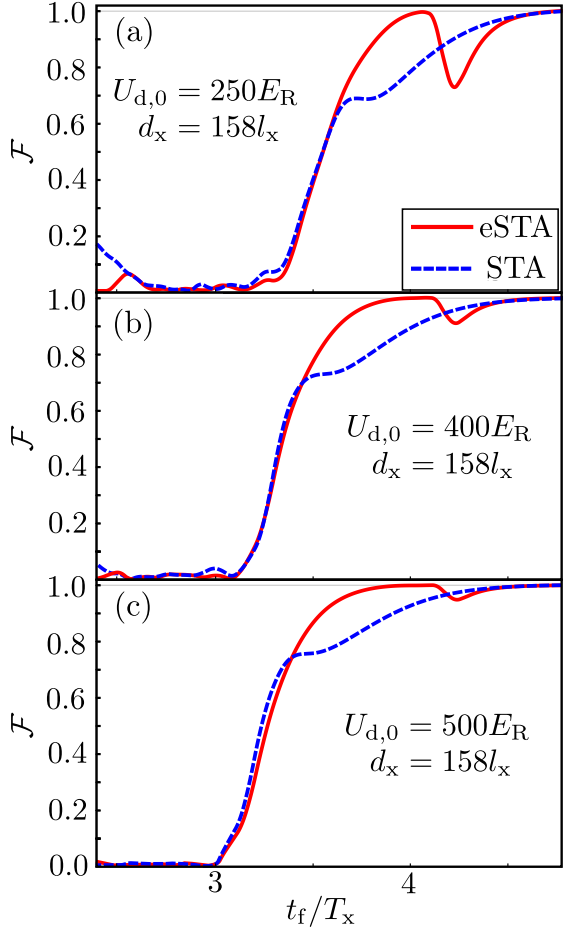}
\caption{\label{FidelityResults2}(Color online) The dependence of the atom-transport fidelity on the
transport time $t_\mathrm{f}$, for a lattice depth $U_\mathrm{d,0}$ of (a) $250\:E_{\textrm{R}}$, 
(b) $400\:E_{\textrm{R}}$ and (c) $500\:E_{\textrm{R}}$. The transport distance $d_\mathrm{x}$ along 
the $x$-axis was set to $158\:l_\mathrm{x}$. For the Gaussian beam along the $z$-axis the transverse 
beam waists were set to $w_\mathrm{x/y,0}=4.2\times 10^3\:l_\mathrm{x}$. The relevant angles are chosen
such that $\beta=3\pi/20$ and $\theta=\phi=\pi/2$. }
\end{figure}

\begin{figure}[t!]
\includegraphics[width=0.85\linewidth]{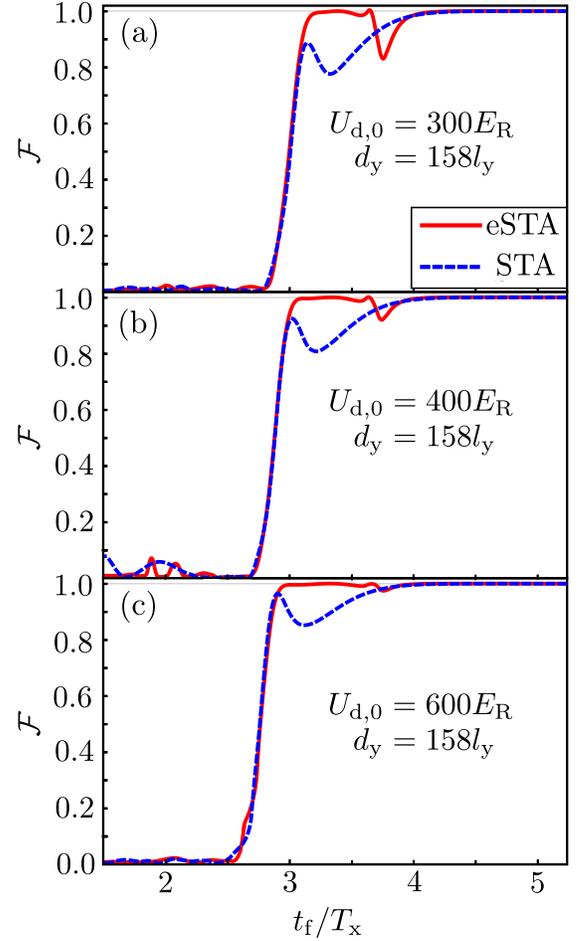}
\caption{\label{FidelityResults3}(Color online) The dependence of the atom-transport fidelity on the transport time $t_\mathrm{f}$,
for a lattice depth $U_\mathrm{d,0}$ of (a) $300\:E_{\textrm{R}}$, (b) $400\:E_{\textrm{R}}$ and (c) $600\:E_{\textrm{R}}$.
The transport distance $d_\mathrm{y}$ along the $y$-axis was set to $158\:l_\mathrm{y}$. For the Gaussian beam along the $z$-axis 
the transverse beam waists were set to $w_\mathrm{x/y,0}=4.2\times 10^3\:l_\mathrm{x}$. The relevant angles are chosen such 
that $\beta=3\pi/20$ and $\theta=\phi=\pi/2$.}
\end{figure}

\begin{figure}[b!]
\includegraphics[width=0.85\linewidth]{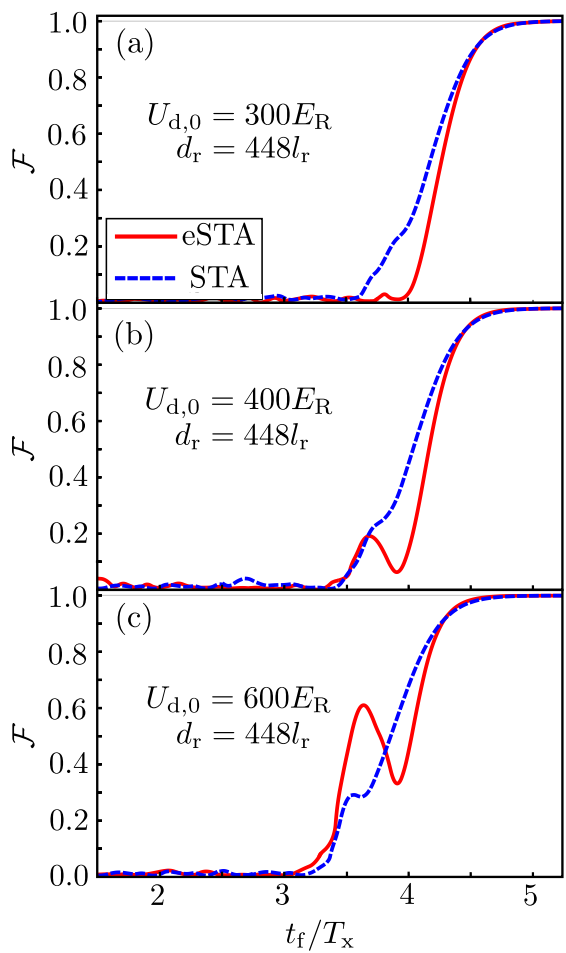}
\caption{\label{FidelityResults4}(Color online) The dependence of the atom-transport fidelity on the transport time $t_\mathrm{f}$,
for a potential strengths $U_\mathrm{d,0}$ (a) of $300\:E_{\textrm{R}}$, (b) $400\:E_{\textrm{R}}$ and (c) $600\:E_{\textrm{R}}$.
The transport distance $d_\mathrm{r}$ along the diagonal was set to $448\:l_\mathrm{r}$. For the Gaussian beam along the $z$-axis 
the transverse beam waists were set to $w_\mathrm{x/y,0}=4.2\times 10^3\:l_\mathrm{x}$. The relevant angles are chosen such that 
$\beta=3\pi/20$ and $\theta=\phi=\pi/2$. }
\end{figure}

\begin{figure}[t!]
\includegraphics[width=0.85\linewidth]{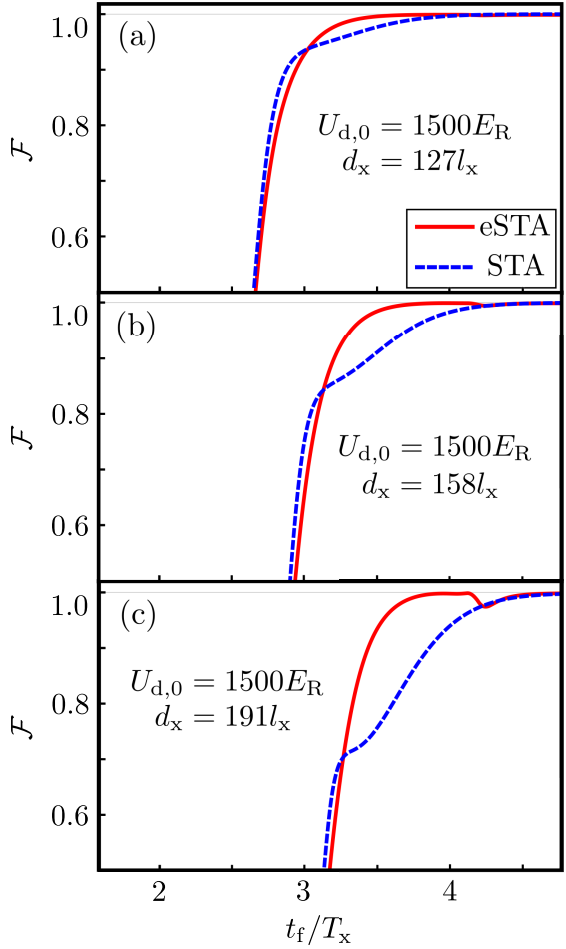}
\caption{\label{FidelityResults5}(Color online) The dependence of the atom-transport fidelity on the 
transport time $t_\mathrm{f}$, for a lattice depth $U_\mathrm{d,0}$ of $1500\:E_{\textrm{R}}$.
The transport distance $d_\mathrm{x}$ along the $x$-axis was set (a) to $127\:l_\mathrm{x}$, (b) 
$158\:l_\mathrm{x}$ and (c) $191\:l_\mathrm{x}$.
For the Gaussian beam along the $z$-axis the transverse beam waists were set to $w_\mathrm{x/y,0}=4.2
\times 10^3\:l_\mathrm{x}$. The relevant angles are chosen such that $\beta=3\pi/20$ and $\theta=\phi=\pi/2$.}
\end{figure}

\begin{figure}[b!]
\includegraphics[width=0.85\linewidth]{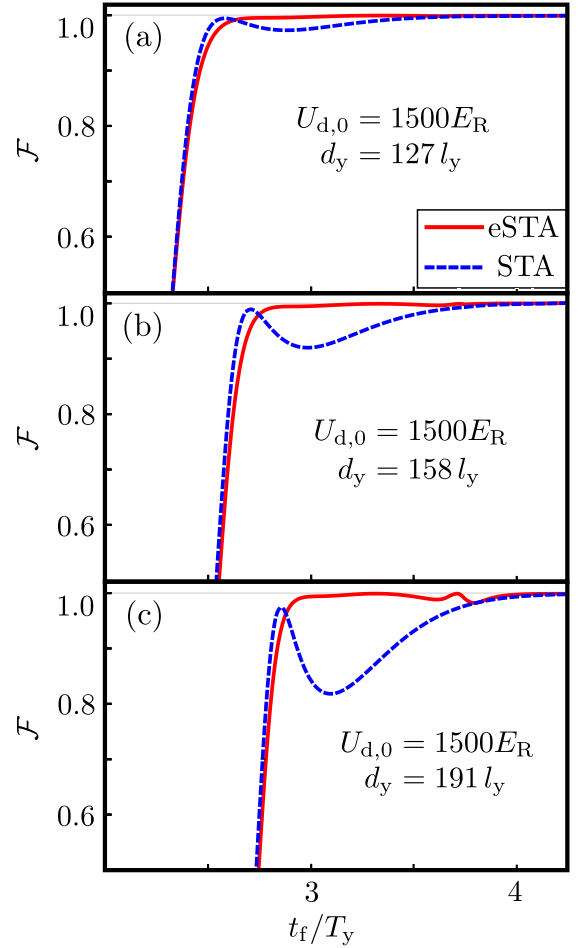}
\caption{\label{FidelityResults6}(Color online) The dependence of the atom-transport fidelity on the 
transport time $t_\mathrm{f}$, for a potential strengths $U_\mathrm{d,0}$ of $1500\:E_{\textrm{R}}$.
The transport distance $d_\mathrm{y}$ along the $y$-axis was set (a) to $127\:l_\mathrm{y}$, (b) $158\:l_\mathrm{y}$ 
and (c) $191\:l_\mathrm{y}$. For the Gaussian beam along the $z$-axis the transverse beam waists were set to 
$w_\mathrm{x/y,0}=4.2\times 10^3\:l_\mathrm{x}$. The relevant angles are chosen such that $\beta=3\pi/20$ 
and $\theta=\phi=\pi/2$.}
\end{figure}

\begin{figure}[t!]
\includegraphics[width=0.85\linewidth]{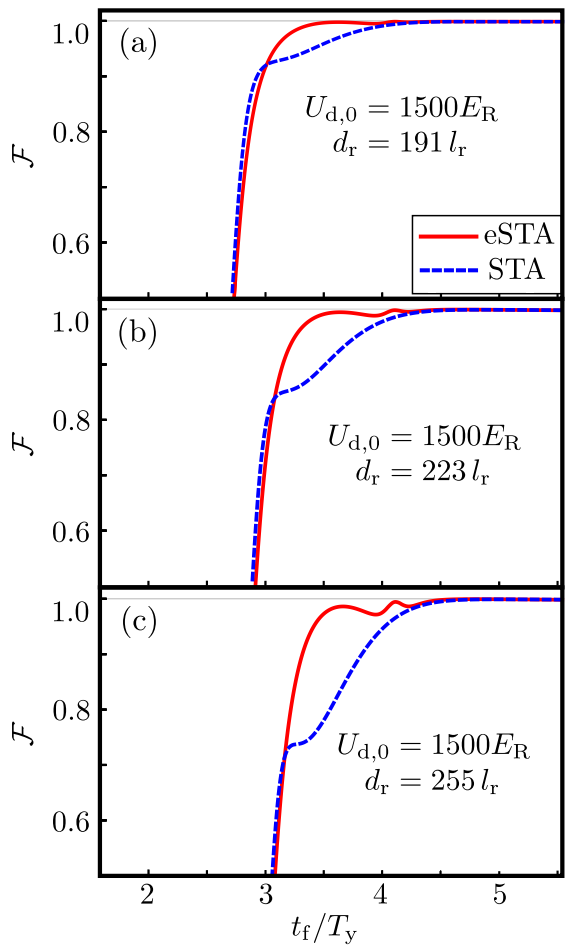}
\caption{\label{FidelityResults7}(Color online) The dependence of the atom-transport fidelity on the transport time $t_\mathrm{f}$, 
for a lattice depth $U_\mathrm{d,0}$ of $1500\:E_{\textrm{R}}$. The transport distance $d_\mathrm{r}$ along the diagonal was set 
to (a) $127\:l_\mathrm{r}$, (b) $158\:l_\mathrm{r}$, and (c) $191\:l_\mathrm{r}$. For the Gaussian beam along the $z$-axis 
the transverse beam waists were set to $w_\mathrm{x/y,0}=4.2\times 10^3\:l_\mathrm{x}$. The relevant angles are chosen such 
that $\beta=3\pi/20$ and $\theta=\phi=\pi/2$. }
\end{figure}

The obtained results for the transport fidelity in the DWOL system under consideration are illustrated 
in Figs.~\ref{FidelityResults1} -- \ref{FidelityResults7}. One common feature of all these results,
regardless of the concrete values of the system parameters (lattice depth, transport distance, beam waists, etc.),
is the breakdown of transport -- manifested by a sharp decrease in the fidelity -- for short transport times
$t_f$. This characteristic breakdown results from the fact that for atomic accelerations above a certain maximal
value $|a_{\textrm{max}}|$, which is proportional to the lattice depth~\cite{Schrader+:01}, the transported
atom cannot be confined by the optical-lattice potential. This coincides with the dissappearance of the 
minima of the potential experienced by the atom in its comoving (non-inertial) reference frame, which differs 
from the potential in the lab frame by a term linear in the transport coordinate~\cite{Hickman+Saffman:20} (thus,
effectively representing a tilted lattice potential). This behavior for short transport times is a generic 
characteristic of coherent atom transport in optical-lattice potentials, i.e. it is not specific only for DWOLs.

The aforementioned collapse actually takes place when the maximal atomic acceleration reached during the transport process,
which will be denoted by $|\tilde{a}_{\textrm{max}}|$ in what follows, exceeds $|a_{\textrm{max}}|$. The value of 
$|\tilde{a}_{\textrm{max}}|$ depends on the concrete moving-lattice trajectory (i.e. the path of the potential minimum), but 
it generally holds that $|\tilde{a}_{\textrm{max}}| \propto t_f^{-2}$. This implies that for larger lattice depths (i.e. for higher 
$|a_{\textrm{max}}|\propto U_0$) the largest atomic acceleration $|\tilde{a}_{\textrm{max}}|$ reached exceeds $|a_{\textrm{max}}|$
at shorter transport times $t_f$. Rephrasing, for deeper lattices (i.e. stronger lattice potentials) the collapse of the fidelity 
takes place for shorter times $t_f$. Our numerical findings are in agreement with this general argument, as can be inferred
from Figs.~\ref{FidelityResults1} -- \ref{FidelityResults4} for gradually increasing potential depths. The characteristic transport
times $t_f$ that correspond to the onset of the aforementioned breakdown are around $3.8\:T_\mathrm{x}$ and $3.3\:T_\mathrm{x}$,
respectively, in Figs.~\ref{FidelityResults1} and \ref{FidelityResults2}. Therefore, they clearly show the anticipated trend of 
decreasing with the increase of the lattice depth. By contrast to this short-time behavior, in the opposite limit of very long 
transport times $t_\mathrm{f}$ both STA and eSTA results correspond to the well-known adiabatic limit. In other words, both methods 
allow one to achieve essentially perfect atom transport ($\mathcal{F}\approx 1$) for $t_\mathrm{f}\gg\omega_\mathrm{u}^{-1}$ ($u=x,y$).

By comparing the two methods used, the superiority of the eSTA method over its STA counterpart for most transport 
times is easily noticeable. This superiority becomes more prominent for deeper lattices, which can be explained by 
the fact that the parameter $\mu_\mathrm{S}$ [cf. Eq.~\eqref{eqCloseHamiltonian}] depends on the characteristic wave 
number $k_\mathrm{L}$ of the potential, while $k_\mathrm{L}$ in turn depends on the lattice depth $U_\mathrm{d,0}$ 
[i.e. $\mu_\mathrm{s}\propto U_\mathrm{d,0}^{-1/2}$]. Thus, larger lattice depths result in smaller values of $\mu_\mathrm{S}$
and, consequently, in a better approximation of the optimization vector pertaining to the eSTA method [cf. Sec.~\ref{eSTAbasics}].
Furthermore, with increasing lattice depth the stability of the eSTA scheme is improving. This can readily be seen 
by the decrease of the oscillatory feature around $4.15\:T_\mathrm{x}$, $3.8\:T_\mathrm{x}$ in Figs. \ref{FidelityResults2} 
and \ref{FidelityResults3}, respectively. However, this only holds true if eSTA already results in a significant improvement 
over STA, otherwise an increase of the lattice depth $U_\mathrm{d,0}$ would lead to a more prominent oscillatory feature 
around $4.15\:T_\mathrm{x}$ and $3.7\:T_\mathrm{x}$ in Figs. \ref{FidelityResults1} and \ref{FidelityResults4}, 
respectively.

To arrive at a more complete physical understanding of coherent atom transport in DWOLs, it is pertinent to also comment 
on the results for the transport fidelity [cf. Figs.~\ref{FidelityResults1} -- \ref{FidelityResults3}] from the standpoint of 
of the characteristic shapes of the corresponding trajectories of a moving DWOL, as obtained using the STA- 
[cf. Fig.~\ref{fig:STAminimum}] and eSTA methods [cf. Fig.~\ref{fig:eSTAminimum}]. What can be inferred by inspecting those 
results is that the typical transport times $t_f$ required for a high-fidelity atom transport correspond to moving-DWOL trajectories 
that do not display oscillatory features. For example, the eSTA-based trajectory that corresponds to $t_f/T_\mathrm{x} = 2$ in 
Fig.~\ref{fig:eSTAminimum}(a) clearly shows oscillating character, but does not permit high-fidelity atom transport in the 
$x$ direction. Similar conclusions can be drawn in connection with atom transport in other directions, discussed in what
follows. Thus, in the DWOL system considered here only non-oscillatory solutions for the moving-DWOL trajectory can enable 
high-fidelity transport.

Aside from the nonseparability of its underlying optical potential $U_\mathrm{D}(x,y,z)$ in the $x-y$ plane 
[cf. Eq.~\eqref{eqDWOLPot}], one of the central features of the DWOL is the anisotropy of this potential. The 
results for atom transport in the $x-y$ plane are therefore strongly direction-dependent and exhibit different 
general behaviour, as can be inferred by comparing Figs.~\ref{FidelityResults2} and \ref{FidelityResults3}.
Only the trivial parameter choices of $\beta=0$ or $\beta=\pi/2$, with $\theta=\pi/2$, would result in a simpler 
form of the potential, which in those cases is similar to that of the ordinary (single-well) optical lattice~\cite{Hauck+:21}.

Apart from the already discussed consequences of the anisotropic character of the underlying optical potential for 
atom transport, another effect specific for DWOLs is worthwhile being considered here. Namely, the presence of the 
intermediate potential barrier between the two wells within a double well has profound consequences for atom transport. 
By analogy to the existence of the maximal acceleration $|a_{\textrm{max}}|$ that a lattice can endure before atom 
transport breaks down (closely related, as explained above, to the dissappearance of minima of the potential in the 
reference frame moving with the transported atom), above a certain acceleration $|a_\mrm{int}|$ -- which is generally 
smaller than $|a_\mathrm{max}|$ -- the intermediate barrier will cease to be a local maximum of the tilted lattice 
potential. Its transport implications are intimately related to the fact that once $|a_\mathrm{int}|$ is exceeded the
wave function of the transported atom becomes a mixture of the wave functions corresponding to adjacent single-well 
minima within the unit cell of a DWOL. Consequently, this leads to a noticeable drop in the transport fidelity.

The consequences of the existence of the intermediate potential barrier can be understood by comparing the fidelities shortly 
before the breakdown for transport in $x$- and $y$ directions [c.f. Figs.~\ref{FidelityResults2} and \ref{FidelityResults3}].
It is noticeable that the peak present around $3\:T_\mathrm{y}$ in Fig.~\ref{FidelityResults3} does not exist for its 
$x$-direction counterpart in Fig.~\ref{FidelityResults2}. Only in the trivial cases $\beta=0$ and $\beta=\pi/2$ the
acceleration $|a_\mathrm{max}|$ remains relevant, because in those cases all barriers between adjacent minima have 
the same height. In all other cases it is possible to exceed $|a_\mathrm{int}|$ without exceeding $|a_\mathrm{max}|$ for 
short enough transport times. Therefore, through the angle $\beta$ one can control the mixing of the wave functions corresponding 
to adjacent minima by controlling the height of the intermediate barrier. Accordingly, by tuning $\beta$ one can control 
the behaviour of the fidelity pertaining to the transport in the $x$ direction for times close to the breakdown time.

The diagonal displacement results in worse results for eSTA in comparison to STA, Fig.~\ref{FidelityResults4}.
However, this is not expected to be a direct problem with the 2D adaptation of the eSTA procedure but rather 
shows that the parameter pair of the distance and lattice depth is not suited for the displacement problem.
Comparing parts (a) and (b) of Fig.~\ref{FidelityResults4} with (c) shows that an increase of the lattice depth 
results in a better performance of the eSTA method. Increasing the potential depth even further would result in 
a superior performance of eSTA over STA. 

Having discussed both general atom-transport effects and those specific for DWOLs, for the sake of completeness
it is pertinent to discuss the STA- and eSTA-based transport for very deep lattices ($U_\mathrm{d,0}\gtrsim 
1000\:E_{\mathrm{R}}$). It should be stressed, that the maximal lattice depth is in principle in multiple 
thousands of the recoil energy $E_{\mathrm{R}}$ and is essentially determined by the available laser power,
as well as the capability to sufficiently tightly focus the laser beam. In particular, 
Figs.~\ref{FidelityResults5} - \ref{FidelityResults7} show different examples of transport in the $x$-, $y$-, 
and diagonal directions for different total transport distances, all of them obtained for $U_\mathrm{d,0}=1500\:E_{\textrm{R}}$.

The $x$- direction transport in the deep-lattice regime, illustrated by Fig.~\ref{FidelityResults5}, shows a significantly 
better performance of eSTA in comparison to STA. In particular, in the close vicinity of the total breakdown around $t_\mathrm{f}
\approx 2.75\:T_\mathrm{x}$ and $t_\mathrm{f}\approx 3.25\:T_\mathrm{x}$, the eSTA-based trajectories still allow one to reach
fidelities above $0.9$. By contrast to that, the fidelity of the STA-based transport falls below $0.9$ for a transport 
distances of $d_\mathrm{x}=158\:l_\mathrm{x}$ [see Fig.~\ref{FidelityResults5}(b)], and even below $0.75$ for distances 
of $d_\mathrm{x}=191\:l_\mathrm{x}$ [cf. Fig.~\ref{FidelityResults5}(c)].

The $y$-direction transport in the deep-lattice regime displays similar behavior to that of a simple optical conveyor belt~\cite{Hauck+:21},
in that the local maxima reach nearly perfect fidelity before the occurrence of the transport breakdown. This is not 
completely surprising, however, because the two systems share a similar periodicity along the $y$ direction. Furthermore, 
our results strongly suggest that even for very large lattice depths the eSTA procedure becomes more unstable for longer total 
transport distances. This trend is manifested by the growing oscillations in Fig.~\ref{FidelityResults6}, resulting in
a worse performance of eSTA than STA for final times around $t_\mathrm{f}\approx 3.7\:T_\mathrm{y}$. These oscillation are
also clearly visible in Fig.~\ref{FidelityResults5}(c) around $t_\mathrm{f}\approx 4.15\:T_\mathrm{x}$.

The results obtained for diagonal atom transport are even better for eSTA in comparison to STA and previously discussed 
parameter sets, this being manifested by (significantly) better or equal fidelity values for all probed final times above 
the breakdown time $t_\mathrm{f}\approx 3\:T_\mathrm{x}$. Even the oscillatory behaviour, which can be noticed in Fig.~\ref{FidelityResults7} 
(b) and (c) around $t_\mathrm{f}\approx 3.9\:T_\mathrm{x}$, results in higher fidelity values for eSTA in comparison to STA.

For the sake of completeness, it is worthwhile pointing out that -- in addition to outperforming STA-based schemes 
in terms of achievable transport fidelity -- the eSTA method also shows increased stability (i.e. robustness against errors)
compared to its STA counterpart. That has quite recently been demonstrated on the example of atomic transport in conventional 
1D optical lattices in Ref.~\cite{Whitty+:22}. In that recent paper, a general heuristic argument that leads to this last 
conclusion was complemented by an extensive numerical investigation of the robustness of STA and eSTA schemes with respect 
to classical Gaussian white noise in the lattice depth and atomic position, as well as systematic errors in various system 
parameters (lattice depth, laser wave number, etc.).

While our principal goal in the present work was to devise control protocols that lend themselves to straightforward experimental 
implementations of fast atom transport, it is worthwhile to note that the obtained eSTA-based control protocols can serve as a convenient 
starting point for further numerical optimization. In this context, it is worth mentioning that the optimal-control approach to 
atom transport has recently been demonstrated on the example of a 1D optical lattice and proved capable of reaching the corresponding 
quantum speed limit~\cite{Lam+:21}. 
It is interesting to point out that -- despite not being obtained using optimal-control methods -- the transport times we found
in the present work compare well with those of Ref.~\cite{Lam+:21}, even though they correspond to a different optical-lattice 
potential and transport distances. To be more specific, the transport distance in Ref.~\cite{Lam+:21} was fixed 
to be $15$ times larger than the linear dimensions of the atomic wave packet, while lattice depths up to $300\:E_{\textrm{R}}$ were
considered. The resulting shortest transport times were found to be roughly equal to the oscillation period corresponding to 
the harmonically approximated lattice potential. On the other, the transport distances discussed in the present work were at least
an order of magnitude larger than those in Ref.~\cite{Lam+:21} (i.e. more than $150$ times larger than the size of the atomic wave 
packet) and we considered lattice depths up to $1500\:E_{\textrm{R}}$. It is interesting to note that the transport times in 
Fig.~\ref{FidelityResults2}, which we obtained in the similar range of lattice depths as in Ref.~\cite{Lam+:21} and for a roughly 
ten times larger distance ($d_\mathrm{x}=158\:l_\mathrm{x}$), are between $3.5$ and $4\:T_\mathrm{x}$. Another argument in favor 
of the conclusion that transport times found using the eSTA method are close to the quantum speed limit has quite recently been 
provided for a moving 1D optical lattice~\cite{Whitty+:22}.

\section{Quantum-technology implications} \label{QuantTechImpl}
Coherent atom transport in moving optical lattices and tweezers, typically modelled using the STA family of quantum-control methods 
(including the eSTA method), is a prerequsite for a number of applications in the realm of emerging quantum technologies. In what follows, 
we dedicate special attention to the implications of fast (nonadiabatic) atom transport in DWOLs for two of those application areas. Namely,
we first discuss the role of such transport in neutral-atom QC, more precisely the realization of entangling two-qubit gates based 
on collisional interactions (Sec.~\ref{TransNeutralAtomQC}). We then discuss the implications of spin-dependent atom transport in 
DWOLs in quantum sensing -- more specifically yet, measurements of homogeneous constant forces using guided-atom interferometry 
(Sec.~\ref{SpiDepTrans}).

\subsection{Coherent atom transport in DWOLs as a prerequisite for neutral-atom QC} \label{TransNeutralAtomQC}
While long-range dipole-dipole interactions between Rydberg atoms are currently the most widely used physical 
mechanism for realizing entangling two-qubit gates for neutral-atom-based QC~\cite{Henriet+:20,Morgado+Whitlock:21,
ShiREVIEW:22}, short-range (contact) interactions between atoms in their electronic ground states represent 
the main ingredient of an alternative approach~\cite{Jaksch+:99,Briegel+:00,Weitenberg+:11,MohrJensen+:19}.
In this approach, a pair of atomic states (e.g. one from each hyperfine-split ground-state manifold of $^{87}$Rb)
play the role of two logical qubit states $|0\rangle$ and $|1\rangle$. A pair of atoms acquires a time-dependent 
phase according to $|jj'\rangle\rightarrow e^{iU_{jj'}t/\hbar}|jj'\rangle$ ($j,j'=0,1$), where this phase is 
determined by the interaction strength $U_{jj'}$. The short-range nature of the atom-atom coupling ensures that 
only the desired pair of qubits participates in the entangling operation, thus eliminating the unwanted coupling 
to other qubits and the environment.

The principal idea behind the collisional-gate approach to neutral-atom QC is to merge two atoms, which were initially 
spatially separated, into a common optical trap such that their wave functions have significant overlap. This gives rise to 
a controlled atomic collision, which is either based on spin-dependent interaction, spin-dependent transport, or spin-exchange
interactions~\cite{Weitenberg+:11}. Following a time interval of fixed duration, which is set by the contact-interaction 
strength in the merged state, the two atoms can again be spatially separated. This is the characteristic three-stage 
merge-wait-separate sequence that underlies the realization of collisional two-qubit gates.

One typical two-qubit gate enabled by collisional spin-exchange interactions is the 
square-root-of-SWAP ($\sqrt{\textrm{SWAP}}$)~\cite{Weitenberg+:11,MohrJensen+:19,
Nemirovsky+Sagi:21}, an entangling counterpart of the $\textrm{SWAP}$ operation 
that exchanges the states of two qubits. For qubits indexed by $n$ and $n+1$,
this gate is given by
\begin{equation}
\sqrt{\textrm{SWAP}}=e^{i\frac{\pi}{8}}e^{-i\frac{\pi}{8}(X_n\otimes X_{n+1}
+Y_n\otimes Y_{n+1}+Z_n\otimes Z_{n+1})} \:,
\end{equation}
where $X_n, Y_n, Z_n$ are the Pauli operators of different qubits. Regardless of the specific 
QC platform used, $\sqrt{\textrm{SWAP}}$ represents the natural two-qubit gate in the presence
of Heisenberg-type exchange interactions between adjacent qubits~\cite{Heule++:10,Heule++:11,Stojanovic:19}; 
together with single-qubit rotations it forms a universal gate set for QC~\cite{DiVincenzo+:00}. 
Owing to their specific geometry and the dynamic-control capability (e.g. coherent splitting 
of atoms from single wells into double wells)~\cite{Strabley:06}, DWOLs were utilized for the
experimental realization of the $\sqrt{\textrm{SWAP}}$ gate~\cite{Anderlini:07} based 
on the mechanism proposed in Ref.~\cite{Hayes+:07}.

One of the crucial prerequisites for the collisional-interaction approach to realizing two-qubit gates -- such as 
$\sqrt{\textrm{SWAP}}$ -- with neutral atoms is to time-efficiently bring two atoms to the desired optical-lattice 
sites; in the context of DWOLs, this amounts to bringing atoms to adjacent single wells within a double well. 
Thus, high-fidelity nonadiabatic single-atom transport is of pivotal importance for the practical viability 
of this approach~\cite{Weitenberg+:11}. Moreover, because collisional gates are sensitive to the vibrational state, 
it is essential to avoid vibrational excitation at the end of transport. This last requirement speaks in favor of 
using schemes based on STA in modelling such fast coherent transport (recall the discussion in Sec.~\ref{Intro}). 
The eSTA method, which provides a consistent improvement compared to STA in terms of achievable transport fidelities 
(as discussed in Sec.~\ref{ResultsDiscussion}), being at the same time more robust to decoherence and noise~\cite{Whitty+:22},
appears as the likely method of choice for designing the desired transport trajectories. 

\subsection{Spin-dependent atom transport in DWOLs for quantum sensing via guided-atom interferometry} \label{SpiDepTrans}
In recent years atom interferometry~\cite{Cronin+:09} has facilitated significant progress towards 
quantum-enhanced sensors. The principal idea underlying the operation of an atom interferometer is to first 
split and then -- after some time -- recombine atomic wave function. By detecting the differential phase accumulated 
during the separation period, one can extract small potential differences between the arms of an interferometer.
While still being less developed than its counterpart with freely falling atoms, interferometry based on trapped 
atoms is advancing rapidly.

One specific direction within this field that has attracted considerable attention in recent years is STA-mediated 
guided interferometry that allows measurements of homogeneous constant forces~\cite{Navez+:16,Dupont-Nivet+:16}. 
In compact atom interferometers envisioned to be used for this purpose the atom is driven by spin-dependent 
trapping potentials that are assumed to move in opposite directions; these spin-dependent potentials are complemented 
by linear and time-dependent potentials that can compensate the acceleration of the trap. While the 
pioneering work in this direction proposed the use of harmonic trapping potentials~\cite{Dupont-Nivet+:16},
it was subsequently understood that one can benefit from the use of moving optical lattices (i.e. an anharmonic 
potential)~\cite{Rodriguez-Prieto+:20}, as the latter allow the atomic wave function to be localized with nanoscale
spatial resolution. [Needless to say, the eSTA method can be used as an alternative to STA in designing moving-lattice
trajectories.] As a result, precise measurements at ultrashort spatial scales can be performed. 

In order to create two spin-dependent moving optical lattices, one makes use of a specially chosen wavelength of light 
that allows atoms in the state $|\uparrow\rangle$ to be trapped only by the right-handed circularly polarized light, while 
atoms in the state $|\downarrow\rangle$ are predominantly trapped by the left-handed circularly polarized light~\cite{Lam+:21}. 
To be more specific, to coherently transport atoms in this way (selectively in either one of the two relevant spin 
states) one makes use of a polarization-synthesized beam, where the phases and amplitudes of its left- and right-handed 
circularly polarized components are steered with high precision. When the polarization-synthesized beam interferes with 
a counterpropagating reference beam of fixed linear polarization, two superposed standing waves are created. 

The envisioned Ramsey-type interferometer makes use of atoms that have two internal states -- the spin-up state $|\uparrow\rangle$
and the spin-down state $|\downarrow\rangle$. The state of the atom at time $t$ is given by $\xi_{\uparrow}|\uparrow
\rangle\psi_{\uparrow}(x,t)+\xi_{\downarrow}|\downarrow\rangle \psi_{\downarrow}(x,t)$, where $\psi_{\uparrow\downarrow}
(x,t)\equiv\langle x|\psi_{\uparrow\downarrow}(t)\rangle$ are the coordinate parts of the total atom wave functions in 
the two internal states. Assuming that one starts with the initial spin-up state of the system, a $\pi/2$ pulse gives 
rise to a coherent superposition of spin-up and spin-down states with equal weights (i.e. $\xi_{\uparrow}=\xi_{\downarrow}=1/\sqrt{2}$).
Then the two internal states are spatially separated and recombined. As a consequence of the presence of spin-dependent 
potentials, the two spin components evolve differently. At the final time $t=t_f$, where $t=0$ corresponds to the end of 
the $\pi/2$ pulse, the differential phase $\Delta\varphi(t_f)$ is given by the 
complex argument of the relevant overlap $\langle\psi_{\downarrow}(t_f)|\psi_{\uparrow}(t_f)\rangle$, i.e. $\langle
\psi_{\downarrow}(t_f)|\psi_{\uparrow}(t_f)\rangle=e^{i\Delta\varphi(t_f)}|\langle\psi_{\downarrow}(t_f)|\psi_{\uparrow}
(t_f)\rangle|$. A second $\pi/2$ pulse allows one to extract the populations 
\begin{equation}
P_{\uparrow\downarrow}(t)=\frac{1}{2}\:\Big(1\pm\textrm{Re}[\langle
\psi_{\downarrow}(t_f)|\psi_{\uparrow}(t_f)\rangle]\Big) \:.
\end{equation}
In the case of maximal visibility $|\langle\psi_{\downarrow}(t_f)|\psi_{\uparrow}(t_f)\rangle|=1$, which can be 
realized using STA-mediated guided interferometry~\cite{Rodriguez-Prieto+:20}, the populations are given by
\begin{equation}
P_{\uparrow\downarrow}(t)=\frac{1}{2}\:\Big(1\pm\cos[\Delta\varphi(t_f)]\Big) \:.
\end{equation}
In case the differential phase is proportional to a constant force $F_0$ [i.e. $\Delta\varphi(t_f)=SF_0$],
$F_0$ can be determined from the populations provided that the sensitivity $S$ is known. Importantly,
in the envisaged scheme for force measurements the differential phase does not depend on the initial
atomic motional state, hence it is not necessary to prepare atoms in their perfect ground state. This
speaks in favor of the flexibility of the proposed scheme.

Having described the basics of the proposed interferometric scheme for constant-force 
measurements, we now point out the reasons as to why DWOLs should be eminently suitable
for its experimental implementation. Firstly, spin-dependent DWOLs were demonstrated 
soon after the original realization of this type of optical lattices~\cite{LeeDW:07}.
Secondly, the optical potential of DWOLs has an unusual property, highly beneficial
in this context. Namely, the expansion of this potential in the $x$-direction, 
up to quadratic order, contains a nonzero term linear in $x$ [cf. Eq.~\ref{eqApproxPottDWOL}], 
a property that sets DWOLs apart from conventional (single-well) optical lattices. 
As a result, the Hamiltonian governing the interferometer between the two $\pi/2$ pulses 
(i.e. the one describing the splitting and recombination parts of the interferometric 
sequence), which in the generic situation would contain the term $-F_{0}x$, in the case of 
DWOLs involves a different term linear in $x$. The modified linear term is given by 
$-\widetilde{F}_{0}x$, where $\widetilde{F}_{0}\equiv F_{0}+ma_x$, 
with $a_x$ being proportional to the optical-potential amplitude $U_\mathrm{d,0}$ [cf. Eq.~\ref{eqDWOLacceleration}]. 
In other words, the proposed force-measurement method in the case of DWOLs can be seen 
as a procedure for extracting the value of $\widetilde{F}_0$, given by a sum of the 
sought-after force $F_0$ and a constant offset term $ma_x$, which -- being proportional 
to $U_\mathrm{d,0}$ -- can be made rather large. This, in turn, drastically reduces
the relative error in the experimental measurement of the unknown constant force. Therefore,
the use of DWOLs should allow much more accurate measurements of constant homogeneous
forces than conventional (single-well) optical lattices.
\section{Summary and Conclusions} \label{SummaryConclusions}
To summarize, in this paper we investigated coherent single-atom transport in a double-well optical
lattice, a system characterized by a nonseparable optical potential in the $x-y$ plane. We first
proposed specific configurations of acousto-optic modulators that give rise to the moving-lattice
effect in this system, thus enabling atom transport in an arbitrary direction. We then
designed appropriate moving-lattice trajectories using both shortcuts to adiabaticity (STA) and their 
recently proposed enhanced version, known as eSTA. Through numerical solution of the time-dependent 
Schr\"{o}dinger equation in the comoving frame, based on the previously obtained STA and eSTA moving-lattice 
trajectories, we evaluated the efficiency of the resulting single-atom transport as quantified by 
the transport fidelity. 

We showed that -- while the STA method enables somewhat faster transport in shallow double-well optical lattices 
(i.e. for small lattice depths) -- the eSTA method outperforms it consistently for sufficiently deep lattices 
(more precisely, for lattice depths above around hundred recoil energies). We also identified the atom-transport 
implications of the specific geometric structure of double-well optical lattices, as well as of their characteristic 
spatial anisotropy. In addition, we discussed the implications of the proposed transport schemes in neutral-atom 
quantum computing, i.e. the realization of collisional two-qubit entangling gates. Finally, we pointed out that
guided-atom interferometry enabled by spin-dependent transport in double-well optical lattices may allow accurate 
measurements of constant homogeneous forces, an important application in the realm of quantum sensing.  

In contrast to most previous theoretical studies of coherent single-atom transport, which were based on simplified
scenarios, in the present paper we investigated this phenomenon based on the full, nonseparable double-well optical 
potential. Therefore, the results we obtained can be corroborated in future atom-transport experiments in such lattices. 
Moreover, our study also complements nicely the recent body of work on quantum-state control of Bose-Einstein condensates 
in optical lattices~\cite{Amri+:19,Dupont+:21}. Finally, our work may motivate further attempts to realistically 
model single-atom transport in complex optically-trapped neutral-atom systems, including those that could potentially 
be based on the recently proposed formalism of open-system STA~\cite{WuMa+:21}.

\begin{acknowledgments}
V. M. S. acknowledges useful discussions on the experimental realizations of double-well optical 
lattices with J. V. Porto. This research was supported by the Deutsche Forschungsgemeinschaft 
(DFG) -- SFB 1119 -- 236615297.
\end{acknowledgments}

\onecolumngrid

\appendix
\section{Derivation of the expression for $G_n$} \label{SecDWOPGn}
In what follows, we derive an expression that can be used as the basis for the evaluation of the first (scalar) auxiliary
function $G_{\mathbf{n}}$ [cf. Eq.~\eqref{eqExpressionG}] in the problem at hand. For the sake of brevity,
the multi-indices $\mathbf{n}\equiv (n_x,n_y,n_z)$ and $\mathbf{n}_r\equiv (n_x,n_y)$ are used.

The transport modes (for transport within the $x-y$ plane) of our simplified DWOL Hamiltonian $H_{\mathrm{D},0}(t)$ 
[cf. Eq.~\eqref{simpleDWOL_Hamiltonian}], which has the form characteristic of a 3D harmonic oscillator, in the coordinate 
representation are given by 
\begin{equation}\label{eqSpatialTransportModeDWOL}
\begin{split}
\braket{\boldsymbol{r}|\Psi_\mathbf{n}}
&
=\exp \left[ -\frac{i}{\hbar} \left( E_\mathbf{n} t + \frac{m}{2} \int_0^t \mathrm{d}t'
\left[\dot{q}_\mathrm{c,x}(t')^2 + \dot{q}_\mathrm{c,y}(t')^2\right]\right) \right]\exp \left(
im \left[\dot{q}_\mathrm{c,x}(t) x + \dot{q}_\mathrm{c,y}(t) y\right]/ \hbar \right)
\\
&
\hspace{0.5cm} \times
\left(  2^{n-3} n_\mathrm{x}! n_\mathrm{y}! n_\mathrm{z}! \pi^3 \right)^{-1/2}
\mathrm{H}_{n_\mathrm{x}} \left(\frac{x-q_\mathrm{c,x}(t) + \frac{\pi}{2 k_\mathrm{L}}}{l_\mathrm{x} \sqrt{2}} \right)
\mathrm{H}_{n_\mathrm{y}} \left(\frac{y-q_\mathrm{c,y}(t) + \frac{\pi}{2 k_\mathrm{L}}}{l_\mathrm{y} \sqrt{2}} \right)
\mathrm{H}_{n_\mathrm{z}} \left( \frac{z}{l_\mathrm{z} \sqrt{2}} \right)
\\
&
\hspace{0.5cm} \times
\left(l_\mathrm{x} \, l_\mathrm{y} \, l_\mathrm{z}\right)^{-1/2} \,
\exp \left( -\frac{[x - q_\mathrm{c,x}(t) + \frac{\pi}{2 k_\mathrm{L}}]^2}{4l_\mathrm{x}^2} \right) \,
\exp \left( -\frac{[y - q_\mathrm{c,y}(t) + \frac{\pi}{2 k_\mathrm{L}}]^2}{4l_\mathrm{y}^2} \right) \, \exp
\left(-\frac{z^2}{4l_\mathrm{z}^2} \right),
\end{split}
\end{equation}
where $E_\mathbf{n}\equiv \hbar\omega_\mathrm{x}(n_\mathrm{x} + 1/2)
+ \hbar\omega_{\mathrm{y}}(n_{\mathrm{y}} + 1/2)+\hbar\omega_{\mathrm{z}}(n_{\mathrm{z}} + 1/2)
-m a_{\mathrm{x}}^2/(2\omega_{\mathrm{x}}^2)$ is the $n$-th energy eingenvalue and $l_{\mathrm{j}}\equiv\sqrt
{\hbar/\left(2 m \omega_j\right)}$ is the characteristic length scale along the direction $j$ ($j=x,y,z$).
The energies $E_{\mathbf{n}}$ contain a constant-energy offset due to the presence of a linear term in $x$
in the approximated potential of Eq.~\eqref{eqApproxPottDWOL}.

By inserting these last transport modes of a 3D harmonic Hamiltonian into Eq.~\eqref{eqExpressionG}, we obtain
\begin{equation}\label{eqGStartDWOL}
\begin{split}
G_\mathbf{n} =
& \int_0^{t_\mathrm{f}} \mathrm{d}t \int_{-\infty}^{\infty} \mathrm{d}Z \int_{-\infty}^{\infty} \mathrm{d}Y
\int_{-\infty}^{\infty} \mathrm{d}X \, \frac{\exp\left[ i
\left( \omega_\mathrm{x} n_\mathrm{x} + \omega_\mathrm{y} n_\mathrm{y} + \omega_\mathrm{z} n_\mathrm{z}\right)
t \right]}{\sqrt{2^n n_\mathrm{x}! n_\mathrm{y}! n_\mathrm{z}! \pi^3}} \,
\\
&
\times  \mathrm{H}_{n_\mathrm{x}} \left[ X_\mathrm{C}(t) \right] \mathrm{H}_{n_\mathrm{y}} \left[ Y_\mathrm{C}(t)
\right] \mathrm{H}_{n_\mathrm{z}} \left( Z \right) \, \exp \left[ - X_\mathrm{C}^2(t) \right] \exp
\left[ - Y_\mathrm{C}^2(t) \right] \exp \left( - Z^2 \right)
\\
&
\times \Bigg\{
U_\mathrm{D}\left[\sqrt{2} \, X_0(t) \, l_\mathrm{x}, \sqrt{2} \, Y_0(t) \, l_\mathrm{y}, \sqrt{2} \, Z \, l_\mathrm{z}\right]
-\frac{\hbar}{2 }\left[ \omega_\mathrm{x} X_0^2(t)  + \omega_\mathrm{y} Y_0^2(t) + \omega_\mathrm{z}
Z^2\right]
\\
&
\hspace{0.5cm}
-
\frac{1}{ \sqrt{2} } \left( 2 \, m \, a_\mathrm{x} \, l_\mathrm{x}-
\hbar \omega_\mathrm{x} \, \pi \, \frac{k_\mathrm{L}^{-1}}{2 l_\mathrm{x}} \right)
\, X_0(t)
-
V_\mathrm{d,0}
\Bigg\}  \:.
\end{split}
\end{equation}
In the equation above, we introduced the dimensionless coordinates
$X = x / ( l_\mathrm{x} \sqrt{2} ) $,
$Y = y / ( l_\mathrm{y} \sqrt{2} ) $,
$Z = z / ( l_\mathrm{z} \sqrt{2} ) $
and new functions
$ X_0(t) := X - q_\mathrm{0,x}(t) / ( l_\mathrm{x}\sqrt{2}) $,
$ Y_0(t) := Y - q_\mathrm{0,y}(t) / ( l_\mathrm{y}\sqrt{2}) $,
$ X_\mathrm{C}(t) := X -  \tilde{q}_\mathrm{c,x}(t) / (l_\mathrm{x}\sqrt{2})$ and
$ Y_\mathrm{C}(t) := Y -  \tilde{q}_\mathrm{c,y}(t) / (l_\mathrm{y}\sqrt{2})$.
The displaced classical path of the potential minima $\tilde{q}_\mathrm{c,x/y}(t) =
q_\mathrm{c,x/y}(t) - \pi / (2 k_\mathrm{L})$ was introduced into the above equation to
simplify the upcoming integrations.
The general solution of Eq.~\ref{eqGStartDWOL} will be obtained by treating the different terms
and integrations separately from each other.

\subsection{``Out-of-plane'' integration}\label{SubSecDWOLZInt}
The integration will be started by considering the relatively simple calculation along the $z$ direction.
Since the full $Z$-dependent integration cannot be done with $Z$-dependent waists, we will use the approximation
\begin{equation}
\sqrt{1+\left(\frac{Z l_\mathrm{z}}{Z_{\mathrm{R},u}}\right)^2}\approx 1 \quad (u=x,y) \:,
\end{equation}
valid for $Z_{\mathrm{R},u}\gg l_\mathrm{z}$.
The approximation above can straightforwardly be verified by comparing the scales on which the exponential function
and the waists change significantly. Ordering the different terms in Eq.~\eqref{eqGStartDWOL} results in the integral
\begin{equation}\label{eqGDWOLstart1}
I_{n_\mathrm{z}}^{\textrm{DW}}=\int_{-\infty}^{\infty}\mathrm{d}Z\:\mathrm{H}_{n_\mathrm{z}}\left(Z\right)
\, \exp\left( - Z^2 \right)\left[A + 2 \, B \, \cos^2( \sqrt{2} \, k_\mathrm{z} Z \,
l_\mathrm{z})+ C \, \cos(\sqrt{2} \, k_\mathrm{z} Z \, l_\mathrm{z}) + D Z^2\right] \:,
\end{equation}
with the prefactors $A,B,C,D$, which themselves can be identified with $x$- and $y$-dependent terms from Eq.~\eqref{eqGStartDWOL}.
For the sake of readability, we refrain from explicitly writing down the long expressions for these prefactors.

To calculate the above integral we restrict ourselves to the integration of a simplified integral, given by
\begin{eqnarray}\label{eqIntegrationZ1Approx}
I_\mathrm{z,1}^{\mathbf{n}}&=& \int_{-\infty}^{\infty} \mathrm{d}Z \, \mathrm{H}_\mathrm{n_z}\left( Z \right) \, \exp
\left( - Z^2 \right)  \cos^2\left( \sqrt{2} \,  k  \, l_\mathrm{z} \, Z \right) \\
&=& \frac{1}{4} \, \sum_{k_1+2k_2=n_\mathrm{z}} \frac{n_\mathrm{z}!}{k_1! k_2!} (-1)^{k_1 + k_2} 2^{k_1}
\int_{-\infty}^{\infty} \mathrm{d}Z \, Z^{k_1} \, \exp \left( - Z^2 \right) \, \left[\exp
\left(i \, 2^{3/2} \, k \, l_\mathrm{z} \, Z\right) + \exp \left(-i\, 2^{3/2} \, k
\, l_\mathrm{z} \, Z\right)+ 2 \right] \nonumber \:.
\end{eqnarray}
In the above calculations, use has been made of the Fa\'{a} di Bruno representation of Hermite polynomials~\cite{Weisstein:21}
\begin{equation}\label{eqHermitFaadibruno}
\mathrm{H}_m(x)=(-1)^m\sum_{k_1+2k_2=m} \frac{m!}{k_1! k_2!}
(-1)^{k_1 + k_2} \left( 2 \mathnormal{x} \right)^{k_1} \:.
\end{equation}

The integral in Eq.~\eqref{eqIntegrationZ1Approx} can be calculated by ordering the different terms and
making use of the following result:
\begin{equation}
\label{eqIntegralRelation}
\int_{-\infty}^\infty \mathrm{d}x \, x^n \, \exp \left( - \mathrm{a} x^2 + \mathrm{b} x +\mathrm{c} \right)
= \exp \left( \frac{b^2}{4a} + c \right) \sum_{k=0}^{\lfloor n/2 \rfloor}
\begin{pmatrix}
n \\
2k
\end{pmatrix}
\left(\mathrm{\frac{b}{2a}}\right)^{n-2k}
\frac{ \Gamma \left( k + 1/2 \right) }{\mathrm{a}^{k + 1/2}}.
\end{equation}
Therefore, we just state the final result, which is given by
\begin{equation}\label{eqIntegrationZ1ApproxFinal}
\begin{split}
I_\mathrm{z,1}^{\mathbf{n}}
&
= \frac{1}{4} \, \sum_{k_1 + 2 k_2 = n_\mathrm{z}} \frac{n_\mathrm{z}!}
{k_1! k_2!} (-1)^{k_1 + k_2}2^{k_1}\,\sum_{\lambda=0}^{\lfloor k_1/2 \rfloor}
\begin{pmatrix}
k_1 \\
2 \lambda
\end{pmatrix}
\Gamma \left( \lambda + 1/2 \right) \, \exp\left(- 2\, k^2 \, l_\mathrm{z}^2 \right)
\\
&
\times \Bigg[
\left(i \, 2^{3/2} \, k \, l_\mathrm{z} \right)^{k_1 - 2 \lambda}
+\left(-i \, 2^{3/2} \, k \, l_\mathrm{z}\right)^{k_1 - 2 \lambda}
+ 2 \, \delta_{k_1,2\lambda}\Bigg] \:.
\end{split}
\end{equation}
With the above solution it is straightforward to complete the evaluation of the integral in Eq.~\eqref{eqGDWOLstart1}.
Thus, we just state the final result
\begin{equation}\label{eqGDWOLstart2}
\begin{split}
I_{n_\mathrm{z}}^{\textrm{DW}}
&
=\left( A + B \right) \sqrt{\pi} \, \delta_{n_\mathrm{z},0}
+ \frac{B}{2} \, \left(2^{3/2} \, i \, k_\mathrm{z} \, l_\mathrm{z}\right)^{n_\mathrm{z}}
\exp\left( -2 \, k_\mathrm{z}^2 \, l_\mathrm{z}^2\right) \, \sqrt{\pi} \, \delta_{n_\mathrm{z},
\mathbb{N}_\mathrm{g}}
\\
&
\hspace{0.5cm}
+ \frac{C}{2} \, \left(i \,  \sqrt{2} \, k_\mathrm{z} \, l_\mathrm{z}\right)^{n_\mathrm{z}}
\exp\left( -\frac{k_\mathrm{z}^2}{2} \, l_\mathrm{z}^2\right) \, \sqrt{\pi} \, \delta_{n_\mathrm{z},
\mathbb{N}_\mathrm{g}}
+ D \sqrt{\pi} \left( \frac{1}{2} \delta_{n_\mathrm{z},0} + 2 \delta_{n_\mathrm{z},2}\right) \:,
\end{split}
\end{equation}
where $\mathbb{N}_{\mathrm{g}}$ stands for the set of even positive integers.

\subsection{``In-plane'' integration} \label{SubSecDWOLXInt}
In the following we consider integrals over the $x$ coordinate. The corresponding integrals in $y$ can straightforwardly
be evaluated in an analogous manner i.e. by replacing the parameters for the $x$ components of the potentials, such as
$q_\mathrm{c,x}(t)$ and $q_\mathrm{0,x}(t)$, with their $y$-coordinate counterparts.

The first integral to be considered is given by
\begin{equation}\label{eqIntegrationX1DWOLpre}
I_{n_\mathrm{x}}^\mathrm{D}\left(k_\mathrm{L}, t \right) = \int_{-\infty}^{\infty} \mathrm{d}X  \, \mathrm{H}_{n_\mathrm{x}}
\left[ X_\mathrm{C}(t) \right]  \, \exp\left[ - X_\mathrm{C}^2(t) \right]\cos \left[ 2^{3/2} \, k_\mathrm{L}
\, l_\mathrm{x} \, X_0(t) \right]
\end{equation}
and can be computed by rewriting the cosine function in terms of exponential functions and making use of the
Fa\'{a} di Bruno representation of Hermite polynomials [cf. Eq.~\eqref{eqHermitFaadibruno}]. In this manner,
the following solution is readily obtained:
\begin{equation}\label{eqIntegrationX1DWOL}
\begin{split}
I_{n_\mathrm{x}}^\mathrm{D}\left( k_\mathrm{L}, t \right)
&
= (-1)^{n_\mathrm{x}} \frac{1}{2}  \sum_{k_1+2k_2=n_\mathrm{x}} \frac{n_\mathrm{x}!}{k_1! k_2!} (-1)^{k_1 + k_2} \, 2^{k_1}
\,  \exp\left(-2 \,k_\mathrm{L}^2 l_\mathrm{x}^2\right)
\,
\sum_{l=0}^{k_1}
\begin{pmatrix}
k_1\\
l
\end{pmatrix}
\left[-\frac{\tilde{q}_\mathrm{c,x}(t)}{\sqrt{2}l_\mathrm{x}}\right]^{k_1-l}
\\
&
\hspace{0.5cm} \times
\sum_{\lambda=0}^{\lfloor l/2 \rfloor}
\begin{pmatrix}
l \\
2 \lambda
\end{pmatrix}
\Gamma \left( \lambda + 1/2 \right)
\,
\Bigg\{ \exp\left( 2 i k_\mathrm{L} \left[ \tilde{q}_\mathrm{c,x}(t)-q_\mathrm{0,x}(t) \right]
\right) \left[ \frac{\tilde{q}_\mathrm{c,x}(t)}{\sqrt{2}\:l_\mathrm{x}} + i \, \sqrt{2} \, k_\mathrm{L}
l_\mathrm{x} \right]^{l-2 \lambda}
\\
&
\hspace{0.5cm}+ \exp \left( -2 i k_\mathrm{L} \left[ \tilde{q}_\mathrm{c,x}(t)-q_\mathrm{0,x}(t) \right]
\right) \left[\frac{\tilde{q}_\mathrm{c,x}(t)}{\sqrt{2}\:l_\mathrm{x}} - i \, \sqrt{2} \, k_\mathrm{L}
l_\mathrm{x}\right]^{l-2 \lambda}\Bigg\}  \:.
\end{split}
\end{equation}

The sine-function counterpart of the integral in Eq.~\eqref{eqIntegrationX1DWOLpre} is readily obtained
by following analogous steps. The final solution is given by
\begin{equation}\label{eqIntegrationX1DWOL15}
\begin{split}
I_{n_\mathrm{x}}^\mathrm{DS}\left( k_\mathrm{L} , t \right)
&
= \int_{-\infty}^{\infty} \mathrm{d}X  \mathrm{H}_{n_\mathrm{x}} \left[ X_\mathrm{C}(t) \right]
\, \exp \left[ - X_\mathrm{C}^2(t) \right] \sin \left[ 2^{3/2} \, k_\mathrm{L} \,
l_\mathrm{x} \, X_0(t) \right]
\\
&
= (-1)^{n_\mathrm{x}+1} \frac{i}{2}  \sum_{k_1+2k_2=n_\mathrm{x}} \frac{n_\mathrm{x}!}{k_1! k_2!}
(-1)^{k_1 + k_2} 2^{k_1}
\sum_{l=0}^{k_1}
\begin{pmatrix}
k_1\\
l
\end{pmatrix}
\left[- \frac{\tilde{q}_\mathrm{c,x}(t)}{\sqrt{2}l_\mathrm{x}}\right]^{k_1-l}
\\
&
\hspace{0.5cm} \times
\sum_{\lambda=0}^{\lfloor l/2 \rfloor}
\begin{pmatrix}
l \\
2 \lambda
\end{pmatrix}
\Gamma \left( \lambda + 1/2 \right) \,
\Bigg\{  \exp \left( 2 i k_\mathrm{L} \left[ \tilde{q}_\mathrm{c,x}(t)-q_\mathrm{0,x}(t) \right] \right)
\left[ \frac{\tilde{q}_\mathrm{c,x}(t)}{\sqrt{2}\:l_\mathrm{x}} + i \, \sqrt{2} \, k_\mathrm{L} l_\mathrm{x}
\right]^{l-2 \lambda}
\\
&
\hspace{0.5cm}- \exp \left( -2 i k_\mathrm{L} \left[ \tilde{q}_\mathrm{c,x}(t)-q_\mathrm{0,x}(t) \right]
\right) \left[\frac{\tilde{q}_\mathrm{c,x}(t)}{\sqrt{2}\:l_\mathrm{x}} - i \, \sqrt{2}\, k_\mathrm{L} l_\mathrm{x}
\right]^{l-2 \lambda}\Bigg\}\exp\left( - 2\,k_\mathrm{L}^2 l_\mathrm{x}^2\right)  .
\end{split}
\end{equation}
As was to be expected, the only differences with respect to Eq.~\eqref{eqIntegrationX1DWOL} are the change of sign
in the brackets and the additional prefactor of $(-i)$.

Let us now consider the $X_0(t)$ dependent exponential term that appears within the potential used in Eq.~\eqref{eqGStartDWOL},
for which we first evaluate the following integral:
\begin{equation}\label{eqIDWOL2}
\begin{split}
I_{n_\mathrm{x}}^\mathrm{D2}(t)
&
=
\int_{-\infty}^{\infty} \mathrm{d}X  \, \mathrm{H}_{n_\mathrm{x}} \left[ X_\mathrm{C}(t) \right]  \, \exp
\left[ - X_\mathrm{C}^2(t) \right] \exp \left( - 4\:  \frac{ X_0^2(t) \, l_\mathrm{x}^2}{ w_\mathrm{0,x}^2 } \right)
\\
&
=
\int_{-\infty}^{\infty} \mathrm{d}X \, \sum_{\lambda=0}^{n_\mathrm{x}}
\begin{pmatrix}
n_\mathrm{x}\\
\lambda
\end{pmatrix}
\mathrm{H}_{\lambda} \left( X\right) \left[ -\sqrt{2}\:\frac{\tilde{q}_\mathrm{c,x}(t) }{ l_\mathrm{x} }
\right]^{n_\mathrm{x}-\lambda}\, \exp \left[ - \left(1 + \frac{4 l_\mathrm{x}^2}{ w_\mathrm{0,x}^2}
\right) X^2 \right]
\\
&
\hspace{0.5cm} \times
\exp \left( \left[ \frac{\tilde{q}_\mathrm{c,x}(t)}{2 l_\mathrm{x}} + 2 q_\mathrm{0,x}(t) \frac{l_\mathrm{x}}
{w_\mathrm{0,x}^2} \right] \sqrt{2}\:X \right) \exp\left[ - \frac{q_\mathrm{c,x}^2(t)}{2 l_\mathrm{x}^2}
- \frac{2 q_\mathrm{0,x}^2(t)  }{ w_\mathrm{0,x}^2 }\right] \:.
\end{split}
\end{equation}
Up until this point we ordered the terms in the argument of the exponential function in terms of
a polynomial in $X$ and used the identity~\cite{ChowBOOK:00}
\begin{equation}\label{eqHermitRelations2}
\mathrm{H}_m(x+y) = \sum_{k=0}^m
\begin{pmatrix}
m \\
k
\end{pmatrix}
\mathrm{H}_k(x)\left( 2 y \right)^{m-k}
\end{equation}
to separate the terms in the argument of the Hermite polynomial. Now we are able to use the integral relation
of Eq.~\eqref{eqIntegralRelation}, along with the Fa\'{a} di Bruno representation of Hermite polynomials [cf.
Eq.~\eqref{eqHermitFaadibruno}], to obtain the following result:
\begin{equation}
\label{eqSolutionIDWOL2}
\begin{split}
I_{n_\mathrm{x}}^\mathrm{D2}(t)
&
=
\sum_{\lambda=0}^{n_\mathrm{x}}
\begin{pmatrix}
n_\mathrm{x}\\
\lambda
\end{pmatrix}
\sum_{k_1+2 k_2 = \lambda}(-1)^{n_\mathrm{x} + k_1+k_2}\frac{\lambda!}{k_1!k_2!} \, 2^{k_1} \,
\exp \left(- \frac{2 \left[ q_\mathrm{0,x}^2(t)-q_\mathrm{c,x}^2(t)\right]}
{4 l_\mathrm{x}^2 + w_\mathrm{0,x}^2} \right)
\\
&
\hspace{0.5cm} \times
\left[\sqrt{2}\:\frac{\tilde{q}_\mathrm{c,x}(t)}{l_\mathrm{x} }\right]^{n_\mathrm{x}-\lambda}
\sum_{\sigma=0}^{\lfloor k_1/2 \rfloor}
\begin{pmatrix}
k_1 \\
2\sigma
\end{pmatrix}
\left[
\frac{4 l_\mathrm{x}^2 q_\mathrm{0,x}(t) + \tilde{q}_\mathrm{c,x}(t) w_\mathrm{0,x}^2}{2^{5/2} l_\mathrm{x}^3 + \sqrt{2}\:
l_\mathrm{x} w_\mathrm{0,x}^2}\right]^{k_1-2\sigma}\frac{ \Gamma \left( \sigma + 1/2 \right) }{\left(1 +
\frac{4 l_\mathrm{x}^2}{ w_\mathrm{0,x}^2} \right) }^{\sigma + 1/2} \:.
\end{split}
\end{equation}

The terms in Eq.~\eqref{eqGStartDWOL} that originate from the contribution $\tilde{U}_{\textrm{cr}}$ to the full DWOL potential
[cf. Eq.~\eqref{eqDWOLPot}] result in the same integral as in Eq.~\eqref{eqIDWOL2} above but with an additional $X_0(t)$-dependent
cosine term and without a factor of two in the argument of the exponential function:
\begin{equation}
\begin{split}
I_{n_\mathrm{x}}^\mathrm{D3}(t)
&
=
\int_{-\infty}^{\infty} \mathrm{d}X  \, \mathrm{H}_{n_\mathrm{x}} \left[ X_\mathrm{C}(t) \right]  \, \cos
\left[ \sqrt{2} \, k_\mathrm{L} \, X_0(t) \, l_\mathrm{x}\right]\, \exp \left[ - X_\mathrm{C}^2(t) \right]
\exp \left( - 2\:\frac{ X_0^2(t) \, l_\mathrm{x}^2}{ w_\mathrm{0,x}^2}\right) \:.
\end{split}
\end{equation}
By repeating analogous steps as in the derivation of Eq.~\eqref{eqSolutionIDWOL2}, we obtain the following solution:
\begin{equation}\label{eqD3DWOL}
\begin{split}
I_{n_\mathrm{x}}^\mathrm{D3}(t)
&
=
\frac{1}{2}
\sum_{\lambda=0}^{n_\mathrm{x}}  (-1)^{\lambda}
\begin{pmatrix}
n_\mathrm{x}\\
\lambda
\end{pmatrix}
\sum_{k_1+2 k_2 = \lambda}
(-1)^{k_1+k_2}
\frac{\lambda!}{k_1!k_2!} \, 2^{k_1}
\left[ -\sqrt{2}\:\frac{\tilde{q}_\mathrm{c,x}(t) }{ l_\mathrm{x} }\right]^{n_\mathrm{x}-\lambda}
\\
&
\hspace{0.5cm} \times
\exp \left(
- \frac{4 \left[ q_\mathrm{0,x}(t) - \tilde{q}_\mathrm{c,x}(t)\right]^2 +2\, k_\mathrm{L}^2 l_\mathrm{x}^2\,
w_\mathrm{0,x}^2}{4 \left(2l_\mathrm{x}^2 + w_\mathrm{0,x}^2 \right) }
\right)
\sum_{\sigma=0}^{\lfloor k_1/2 \rfloor}
\begin{pmatrix}
k_1 \\
2\sigma
\end{pmatrix}
\frac{ \Gamma \left( \sigma + 1/2 \right) }{ \left(1 + \frac{4\, l_\mathrm{x}^2}{ w_\mathrm{0,x}^2}
\right)^{\sigma + 1/2}}
\\
&
\hspace{0.5cm} \times
\Bigg\{ \exp \left(-\frac{ i \, k_\mathrm{L}\, \left[ q_\mathrm{0,x}(t) -
\tilde{q}_\mathrm{c,x}(t) \right] \, w_\mathrm{0,x}^2}
{\left(2 l_\mathrm{x}^2 + w_\mathrm{0,x}^2 \right) }
\right)
\left[
\frac{\frac{2 l_\mathrm{x} q_\mathrm{0,x}(t)}{w_\mathrm{0,x}^2} + i\, k_\mathrm{L} \, l_\mathrm{x} +
\frac{\tilde{q}_\mathrm{c,x}(t)}{l_\mathrm{x}}}{\sqrt{2}\: \left( 1 + \frac{2 \, l_\mathrm{x}^2}{w_\mathrm{0,x}^2} \right)}
\right]^{k_1-2\sigma}
\\
&
\hspace{0.5cm} + \hspace{0.25cm}
\exp\left(\frac{+ i \, k_\mathrm{L}\, \left[ q_\mathrm{0,x}(t) - \tilde{q}_\mathrm{c,x}(t) \right] \, w_\mathrm{0,x}^2 }
{\left( 2 l_\mathrm{x}^2 + w_\mathrm{0,x}^2 \right) }\right)\left[
\frac{\frac{2 l_\mathrm{x} q_\mathrm{0,x}(t)}{w_\mathrm{0,x}^2} - i \, k_\mathrm{L} \, l_\mathrm{x} +
\frac{\tilde{q}_\mathrm{c,x}(t)}{l_\mathrm{x}}}{\sqrt{2}\:\left( 1 + \frac{ 2 \, l_\mathrm{x}^2}{w_\mathrm{0,x}^2} \right)}
\right]^{k_1-2\sigma}
\Bigg\} \:.
\end{split}
\end{equation}

The next step is to consider the integral similar to the one discussed above, in which 
the cosine is substituted by a sine function. The integral in question is given by
\begin{equation}
I_{n_\mathrm{x}}^\mathrm{D3S}(t)=
\int_{-\infty}^{\infty} \mathrm{d}X  \mathrm{H}_{n_\mathrm{x}} \left[ X_\mathrm{C}(t) \right]  \,
\sin \left[\sqrt{2} \, k_\mathrm{L} \, X_0(t) \, l_\mathrm{x}\right]\, \exp \left[
- X_\mathrm{C}^2(t) \right] \exp \left( - 2 \:\frac{ X_0^2(t) \, l_\mathrm{x}^2}{ w_\mathrm{0,x}^2} \right).
\end{equation}
By repeating similar steps as in the derivation of Eq.~\eqref{eqD3DWOL}, 
we obtain the solution of the last integral in the following form:
\begin{equation}
\begin{split}
I_{n_\mathrm{x}}^\mathrm{D3S}(t)
&
=
\int_{-\infty}^{\infty} \mathrm{d}X  \mathrm{H}_{n_\mathrm{x}} \left[ X_\mathrm{C}(t) \right]  \, \sin
\left[\sqrt{2} \, k_\mathrm{L} \, X_0(t) \, l_\mathrm{x}\right]\, \exp \left[ - X_\mathrm{C}^2(t) \right]
\exp \left( - 2\: \frac{ X_0^2(t) \, l_\mathrm{x}^2}{ w_\mathrm{0,x}^2 }  \right)
\\
&
=
\frac{i}{2}
\sum_{\lambda=0}^{n_\mathrm{x}}  (-1)^{\lambda+1}
\begin{pmatrix}
n_\mathrm{x}\\
\lambda
\end{pmatrix}
\sum_{k_1+2 k_2 = \lambda}
(-1)^{k_1+k_2}
\frac{\lambda!}{k_1!k_2!} \, 2^{k_1}
\left[ -\sqrt{2}\:\frac{\tilde{q}_\mathrm{c,x}(t) }{ l_\mathrm{x} }\right]^{n_\mathrm{x}-\lambda}
\\
&
\hspace{0.5cm} \times
\exp \left(
- \frac{2 \left[ q_\mathrm{0,x}(t) - \tilde{q}_\mathrm{c,x}(t)\right]^2 + k_\mathrm{L}^2 l_\mathrm{x}^2\,
w_\mathrm{0,x}^2} {2 \left(2 l_\mathrm{x}^2 + w_\mathrm{0,x}^2 \right) }
\right)
\sum_{\sigma=0}^{\lfloor k_1/2 \rfloor}
\begin{pmatrix}
k_1 \\
2\sigma
\end{pmatrix}
\frac{ \Gamma \left( \sigma + 1/2 \right) }{ \left(1 + \frac{4 l_\mathrm{x}^2}{ w_\mathrm{0,x}^2}
\right)^{\sigma + 1/2}}
\\
&
\hspace{0.5cm} \times
\Bigg\{
\exp \left(- \frac{ i \, k_\mathrm{L}\, \left[ q_\mathrm{0,x}(t) - \tilde{q}_\mathrm{c,x}(t) \right]
\, w_\mathrm{0,x}^2} {\left(2 l_\mathrm{x}^2 + w_\mathrm{0,x}^2 \right) } \right)
\left[ \frac{\frac{2 l_\mathrm{x} q_\mathrm{0,x}(t)}{w_\mathrm{0,x}^2} + i \, k_\mathrm{L} \, l_\mathrm{x} +
\frac{\tilde{q}_\mathrm{c,x}(t)}{l_\mathrm{x}}}{\sqrt{2} \:\left( 1 + \frac{2 l_\mathrm{x}^2}{w_\mathrm{0,x}^2} \right)}
\right]^{k_1-2\sigma}
\\
&
\hspace{0.5cm} - \hspace{0.25cm}
\exp \left(\frac{ i \, k_\mathrm{L}\, \left[ q_\mathrm{0,x}(t) - \tilde{q}_\mathrm{c,x}(t) \right] \, w_\mathrm{0,x}^2 }
{2 \left( l_\mathrm{x}^2 + w_\mathrm{0,x}^2 \right) }
\right)
\left[
\frac{\frac{2 l_\mathrm{x} q_\mathrm{0,x}(t)}{w_\mathrm{0,x}^2} - i \, k_\mathrm{L} \, l_\mathrm{x} +
\frac{\tilde{q}_\mathrm{c,x}(t)}{l_\mathrm{x}}}{\sqrt{2}\: \left( 1 + \frac{2 l_\mathrm{x}^2}{w_\mathrm{0,x}^2} \right)}
\right]^{k_1-2\sigma}
\Bigg\}  \:.
\end{split}
\end{equation}

The integration over the $\theta$-dependent trigonometric function emergent through the potential of Eq.~\eqref{eqGStartDWOL}
can straightforwardly be carried out using the following familiar identity:
\begin{equation}\label{eqTrigRelCosTheta}
\sin\left[\sqrt{2} \, k_\mathrm{L} l_\mathrm{x} X_0(t)-\theta\right]=\sin\left[\sqrt{2} \,
k_\mathrm{L} l_\mathrm{x} X_0(t)\right] \cos\theta - \cos\left[\sqrt{2} \, k_\mathrm{L}
l_\mathrm{x} X_0(t)\right] \sin \theta \:.
\end{equation}
Thus, the integral can be reduced to the integrals involving trigonometric functions that were already evaluated
above [cf. Eq.~\eqref{eqIntegrationX1DWOL} and \eqref{eqIntegrationX1DWOL15}].

The integration involving the $X_0^2(t)$-dependent terms of Eq.~\eqref{eqGStartDWOL} can be carried out
by making use of the identity
\begin{equation}
\label{eqRelationHermit1}
x \, \mathrm{H}_m(x)=\frac{1}{2}\mathrm{H}_{m+1}(x) + m \, \mathrm{H}_{m-1}(x),
\end{equation}
as well as the orthogonality of Hermite polynomials. Thus, the integral in $X$ is given by
\begin{equation}\label{eqGInetgrationZ2}
\begin{split}
I_{\mathbf{n},\mathrm{x}}^\mathrm{D4}(t)
&
= \frac{\hbar\omega_\mathrm{x}}{2}\:\int_{-\infty}^{\infty} \mathrm{d}X \, \exp\left[ - X_\mathrm{C}^2(t) \right]
\, X_0^2(t) \, \mathrm{H}_{n_\mathrm{x}} \left[ X_\mathrm{C}(t) \right] \delta_{n_\mathrm{z},0} \, \delta_{n_\mathrm{y},0}
\\
&
= \frac{\hbar\omega_\mathrm{x}}{2}\: \int_{-\infty}^\infty \mathrm{d}X \sum_{m=0}^\infty \frac{\mathrm{H}_{m} \left( X \right)}{m!}
\left[ \frac{\tilde{q}_\mathrm{c,x}(t)}{\sqrt{2} \, l_\mathrm{x}}\right]^m \exp \left( -X^2 \right) \, \delta_{n_\mathrm{z},0}
\, \delta_{n_\mathrm{y},0} \\
&
\hspace{0.5cm} \times  \left[X^2 - \sqrt{2}\: X \frac{q_\mathrm{0,x}(t)}{l_\mathrm{x}} +  \frac{q_\mathrm{0,x}^2(t)}{2 \,
l_\mathrm{x}^2} \right] \, \sum_{l=0}^{n_\mathrm{x}}
\begin{pmatrix}
n_\mathrm{x}\\
l
\end{pmatrix}
\mathrm{H}_{l} \left( X \right) \left[- \sqrt{2} \:\frac{\tilde{q}_\mathrm{c,x}(t)}{l_\mathrm{x}} \right]^{n_\mathrm{x}-l}
\\
&
= \frac{\hbar}{2} \, \sqrt{\pi} \, \omega_\mathrm{x} \, \delta_{n_\mathrm{z},0} \, \delta_{n_\mathrm{y},0} \left[ \sqrt{2}
\, \frac{\tilde{q}_\mathrm{c,x}(t)}{l_\mathrm{x}} \right]^{n_\mathrm{x}} \,
\\
&
\hspace{0.5cm} \times \sum_{l=0}^{n_\mathrm{z}}
\begin{pmatrix}
n_\mathrm{x}\\
l
\end{pmatrix}
(-1)^{n_\mathrm{x} - l}\Bigg\{\frac{1}{4} \, l \, (l-1) \, \left[ \frac{\tilde{q}_\mathrm{c,x}(t)}{\sqrt{2}\:
l_\mathrm{x}} \right]^{-2}-l  \, q_\mathrm{0,x}(t) \, \tilde{q}_\mathrm{c,x}(t)^{-1}+ l \Bigg\} \:,
\end{split}
\end{equation}
where we have made use of the binomial theorem and the condition $n>0$. The final expression for the above integral
is given by
\begin{equation}
\begin{split}\label{eqHarmoicTreatmentDWOL1}
I_\mathrm{\mathbf{n},x}^\mathrm{D4}(t)
&
= \frac{\hbar}{2} \omega_\mathrm{x} \int_{-\infty}^{\infty} \mathrm{d}X \, \exp \left[ - X_\mathrm{C}^2(t) \right]
\, X_0^2(t) \, \mathrm{H}_{n_\mathrm{x}} \left[ X_\mathrm{C}(t) \right] \delta_{n_\mathrm{y},0} \, \delta_{n_\mathrm{z},0}
\\
&
= \hbar \, \omega_\mathrm{x} \, \sqrt{\pi}  \, \delta_{n_\mathrm{y},0} \, \delta_{n_\mathrm{z},0} \left[\delta_{n_\mathrm{x},1}
\, \frac{\tilde{q}_\mathrm{c,x}(t)-q_\mathrm{0,x}(t)}{\sqrt{2}\:l_\mathrm{x}} + \delta_{n_\mathrm{x},2}\right].
\end{split}
\end{equation}

The last term that we need to integrate is the term linear in $X_0(t)$. The corresponding solution can straightforwardly be
obtained through the following evaluation:
\begin{equation}\label{eqGInetgrationZ2}
\begin{split}
I_\mathrm{x}^\mathrm{e}
&
= \int_{-\infty}^{\infty} \mathrm{d}X \, \exp \left[ - X_\mathrm{C}^2(t) \right] \, X_0(t)
\, \mathrm{H}_{n_\mathrm{x}}
\left[ X_\mathrm{C}(t) \right] \delta_{n_\mathrm{x},0} \, \delta_{n_\mathrm{y},0}
\\
&
= \int_{-\infty}^\infty\mathrm{d}X \sum_{m=0}^\infty\frac{1}{m!}\,\left[\frac{\tilde{q}_\mathrm{c,x}(t)}{\sqrt{2}l_\mathrm{x} }
\right]^m
\sum_{l=0}^{n_\mathrm{x}}
\begin{pmatrix}
n_\mathrm{x}\\
l
\end{pmatrix}
\mathrm{H}_{l} \left( X \right) \left[- \sqrt{2}\:\frac{\tilde{q}_\mathrm{c,x}(t) }{l_\mathrm{x} } \right]^{n_\mathrm{x}-l}
\\
&
\hspace{0.5cm} \times \Bigg[
\frac{1}{2} \, \mathrm{H}_{m + 1} \left( X \right) - \frac{q_\mathrm{0,x}(t) }{\sqrt{2}\:l_\mathrm{x} }\, \mathrm{H}_{m}
\left( X \right) + m \mathrm{H}_{m - 1} \left( X \right)
\Bigg]
\exp \left( -X^2 \right) \, \delta_{n_\mathrm{x},0} \, \delta_{n_\mathrm{y},0}
\\
&
= \sqrt{\pi} \, \delta_{n_\mathrm{x},0} \, \delta_{n_\mathrm{y},0} \left[ \sqrt{2} \,\frac{\tilde{q}_\mathrm{c,x}(t) }{l_\mathrm{x}}
\right]^{n_\mathrm{x}-1} \,
\sum_{l=0}^{n_\mathrm{x}}
\begin{pmatrix}
n_\mathrm{x}\\
l
\end{pmatrix}
(-1)^{n_\mathrm{x} - l}
l\:,
\end{split}
\end{equation}
where use has been made of the same steps utilized above for the evaluation of the integral in Eq.~\eqref{eqHarmoicTreatmentDWOL1}.
Using mathematical induction it can further be demonstrated that only contributions with $n_\mathrm{x}=1$ are
not equal to zero. Thus, the final result reads
\begin{equation}
I_\mathrm{x}^\mathrm{e} = \sqrt{\pi} \, \delta_{n_\mathrm{x},1} \, \delta_{n_\mathrm{y},0}
\delta_{n_\mathrm{z},0} \:.
\end{equation}

Putting the different solutions for the integrals of Eq.~\eqref{eqTrigRelCosTheta} together and using the
orthogonality of Hermite polynomials, the first auxiliary function can be reduced to the time integral
\begin{equation}\label{eqGEndDWOL1}
\begin{split}
G_\mathbf{n} =
& -\int_0^{t_\mathrm{f}} \mathrm{d}t \, U_\mathrm{d,0} \, \frac{\exp \left[ i \left( \omega_\mathrm{x}
n_\mathrm{x} + \omega_\mathrm{y} n_\mathrm{y} + \omega_\mathrm{z} n_\mathrm{z}\right) t \right]}{\sqrt{2^n n_\mathrm{x}!
n_\mathrm{y}! n_\mathrm{z}! \pi}}
\Bigg\{
\zeta_{\parallel,\boldsymbol{n}}(t)
+
\zeta_{\perp,\boldsymbol{n}}(t)
+
\zeta_{\mathrm{z},\boldsymbol{n}}(t)
+
\zeta_{\mathrm{cr},\boldsymbol{n}}(t)
\\
&
+\frac{\hbar \sqrt{\pi}}{U_\mathrm{d,0}} \Bigg[ \omega_\mathrm{x} \delta_{n_\mathrm{y},0} \, \delta_{n_\mathrm{z},0}
\left[\delta_{n_\mathrm{x},1}  \, \frac{\tilde{q}_\mathrm{c,x}(t)-q_\mathrm{0,x}(t)}{\sqrt{2} l_\mathrm{x}} + \delta_{n_\mathrm{x},2}
\right]+ \omega_\mathrm{y} \delta_{n_\mathrm{x},0} \, \delta_{n_\mathrm{z},0} \left[\delta_{n_\mathrm{y},1} \,
\frac{\tilde{q}_\mathrm{c,y}(t)-q_\mathrm{0,y}(t)}{\sqrt{2} l_\mathrm{y}} + \delta_{n_\mathrm{y},2}\right]
\\
&
+ \omega_\mathrm{z} \left( \frac{1}{2}\delta_{n_\mathrm{z},0} + 2 \delta_{n_\mathrm{z},2}\right) \Bigg]
+
\frac{1}{ \sqrt{2} \, U_\mathrm{d,0}}\left( 2 \, m \, a_\mathrm{x} \, l_\mathrm{x}+
\hbar \omega_\mathrm{x} \, \pi \, \frac{k_\mathrm{L}^{-1}}{2 l_\mathrm{x}} \right) \, \pi \, \delta_{n_\mathrm{x},1}
\, \delta_{n_\mathrm{y},0} \, \delta_{n_\mathrm{z},0}
\Bigg\} \:,
\end{split}
\end{equation}
where $\zeta_{\parallel,\boldsymbol{n}}(t)$, $\zeta_{\perp,\boldsymbol{n}}(t)$,
$\zeta_{\mathrm{z},\boldsymbol{n}}(t)$, and $\zeta_{\mathrm{cr},\boldsymbol{n}}(t)$
are functions of time given by:
\begin{equation}\label{eqGEndDWOLfunc1}
\begin{split}
\zeta_{\parallel,\boldsymbol{n}}(t)
=
&
\cos^2 \left( \frac{\beta}{2} \right)  \, \Bigg[ I_{n_\mathrm{y}}^\mathrm{D}\left( k_\mathrm{L} , t \right)
\delta_{n_\mathrm{x},0} - I_{n_\mathrm{x}}^\mathrm{D}\left( k_\mathrm{L} , t \right)  \delta_{n_\mathrm{y},0}\Bigg]
\delta_{n_\mathrm{z},0}\:,
\\
\zeta_{\perp,\boldsymbol{n}}(t)
=
&
\sin^2 \left( \frac{\beta}{2} \right) \Bigg\{
I_{n_\mathrm{y}}^\mathrm{D}\left( k_\mathrm{L} , t \right) \delta_{n_\mathrm{x},0}
- \cos \left(2 \, \theta \right) \, I_{n_\mathrm{x}}^\mathrm{D}\left( k_\mathrm{L} , t \right) \,  \delta_{n_\mathrm{y},0}
+ \frac{4}{\sqrt{\pi}} \, I_{n_\mathrm{y}}^\mathrm{D}\left( k_\mathrm{L}/2 , t \right)
\Big[\sin \theta  \: I_{n_\mathrm{x}}^\mathrm{D}\left( k_\mathrm{L}/2 , t \right)
\\
&
- \cos \theta \: I_{n_\mathrm{x}}^\mathrm{DS}\left( k_\mathrm{L}/2 , t \right)\Big]
- I_{n_\mathrm{x}}^\mathrm{DS}\left( k_\mathrm{L} , t \right) \, \sin \left( 2 \theta \right) \delta_{n_\mathrm{y},0}
\Bigg\} \delta_{n_\mathrm{z},0}\:,
\\
\zeta_{\mathrm{z},\boldsymbol{n}}(t)
=
&
\frac{ \xi_\mathrm{z} }{ \sqrt{\pi}} \, I_{n_\mathrm{x}}^\mathrm{D2} (t) \, I_{n_\mathrm{y}}^\mathrm{D2} (t) \Bigg[
\frac{1}{2}\delta_{n_\mathrm{z},0}
+ 2 \, \left(2^{3/2} \, i \, k_\mathrm{z}  \, l_\mathrm{z}\right)^{n_\mathrm{z}} \exp\left( -2 \,
k_\mathrm{z}^2 l_\mathrm{z}^2\right) \, \delta_{n_\mathrm{z}, \mathbb{N}_\mathrm{g}} \Bigg]\:,
\\
\zeta_{\mathrm{cr},\boldsymbol{n}}(t)
=
&
2 \, \sqrt{\frac{\xi_\mathrm{z}}{\pi}} \, \cos \left(\frac{\beta}{2}\right)
\Bigg[\cos \left(\frac{\phi}{2}\right) \, I_{n_\mathrm{x}}^\mathrm{D2}(t) \, I_{n_\mathrm{y}}^\mathrm{D3}(t)
- \sin \left(\frac{\phi}{2}\right) \, I_{n_\mathrm{x}}^\mathrm{D3S}(t) \, I_{n_\mathrm{y}}^\mathrm{D2}(t) \Bigg]
\left[ \left( i \, \sqrt{2} \, k_\mathrm{z} \, l_\mathrm{z}\right)^{n_\mathrm{z}} \exp \left( -\frac
{k_\mathrm{z}^2}{2} l_\mathrm{z}^2 \right) \, \delta_{n_\mathrm{z}, \mathbb{N}_\mathrm{g}} \right] \: .
\end{split}
\end{equation}
The evaluation of the time integral in Eq.~\eqref{eqGEndDWOL1}, the last step in the calculation of the first
auxiliary function $G_\mathbf{n}$, can only be done numerically.

\section{Derivation of the expression for $\mathbf{K}_\mathbf{n}$}\label{SecDWOPKn}
In the following, we derive an expression that can be used as the basis for a numerical evaluation of the second (vector) auxiliary
function $\mathbf{K}_{\mathbf{n}}$ [cf. Eq.~\eqref{eqExpressionK}] in the problem at hand.

To derive the desired expression for $\mathbf{K}_{\mathbf{n}}$, the gradient with respect to $\boldsymbol{\lambda}$,
and subsequently to $\boldsymbol{\alpha}$, of the Hamiltonian $H_\mathrm{DW}$ is needed. Since the
differentiations are straightforward, we just state the resulting equation for $\mathbf{K}_{\mathbf{n}}$:
\begin{equation}\label{eqKDWOLafterZ}
\begin{split}
\mathbf{K}_\mathbf{n}
&
=  \, \int_0^{t_\mathrm{f}} \, dt \, \int_{-\infty}^\infty \, \mathrm{d}Z \, \int_{-\infty}^\infty \, \mathrm{d}Y \,
\int_{-\infty}^\infty \, \mathrm{d}X
\frac{ 1  }{\sqrt{2^n n_\mathrm{x}! n_\mathrm{y}! n_\mathrm{z}! \pi^3}} \exp \left[ i
\left( n_\mathrm{x} \omega_\mathrm{x} + n_\mathrm{y} \omega_\mathrm{y} + n_\mathrm{z} \omega_\mathrm{z} \right) t \right]
\\
&
\hspace{0.5cm} \times  \mathrm{H}_{n_\mathrm{x}} \left[ X_\mathrm{C}(t) \right] \mathrm{H}_{n_\mathrm{y}} \left[
Y_\mathrm{C}(t) \right] \mathrm{H}_{n_\mathrm{z}} \left( Z \right) \, \exp \left[ - X_\mathrm{C}^2(t) \right]
\exp \left[ - Y_\mathrm{C}^2(t) \right] \exp \left( - Z^2 \right)
\\
&
\hspace{0.5cm} \times
\boldsymbol{\nabla}_{\boldsymbol{\alpha}}
U_\mathrm{D}\left[\sqrt{2} \, X_0(t) \, l_\mathrm{x}, \sqrt{2} \, Y_0(t) \, l_\mathrm{y}, \sqrt{2} \, Z \, l_\mathrm{z}\right]
\Big|_{\boldsymbol{\alpha}=\boldsymbol{0}} \:.
\end{split}
\end{equation}

We start by noting that the integral over $Z$ has the same general form as Eq.~\eqref{eqGDWOLstart1} and, accordingly,
the same solution that is given by Eq.~\eqref{eqGDWOLstart2}. In fact, most of the needed integrals have already been 
carried out in the evaluation of the first auxiliary function (cf. Appendix~\ref{SecDWOPGn}). Even the cosine term
originating from the potential in Eq.~\eqref{eqKDWOLafterZ} can be rewritten using the trigonometric identity 
\begin{equation}\label{eqTrigRelationCos1}
\cos \left[ \sqrt{2} \, k_\mathrm{L} l_\mathrm{x} X_0(t) - \theta \right] = \cos \left[ \sqrt{2} \, k_\mathrm{L}
l_\mathrm{x} X_0(t)\right] \, \cos \theta  + \sin \left[ \sqrt{2} \, k_\mathrm{L} l_\mathrm{x}
X_0(t)\right] \, \sin \theta \:,
\end{equation}
using which we can evaluate the relevant integrals by making use of the solutions in Eq.~\eqref{eqIntegrationX1DWOL} and
Eq.~\eqref{eqIntegrationX1DWOL15}.

However, there are integrals that originate from to the influence of the laser in the $z$ direction, which 
have not been considered previously. The first one among those integrals is given by
\begin{equation}\label{eqDWOLDW2}
\begin{split}
\tilde{I}_{n_\mathrm{x}}^\mathrm{D2}(t)
&=
\int_{-\infty}^{\infty} \mathrm{d}X \, X_0(t) \, \mathrm{H}_{n_\mathrm{x}} \left[ X_\mathrm{C}(t) \right]  \, \exp
\left[ - X_\mathrm{C}^2(t) \right] \exp \left( - 4\: \frac{ X_0^2(t) \, l_\mathrm{x}^2}{ w_\mathrm{0,x}^2} \right)
\\
&
=
\int_{-\infty}^{\infty} \mathrm{d}X \sum_{\lambda=0}^{n_\mathrm{x}}
\,
\begin{pmatrix}
n_\mathrm{x}\\
\lambda
\end{pmatrix}
\mathrm{H}_{\lambda} \left( X\right) \left[ -\sqrt{2}\:\frac{\tilde{q}_\mathrm{c,x}(t) }
{ l_\mathrm{x} }\right]^{n_\mathrm{x}-\lambda}  \,
\exp \left[ - \left(1 + \frac{4 l_\mathrm{x}^2}{ w_\mathrm{0,x}^2} \right) X^2 \right]
\\
&
\hspace{0.5cm} \times
\left[ X - \frac{q_\mathrm{0,x}(t)}{\sqrt{2}\:l_\mathrm{x}} \right]
\exp \left( \left[ \frac{\tilde{q}_\mathrm{c,x}(t)}{2 l_\mathrm{x}} + 2 q_\mathrm{0,x}(t) \frac{l_\mathrm{x}}{w_\mathrm{0,x}^2}
\right] \sqrt{2} \, X \right) \exp \left[ - \frac{q_\mathrm{c,x}^2(t)}{2 l_\mathrm{x}^2} - \frac{2 q_\mathrm{0,x}^2(t)}
{ w_\mathrm{0,x}^2 }\right]
\\
&
=\sum_{\lambda=0}^{n_\mathrm{x}}  (-1)^{\lambda}
\begin{pmatrix}
n_\mathrm{x}\\
\lambda
\end{pmatrix}
\sum_{k_1+2 k_2 = \lambda}
(-1)^{k_1+k_2}
\frac{\lambda!}{k_1!k_2!} 2^{k_1}
\left[ -\sqrt{2}\:\frac{\tilde{q}_\mathrm{c,x}(t) }{ l_\mathrm{x} }\right]^{n_\mathrm{x}-\lambda}  \,
\\
&
\hspace{0.5cm} \times \exp\left(-\frac{2 \left[ q_\mathrm{0,x}(t)-\tilde{q}_\mathrm{c,x}(t)\right]^2}
{2^{3/2} l_\mathrm{x}^2 + w_\mathrm{0,x}^2}\right)
\left[ \tilde{D}_{n_\mathrm{x}} \left( k_1 + 1 ,t \right) -\frac{q_\mathrm{0,x}(t)}{\sqrt{2}\:l_\mathrm{x}}
\tilde{D}_{n_\mathrm{x}} \left( k_1 ,t\right) \right],
\end{split}
\end{equation}
where use was made of the integral- and Hermite-polynomial identities in Eqs.~\eqref{eqIntegralRelation} and \eqref{eqSolutionIDWOL2},
as well as the following function:
\begin{equation}
\tilde{D}_{n_\mathrm{x}} \left( k, t \right) = \sum_{\sigma=0}^{\lfloor k/2 \rfloor}
\begin{pmatrix}
k \\
2\sigma
\end{pmatrix}
\left[
\frac{4 l_\mathrm{x}^2 q_\mathrm{0,x}(t) + \tilde{q}_\mathrm{c,x}(t) w_\mathrm{0,x}^2}{\sqrt{2}\:
l_\mathrm{x}\left(4 l_\mathrm{x}^2 +  w_\mathrm{0,x}^2\right)}\right]^{k-2\sigma}
\frac{ \Gamma \left( \sigma + 1/2 \right) }{\mathrm{ \left(1 + \frac{4 l_\mathrm{x}^2}
{ w_\mathrm{0,x}^2} \right) }^{\sigma + 1/2}} \:.
\end{equation}

The integrations corresponding to the cross terms lead to the following final expressions:
\begin{equation}\label{eqDWOLDW2eqn}
\begin{split}
\tilde{I}_{n_\mathrm{x}}^\mathrm{D3} \left( k_\mathrm{L},t\right)
&
=
\int_{-\infty}^{\infty} \mathrm{d}X  \mathrm{H}_{n_\mathrm{x}} \left[ X_\mathrm{C}(t) \right] \, X_0(t) \,
\cos\left[\sqrt{2} k_\mathrm{L} \, X_0(t) \, l_\mathrm{x}\right]\, \exp \left[ - X_\mathrm{C}^2(t) \right]
\exp \left( - 4 \:\frac{ X_0^2(t) \, l_\mathrm{x}^2}{ w_\mathrm{0,x}^2 }\right)
\\
&
=
\frac{1}{2^{3/2}}
\int_{-\infty}^{\infty} \mathrm{d}X  \sum_{\lambda=0}^{n_\mathrm{x}}
\begin{pmatrix}
n_\mathrm{x}\\
\lambda
\end{pmatrix}
\mathrm{H}_{\lambda} \left( X\right) \left[ -\sqrt{2}  \frac{\tilde{q}_\mathrm{c,x}(t) }
{ l_\mathrm{x} }\right]^{n_\mathrm{x}-\lambda}
\,
\exp \left[ - \left(1 + \frac{4 l_\mathrm{x}^2}{ w_\mathrm{0,x}^2} \right) X^2 \right]
\\
&
\hspace{0.5cm}
\times
\left[ X - \frac{q_0(t)}{\sqrt{2}\:l_\mathrm{x}} \right] \, \exp \left( \left[ \frac{\tilde{q}_\mathrm{c,x}(t)}{2 l_\mathrm{x}} +
2 q_\mathrm{0,x}(t) \frac{l_\mathrm{x}}{w_\mathrm{0,x}^2} \right] \sqrt{2}\:X \right) \exp \left[ - \frac{q_\mathrm{c,x}^2
(t)}{2 l_\mathrm{x}^2} - \frac{2 q_\mathrm{0,x}^2(t)  }{ w_\mathrm{0,x}^2 }\right]
\\
&
\hspace{0.5cm}
\times
\Big[ \exp \left( i \, \sqrt{2}\left[ X - \frac{q_\mathrm{0,x}(t)}{\sqrt{2}\:l_\mathrm{x}} \right] k_\mathrm{L}
\, l_\mathrm{x} \right) + \exp \left( -i \, \sqrt{2}\left[ X - \frac{q_\mathrm{0,x}(t)}{ \sqrt{2}\:l_\mathrm{x}}
\right] k_\mathrm{L} \, l_\mathrm{x} \right)\Big]
\\
&
=
\frac{1}{2}
\sum_{\lambda=0}^{n_\mathrm{x}}  (-1)^{\lambda}
\begin{pmatrix}
n_\mathrm{x}\\
\lambda
\end{pmatrix}
\sum_{k_1+2 k_2 = \lambda}
(-1)^{k_1+k_2}
\frac{\lambda!}{k_1!k_2!} 2^{k_1}
\left[ -\sqrt{2}\:\frac{\tilde{q}_\mathrm{c,x}(t) }{ l_\mathrm{x} }\right]^{n_\mathrm{x}-\lambda}
\\
&
\hspace{0.5cm} \times
\exp \left(- \frac{4 \left[ q_\mathrm{0,x}^2(t) - q_\mathrm{c,x}^2(t)\right] + k_\mathrm{L}^2
l_\mathrm{x}^2 w_\mathrm{0,x}^2}{2 \left( 4 l_\mathrm{x}^2 + w_\mathrm{0,x}^2 \right) }
\right) \left[\tilde{D}^{2}_+(k_1+1,k_\mathrm{L},t) - \frac{q_0(t)}{\sqrt{2}\:l_\mathrm{x}}
\tilde{D}^{2}_+(k_1,k_\mathrm{L},t)\right] \:.
\end{split}
\end{equation}

Similar results are obtained by substituting the cosine with a sine in Eq.~\eqref{eqDWOLDW2eqn}:
\begin{equation}
\begin{split}
\tilde{I}_{n_\mathrm{x}}^\mathrm{D3S} \left( k_\mathrm{L},t\right)
&
=\int_{-\infty}^{\infty} \mathrm{d}X  \mathrm{H}_{n_\mathrm{x}} \left[ X_\mathrm{C}(t) \right]
\, X_0(t) \, \sin\left[\sqrt{2} \, k_\mathrm{L} \, X_0(t) \, l_\mathrm{x}\right]\,
\exp \left[ - X_\mathrm{C}^2(t) \right] \exp \left( - 4\:\frac{ X_0^2(t)
\, l_\mathrm{x}^2}{ w_\mathrm{0,x}^2 } \right)
\\
&
=
-\frac{i}{2}
\sum_{\lambda=0}^{n_\mathrm{x}}  (-1)^{\lambda}
\begin{pmatrix}
n_\mathrm{x}\\
\lambda
\end{pmatrix}
\sum_{k_1+2 k_2 = \lambda}(-1)^{k_1+k_2}\frac{\lambda!}{k_1!k_2!} 2^{k_1}
\times\left[ -\sqrt{2}\:\frac{\tilde{q}_\mathrm{c,x}(t) }{ l_\mathrm{x} }\right]^{n_\mathrm{x}-\lambda}
\\
&
\times\exp \left(-\frac{4 \left[ q_\mathrm{0,x}^2(t) - q_\mathrm{c,x}^2(t)\right] + k_\mathrm{L}^2
l_\mathrm{x}^2 w_\mathrm{0,x}^2}{2\left(4 l_\mathrm{x}^2 + w_\mathrm{0,x}^2 \right) }
\right)\left[\tilde{D}^{2}_-(k_1+1,k_\mathrm{L},t) - \frac{q_0(t)}{\sqrt{2}\:l_\mathrm{x}}
\tilde{D}^{2}_-(k_1,k_\mathrm{L},t)\right] \:.
\end{split}
\end{equation}
In order to be able to succinctly write the solutions of the last integrals, we have 
introduced the following function:
\begin{equation}
\begin{split}
\tilde{D}^{2}_\pm(k,k_\mathrm{L},t)
&
= \sum_{\sigma=0}^{\lfloor k/2 \rfloor}
\begin{pmatrix}
k \\
2\sigma
\end{pmatrix}
\frac{ \Gamma \left( \sigma + 1/2 \right) }{\mathrm{ \left[1 + \frac{4 l_\mathrm{x}^2}
{ w_\mathrm{0,x}^2} \right] }^{\sigma + 1/2}}
\\
&
\hspace{0.5cm} \times
\Bigg \{ \exp \left(- \frac{ i \, k_\mathrm{L}\, \left[ q_\mathrm{0,x}(t) - \tilde{q}_\mathrm{c,x}(t)
\right] \, w_\mathrm{0,x}^2}{\left(4l_\mathrm{x}^2 + w_\mathrm{0,x}^2 \right)}\right)
\left[\frac{4 l_\mathrm{x}^2 q_\mathrm{0,x}(t) + i \, k_\mathrm{L} \, l_\mathrm{x}^2 \,
w_\mathrm{0,x}^2 +  \tilde{q}_\mathrm{c,x}(t) w_\mathrm{0,x}^2}{\sqrt{2}\: l_\mathrm{x}\left(4 l_\mathrm{x}^2
+  w_\mathrm{0,x}^2 \right)}\right]^{k-2\sigma}
\\
&
\hspace{0.5cm}
\pm \exp\left(
\frac{ i \, k_\mathrm{L}\, \left[ q_\mathrm{0,x}(t) - \tilde{q}_\mathrm{c,x}(t)
\right] \, w_\mathrm{0,x}^2 }{\left( 4 l_\mathrm{x}^2
+ w_\mathrm{0,x}^2 \right) }
\right)
\left[
\frac{4 l_\mathrm{x}^2 q_\mathrm{0,x}(t) - i \, k_\mathrm{L} \, l_\mathrm{x}^2
\, w_\mathrm{0,x}^2 +  \tilde{q}_\mathrm{c,x}(t) w_\mathrm{0,x}^2}{\sqrt{2}\: l_\mathrm{x}
\left(4 l_\mathrm{x}^2 +  w_\mathrm{0,x}^2 \right)}
\right]^{k-2\sigma}
\Bigg\} \:.
\end{split}
\end{equation}

Having completed all the necessary spatial integrals, we can finally express the second auxiliary function
in the form of the time integral
\begin{equation}
\begin{split}
\mathbf{K}_\mathbf{n}
&
= - \, \int_0^{t_\mathrm{f}} \, dt \,
\frac{2 \, U_\mathrm{d,0} \, k_\mathrm{L}}{\sqrt{2^n n_\mathrm{x}! n_\mathrm{y}! n_\mathrm{z}! \pi}} \exp \left[ i
\left( n_\mathrm{x} \omega_\mathrm{x} + n_\mathrm{y} \omega_\mathrm{y} + n_\mathrm{z} \omega_\mathrm{z} \right) t \right]
\Bigg[ \tilde{\boldsymbol{\zeta}}_{\parallel,\boldsymbol{n}}(t)
     + \tilde{\boldsymbol{\zeta}}_{\perp,\boldsymbol{n}}(t)
     + \tilde{\boldsymbol{\zeta}}_{\mathrm{z},\boldsymbol{n}}(t)
     + \tilde{\boldsymbol{\zeta}}_{\mathrm{cr},\boldsymbol{n}}(t) \Bigg] \:,
\end{split}
\end{equation}
where $\tilde{\boldsymbol{\zeta}}_{\parallel,\boldsymbol{n}}(t)$, $\tilde{\boldsymbol{\zeta}}_{\perp,\boldsymbol{n}}(t)$,
$\tilde{\boldsymbol{\zeta}}_{\mathrm{z},\boldsymbol{n}}(t)$, and $\tilde{\boldsymbol{\zeta}}_{\mathrm{cr},\boldsymbol{n}}(t)$
are vector functions of time given by:
\begin{equation}\label{eqGEndDWOL2}
\begin{split}
\tilde{\boldsymbol{\zeta}}_{\parallel,\boldsymbol{n}}(t)
=
&
\cos^2\left(\frac{\beta}{2}\right)  \, \delta_{n_\mathrm{z},0} \,  \Bigg[\boldsymbol{\nabla}_{\boldsymbol{\alpha}}
f_\mathrm{y}(\boldsymbol{\alpha}_\mathrm{y} ; t )
\,  I_{n_\mathrm{y}}^\mathrm{DS}\left( k_\mathrm{L},t \right) \delta_{n_\mathrm{x},0}
-
\boldsymbol{\nabla}_{\boldsymbol{\alpha}} f_\mathrm{x}(\boldsymbol{\alpha}_\mathrm{x}; t ) \,
I_{n_\mathrm{x}}^\mathrm{DS}\left( k_\mathrm{L} ,t\right)  \delta_{n_\mathrm{y},0}\Bigg]\Big|_{\boldsymbol{\alpha}=\boldsymbol{0}} \: ,
\\
\tilde{\boldsymbol{\zeta}}_{\perp,\boldsymbol{n}}(t)
=
&
2 \, \delta_{n_\mathrm{z},0} \, \sin^2\left(\frac{\beta}{2}\right)  \Bigg[ \boldsymbol{\nabla}_{\boldsymbol
{\alpha}}
f_\mathrm{y}(\boldsymbol{\alpha}_\mathrm{y}; t ) \, I_{n_\mathrm{y}}^\mathrm{DS}\left( k_\mathrm{L} / 2 , t \right)
\delta_{n_\mathrm{x},0}
\\
&
+
\boldsymbol{\nabla}_{\boldsymbol{\alpha}} f_\mathrm{x}(\boldsymbol{\alpha}_\mathrm{x}; t )  \,
\left[ I_{n_\mathrm{x}}^\mathrm{D}\left( k_\mathrm{L} / 2 , t \right) \, \cos\theta +
I_{n_\mathrm{x}}^\mathrm{DS}\left(k_\mathrm{L} / 2 , t \right) \, \sin\theta \right] \delta_{n_\mathrm{y},0}
\Bigg]\Big|_{\boldsymbol{\alpha}=\boldsymbol{0}} \: ,
\\
\tilde{\boldsymbol{\zeta}}_{\mathrm{z},\boldsymbol{n}}(t)
=
&
2^{3/2} \frac{ \xi_\mathrm{z} }{  \sqrt{\pi} k_\mathrm{L} }
\Bigg[ \boldsymbol{\nabla}_{\boldsymbol{\alpha}} f_\mathrm{x}(\boldsymbol{\alpha}_\mathrm{x}; t ) \,
\frac{ l_\mathrm{x}}{  w_\mathrm{0,x}^2 } \tilde{I}_{n_\mathrm{x}}^\mathrm{D2}(t) I_{n_\mathrm{y}}^\mathrm{D2}(t)
+
\boldsymbol{\nabla}_{\boldsymbol{\alpha}} f_\mathrm{y}(\boldsymbol{\alpha}_\mathrm{y}; t )
\,\frac{ l_\mathrm{y} }{  w_\mathrm{0,y}^2 } \tilde{I}_{n_\mathrm{y}}^\mathrm{D2}(t)  I_{n_\mathrm{x}}^\mathrm{D2}(t) \Bigg]
\Big|_{\boldsymbol{\alpha}=\boldsymbol{0}}
\\
&
\times \left[\frac{1}{2}\:\delta_{n_\mathrm{z},0} + 2 \, \left(2^{3/2} \, i \, k_\mathrm{z} \, l_\mathrm{z}\right)^{n_\mathrm{z}}
\exp\left( -2 k_\mathrm{z}^2 \, l_\mathrm{z}^2\right) \, \delta_{n_\mathrm{z}, \mathbb{N}_\mathrm{g}} \right] \: ,
\\
\tilde{\boldsymbol{\zeta}}_{\mathrm{cr},\boldsymbol{n}}(t)
=
&
\frac{2}{k_\mathrm{L}} \, \sqrt{\frac{\xi_\mathrm{z}}{\pi}} \, \cos \left(\frac{\beta}{2}\right) \left[i \,
\sqrt{2}\: l_\mathrm{x}\left(4 l_\mathrm{x}^2 +  w_\mathrm{0,x}^2 \right)\, k_\mathrm{z} \, l_\mathrm{z}\right]^{n_\mathrm{z}}
\exp\left( -\frac{k_\mathrm{z}^2}{2} \, l_\mathrm{z}^2\right)  \, \delta_{n_\mathrm{z}, \mathbb{N}_\mathrm{g}}
\\
&
\times \Bigg\{
(-2^{3/2}) \Bigg[
- \sin \left(\frac{\phi}{2}\right) \frac{l_\mathrm{x} \boldsymbol{\nabla}_{\boldsymbol{\alpha}}
f_\mathrm{x}(\boldsymbol{\alpha}_\mathrm{x}; t )}{w_\mathrm{0,x}^2}  \, \tilde{I}_{n_\mathrm{x}}^\mathrm{D3S}
\left( k_\mathrm{L} , t \right) \,  I_{n_\mathrm{y}}^\mathrm{D2}(t) \\
&
+\cos\left(\frac{\phi}{2}\right) \frac{l_\mathrm{y} \boldsymbol{\nabla}_{\boldsymbol{\alpha}}
f_\mathrm{y}(\boldsymbol{\alpha}_\mathrm{y}; t )}{w_\mathrm{0,y}^2}  \, \tilde{I}_{n_\mathrm{y}}^\mathrm{D3}
\left(k_\mathrm{L}, t \right) \,  I_{n_\mathrm{x}}^\mathrm{D2} (t)
\\
&
 -
\sin \left(\frac{\phi}{2}\right) \frac{l_\mathrm{y} \boldsymbol{\nabla}_{\boldsymbol{\alpha}}
f_\mathrm{y}(\boldsymbol{\alpha}_\mathrm{y}; t )}{w_\mathrm{0,y}^2} \,
I_{n_\mathrm{x}}^\mathrm{D3S}(t) \, \tilde{I}_{n_\mathrm{y}}^\mathrm{D3} \left( 0 , t \right)
+
\cos \left(\frac{\phi}{2}\right) \frac{l_\mathrm{x} \boldsymbol{\nabla}_{\boldsymbol{\alpha}}
f_\mathrm{x}(\boldsymbol{\alpha}_\mathrm{x}; t )}{w_\mathrm{0,x}^2}  \,
I_{n_\mathrm{y}}^\mathrm{D3}(t) \, \tilde{I}_{n_\mathrm{x}}^\mathrm{D3} \left( 0 , t \right)
\Bigg]
\\
&
+
\Bigg[\cos \left(\frac{\phi}{2}\right) \, k_\mathrm{L} \, \boldsymbol{\nabla}_{\boldsymbol{\alpha}}
f_\mathrm{y}(\boldsymbol{\alpha}_\mathrm{y}; t )\, I_{n_\mathrm{y}}^\mathrm{D3}(t) \, I_{n_\mathrm{x}}^\mathrm{D2}(t)
+
\sin \left(\frac{\phi}{2}\right) \, k_\mathrm{L} \, \boldsymbol{\nabla}_{\boldsymbol{\alpha}}
f_\mathrm{x}(\boldsymbol{\alpha}_\mathrm{x}; t ) \, I_{n_\mathrm{x}}^\mathrm{D3S}(t) \, I_{n_\mathrm{y}}^\mathrm{D2}(t) \Bigg]
\Bigg\}\Big|_{\boldsymbol{\alpha}=\boldsymbol{0}} \: .
\end{split}
\end{equation}
The remaining integral over time can only be evaluated numerically, similarly to the evaluation of the first auxiliary function 
$G_\mathbf{n}$ [cf. Eq.~\eqref{eqGEndDWOL1} in Sec.~\ref{SubSecDWOLXInt}],

\section{Derivation of the expression for $\hat{T}_{\mathbf{r},\mathbf{p}}(\delta t)$} \label{derivationTrp}
In the following we derive the expression for $\hat{T}_{\mathbf{r},\mathbf{p}}(\delta t)$, defined in Eq.~\eqref{TrpDef},
using the Baker-Campbell-Hausdorff (BCH) formula and its special case known as the Weyl identity~\cite{GilmoreBOOK:12}.

We start by noting that $\hat{T}_{\mathbf{r},\mathbf{p}}(\delta t)$ can be rewritten as a product of three terms, 
each of which involves the coordinate- and momentum operators corresponding to one spatial direction:
\begin{equation}\label{Trp3D}
\hat{T}_{\mathbf{r},\mathbf{p}}(\delta t)=\hat{T}_{x,p_x}(\delta t)
\hat{T}_{y,p_y}(\delta t)\hat{T}_{z,p_z}(\delta t) \:.
\end{equation}
Therefore, it is sufficient to derive the expression for one of these operators, for instance 
\begin{equation}\label{exprTxpx1}
\hat{T}_{x,p_x}(\delta t)=\exp\left(-\frac{i}{\hbar}\left[\frac{\hat{p}_x^2}{2m}\:\delta t
+\hat{x}\int_t^{t+\delta t} m\ddot{q}_{0,x}(t') \mathrm{d} t'\right]\right) \:,
\end{equation}
and the remaining two can be derived in an analogous fashion. By realizing that 
\begin{equation}
\int_t^{t+\delta t}\ddot{q}_{0,x}(t')dt'\equiv\dot{q}_{0,x}(t+\delta t)-\dot{q}_{0,x}(t) \:,
\end{equation}
and introducing $\delta\dot{q}_{0,x}(t)\equiv\dot{q}_{0,x}(t+\delta t)-\dot{q}_{0,x}(t)$,
Eq.~\eqref{exprTxpx1} can be rewritten more succinctly as
\begin{equation}\label{exprTxpx2}
\hat{T}_{x,p_x}(\delta t)=\exp\left(-\frac{i}{\hbar}\left[\frac{\hat{p}_x^2}{2m}\:\delta t
+m\delta\dot{q}_{0,x}(t)\:\hat{x}\right]\right) \:.
\end{equation}

At this point, we invoke the BCH formula~\cite{GilmoreBOOK:12}
\begin{equation}\label{BCHfla}
e^{\hat{A}}e^{\hat{B}}=e^{\hat{A}+\hat{B}+\frac{1}{2}\:[\hat{A},\hat{B}]
+\frac{1}{12}\:[\hat{A},[\hat{A},\hat{B}]]+\frac{1}{12}\:[\hat{B},[\hat{B},
\hat{A}]]+\ldots} \:,
\end{equation}
where the ellipses in the exponent on the RHS of the last equation indicate the higher-order 
repeated commutators of $\hat{A}$ and $\hat{B}$. In what follows, we will use the last formula for 
$\hat{A}=-i\delta t\:(2m\hbar)^{-1}\:\hat{p}_x^2$ and $\hat{B}=-im\hbar^{-1}\delta\dot{q}_{0,x}(t)\:\hat{x}$.

We first make use of the fact that $[\hat{p}_x^2,\hat{x}]=-2i\hbar\hat{p}_x$, and that, accordingly,
$[\hat{p}_x^2,[\hat{p}_x^2,\hat{x}]]=0$ and $[\hat{x},[\hat{x},\hat{p}_x^2]]=-4\hbar^2$. Thus,
we first find that $[\hat{A},\hat{B}] = i \hbar^{-1}\:\delta t \delta\dot{q}_{0,x}(t)\hat{p}_x$, which along with 
the basic commutator $[\hat{x},\hat{p}_x]=i\hbar$ implies that the second-order commutators of $\hat{A}$ 
and $\hat{B}$ are given by $[\hat{A},[\hat{A},\hat{B}]] = 0$ and $[\hat{B},[\hat{B},\hat{A}]]=-2i\hbar^{-1}\:\delta t 
[\delta\dot{q}_{0,x}(t)]^2$. The fact that the last commutator does not depend on $\hat{x}$ and $\hat{p}_x$ has 
two important implications. Firstly, all higher-order commutators are equal to zero, therefore for 
our choice of operators $\hat{A}$ and $\hat{B}$ there are no additional terms in the exponent on the RHS of 
Eq.~\eqref{BCHfla}. Secondly, the commutator $[\hat{B},[\hat{B},\hat{A}]]$ itself gives rise only to a time-dependent
global phase factor of the kind that is immaterial for our treatment of atom transport [recall similar 
arguments used for dropping terms in Eq.~\eqref{eqSchroedingerequationTrapFrame} of Sec.~\ref{ComovingFrameFSOM}] 
and can henceforth be disregarded. Therefore, we have 
\begin{equation} 
\hat{T}_{x,p_x}(\delta t) = \exp\left(-\frac{i}{\hbar} \frac{\hat{p}_x^2}{2m}\:\delta t\right) \,
\exp\left[-\frac{im\:\delta\dot{q}_{0,x}(t)}{\hbar}\:\hat{x} \right]
\exp\left(\frac{1}{4 \hbar^2} \left[\hat{p}_x^2,\hat{x}\right]\delta t 
\delta\dot{q}_{0,x}(t)\right)\:,
\end{equation}
which further reduces to
\begin{equation} \label{TxpExpr}
\hat{T}_{x,p_x}(\delta t) = \exp\left(-\frac{i}{\hbar} \frac{\hat{p}_x^2}{2m}\:\delta t\right) \,
\exp\left[-\frac{im\delta\dot{q}_{0,x}(t)}{\hbar}\:\hat{x}\right]
\exp\left[-\frac{i\delta t\delta\dot{q}_{0,x}(t)}{2\hbar}\:\hat{p}_x\right] \:.
\end{equation}

Our goal is to reverse the order of the last two exponential operators on the RHS of the last equation, 
such that we can group all the kinetic terms together. To this end, we make use of the Weyl identity~\cite{GilmoreBOOK:12},
which is a special case of the BCH formula in Eq.~\eqref{BCHfla} for arbitrary operators $\hat{A}$ and $\hat{B}$
that commute with their commutator $[\hat{A},\hat{B}]$ (i.e., $[\hat{A},[\hat{A},\hat{B}]]=[\hat{B},[\hat{A},\hat{B}]]=0$):
\begin{equation}\label{WeylAB}
e^{\hat{A}}e^{\hat{B}}=e^{\hat{A}+\hat{B}}e^{\frac{1}{2}\:[\hat{A},\hat{B}]} \:.
\end{equation}
By exchanging the operators $\hat{A}$ and $\hat{B}$, we also obtain
\begin{equation}\label{WeylBA}
e^{\hat{B}}e^{\hat{A}}=e^{\hat{A}+\hat{B}}e^{-\frac{1}{2}\:[\hat{A},\hat{B}]} \:.
\end{equation}
Finally, by comparing Eqs.~\eqref{WeylAB} and \eqref{WeylBA}, we find that for operators 
$\hat{A}$ and $\hat{B}$ that commute with their commutator, one has
\begin{equation}
e^{\hat{B}}e^{\hat{A}}=e^{\hat{A}}e^{\hat{B}}\:e^{-[\hat{A},\hat{B}]} \:.
\end{equation}
We utilize the last identity for the operators
\begin{eqnarray}
\hat{A} &=& -\frac{im\delta\dot{q}_{0,x}(t)}{\hbar}\:\hat{x}  \:,\\
\hat{B} &=& -\frac{i\delta t\delta\dot{q}_{0,x}(t)}{2\hbar}\:\hat{p}_x \:,
\end{eqnarray}
whose commutator is given by $[\hat{A},\hat{B}]=-im\:(2\hbar)^{-1}\:\delta t\:[\delta\dot{q}_{0,x}(t)]^2$. 
The last commutator not only commutes with both $\hat{A}$ and $\hat{B}$, but also does not depend on 
the operators $\hat{x}$ and $\hat{p}_x$ and is therefore immaterial for our present purposes. Therefore, 
the order of the last two terms on the RHS of Eq.~\eqref{TxpExpr} can be reversed at the expense
of an unimportant global phase factor. Thus, we can finally recast Eq.~\eqref{TxpExpr} as
\begin{equation}
\hat{T}_{x,p_x}(\delta t)=
\exp\left[-\frac{i}{\hbar}\frac{\hat{p}_x^2}{2m}\:\delta t \right] \,
\exp\left[-\frac{i}{2\hbar}\:\hat{p}_x \, \delta t \,  \delta\dot{q}_0(t)\right] \,
\exp\left[-\frac{i}{\hbar}\:m\, \delta\dot{q}_{0,x}(t) \, \hat{x} \right] \:.
\end{equation}
The generalization of the last result to three dimensions, i.e. recovering the full expression 
for $\hat{T}_{\mathbf{r},\mathbf{p}}(\delta t)$ based on Eq.~\eqref{Trp3D}, is straightforward.
We finally obtain 
\begin{eqnarray} \label{TrpExprFinal}
\hat{T}_{\mathbf{r},\mathbf{p}}(\delta t) = \exp\left[-\frac{i}{\hbar} 
\frac{\hat{\mathbf{p}}^2}{2m}\:\delta t \right] \:\exp\left[-\frac{i}{2\hbar}\:\hat{\mathbf{p}}
\cdot\delta\dot{\mathbf{q}}_0(t)\:\delta t\right]\:\exp\left[-\frac{i}{\hbar}\:m
\hat{\mathbf{r}}\cdot\delta\dot{\mathbf{q}}_0(t)\right] \:,
\end{eqnarray}
where $\delta\dot{\mathbf{q}}_0(t)\equiv \dot{\mathbf{q}}_0(t+\delta t)-\dot{\mathbf{q}}_0(t)$.

\twocolumngrid

\end{document}